\documentclass[pdflatex,a4paper,11pt]{article}
\usepackage[english]{babel}
\usepackage{csquotes}
\usepackage{natbib}

\usepackage{bm}
\usepackage{graphicx}
\usepackage{xcolor}

\usepackage{multirow}%
\usepackage{amsmath,amssymb,amsfonts}%
\usepackage{amsthm}%
\usepackage{mathrsfs}%
\usepackage{xcolor}%
\usepackage{textcomp}%
\usepackage{manyfoot}%
\usepackage{booktabs}%
\usepackage{algorithm}%
\usepackage{algorithmicx}%
\usepackage{algpseudocode}%
\usepackage{listings}%

\usepackage{layout}
\usepackage{geometry}
\usepackage[indent=10pt]{parskip}
\usepackage{palatino}
\usepackage{enumitem}
\usepackage[bottom]{footmisc}
\usepackage{hyperref}

\hypersetup{
    colorlinks=true, 
    citecolor=blue, 
    linkcolor=blue, 
    urlcolor=blue 
}

\usepackage{cleveref}
\crefname{appendix}{Appendix}{Appendices}
\Crefname{appendix}{Appendix}{Appendices}
\crefname{figure}{Fig.}{Figs.}
\Crefname{figure}{Fig.}{Figs.}

\usepackage{ctable}
\usepackage[labelsep=none]{caption}
\renewcommand{\figurename}{Figure}
\usepackage{subcaption}

\renewcommand{\tablename}{Table}
\begin{document}

\title{\textsc{Sustainability of cities under declining population and decreasing distance frictions: The case of Japan}}
\author{Tomoya Mori and Daisuke Murakami\thanks{{\bf Mori:} (1) Institute of Economic Research, Kyoto University, Yoshida-Honmachi, Sakyo-Ku, Kyoto, 606-8501 Kyoto, Japan. (2) Research Institute of Economy, Trade and Industry (RIETI), 11th floor 1-3-1, Annex, Ministry of Economy, Trade and Industry, Kasumigaseki, Chiyoda-Ku, 100-8901 Tokyo, Japan. E-mail: mori@kier.kyoto-u.ac.jp. {\bf Murakami:} Institute of Statistical Mathematics, 10-3 Midori-Cho, Tachikawa, 190-8562, Tokyo, Japan. E-mail: dmuraka@ism.ac.jp. 
The earlier versions of this paper were presented at the 37th ARSC meeting in Osaka, the 13th UEA meeting in Copenhagen, Infrastructure and Urban Development in the Developing World in Tokyo, the 18th UEA meeting in Washington D.C., RIETI and Hitotsubashi University. The authors are greatful to Barth\'{e}l\'{e}my Bonadio, Gilles Duranton, Megan Haasbroek, Keisuke Kondo, Chihiro Shimizu, Shinichiro Umeya, Jens Wrona, and seminar and conference participants for their helpful comments. 
This research was conducted as part of the project, “Development of Quantitative Framework for Regional Economy Based on the Theory of Economic Agglomeration,'' undertaken at the Research Institute of Economy, Trade and Industry (RIETI). We acknowledge financial support from the Kajima Foundation, the Mitsubishi Foundation, Grant-in-Aid for Scientific Research (A) 25H00543 and (S) 24H00012, and the Joint Research Program of KIER, Kyoto University. 
}}
\date{\today}

\maketitle
\thispagestyle{empty}
\setcounter{page}{0}

\begin{abstract}
\noindent
This study develops a statistical model that integrates economic agglomeration theory and power-law distributions of city sizes to project future population distribution on 1-km grid cells. We focus on Japan -- a country at the forefront of rapid population decline.
Drawing on official population projections and empirical patterns from past urban evolution in response to the development of high-speed rail and highway networks, we examine how ongoing demographic contraction and expected reductions in distance frictions may reshape urban geography. Our analysis suggests that urban economies will consolidate around fewer and larger cities, each of which will experience a flattening of population density as the decentralization of urban populations accelerates, while rural areas are expected to experience further depopulation as a result of these spatial and economic shifts. By identifying sustainable urban cores capable of anchoring regional economies, our model provides a framework for policymakers to manage population decline while maintaining resilience through optimized infrastructure and resource allocation focused on these key urban centers.
\end{abstract}


\medskip



\maketitle

\thispagestyle{empty}
\clearpage

\setcounter{page}{1}

\section{Introduction}
\label{sec1}

The world is undeniably facing trends of aging, declining birth rates and shrinking populations \citep[][]{UN-2022}.
In recent decades, these demographic shifts have been especially acute in Asian countries where immigration has not significantly offset natural decreases. 
Japan stands at the forefront of this phenomenon: its population peaked at 128 million in 2008 (\citeauthor{MIC2008} (hereafter MIC), \citeyear{MIC2008}) and has declined every year since  (Fig.\,\ref{fig:total-pop}).
As of October 2024, Japan’s total population fell to 123.8 million, marking the fourteenth consecutive year of decline, with this continuous decrease beginning in 2011 (MIC, \citeyear{MIC2024}). 
The total fertility rate, after peaking modestly at 1.45 in 2015, has since fallen to 1.20 in 2023, with little sign of recovery  (Fig.\,\ref{fig:fertility}).%
\footnote{While there is no compelling explanation for the increase in the fertility rate from 1.26 to 1.45 between 2005 and 2015, it is often attributed to the improvement in economic conditions \citep[][]{Fujinami-RF2020} during this period.}\ 
Meanwhile, China and South Korea have also entered phases of population decline, and similar trends are expected in Thailand and the Philippines in the near future \citep{UN-2022}.

This demographic shift is not limited to Asia; many countries outside the region are also experiencing sustained declines in fertility. Across the OECD, the average total fertility rate fell to just 1.58 by 2021, well below the replacement level of 2.1, and in most member countries now ranges between 1.2 and 1.8 \citep[][]{OECD-FDB2023}.%
\footnote{For example, the fertility rate of native-born Germans fell to 1.26 in 2023, according to \href{https://www.destatis.de/EN/Press/2024/07/PE24_274_12.html}{the recent press release} from the German Federal Statistical Office.}\ 
This persistent decline underscores that population aging and shrinkage are becoming global phenomena, with few exceptions such as Israel maintaining fertility rates above replacement level \citep[][]{OECD-FDB2023}.
A recent study by \cite{IHME-Lancet2024} projects that by 2100, only six countries in sub-Saharan Africa will maintain fertility rates above the replacement level, while all others worldwide will fall below this threshold.
As a result, the longstanding assumption that immigration can offset population decline in Europe and the United States is becoming untenable; the countries that have traditionally supplied migrants will themselves face demographic contraction, raising concerns about their own long-term sustainability.

\begin{figure}[h!]
 \centering
 \captionsetup{width=\linewidth}
 
 \includegraphics[width=.55\textwidth]{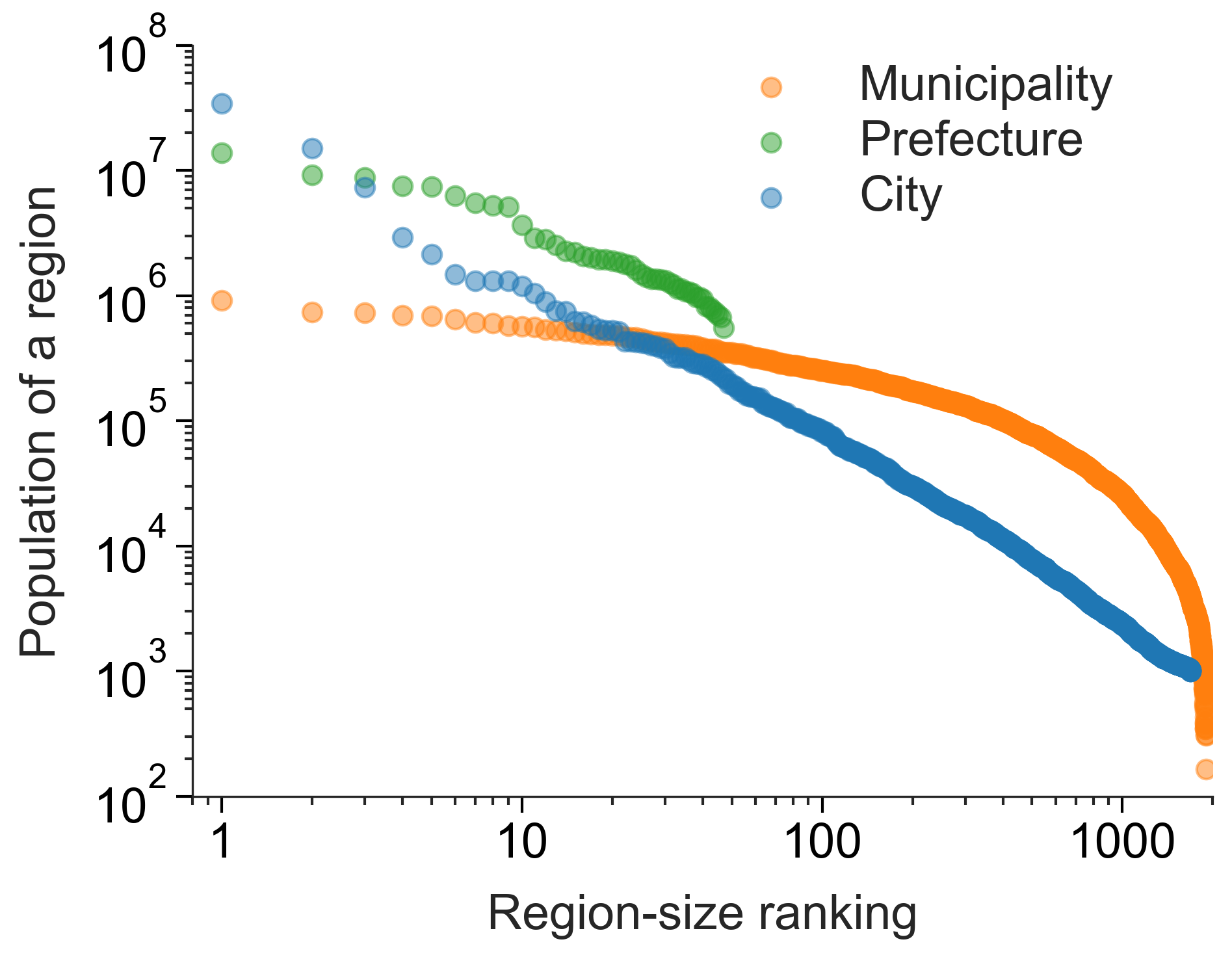}
 \caption{The size distributions of prefectures, municipalities and cities in Japan}
 \caption*{\footnotesize\textit{Note}: Each city in Japan in 2020 was identified as the set of 1km grid cells that (i) have a population density of at least 1,000 per square kilometer, (ii) are geographically contiguous, and (iii) have a total population of at least 1,000. 
Population count data in 30''$\times$45'' grids were obtained from the Grid Square Statistics of the 2020 Population Census of Japan. The graphs include 46 prefectures, 1,907 municipalities, and 1,965 cities in the Japanese archipelago that are connected by road to the four major islands of Japan, Hokkaido, Honshu, Shikoku, and Kyushu in 2020.
The horizontal axis shows the ranking of regions by their population size, and the vertical axis shows their population size.}
 \label{fig:region-size-dist}
\end{figure}

This paper examines how a country's economic geography evolves in an era of rapid population decline.
Given the persistent global trend toward urbanization,%
\footnote{
Under the United Nations' definition of urbanization, the global urbanization rate has risen steadily from 29.6\% in 1950 to 56.2\% in 2020, and is projected to reach 68.4\% by 2050. Many countries in Europe, the United States, and Japan have already exceeded 70\% as of 2020 and are expected to reach 80--93\% by 2050 \citep[][]{UN-2018}.}\ 
it is natural to focus on the dynamnics of urban agglomeration. 
Notably, a substantial body of empirical research demonstrates that the distribution of city sizes within a country generally follows a power law \citep[e.g.,][]{Duranton-Puga-HB2014,Gabaix-Ioannides-HB2004}, a pattern not observed for other regional aggregates such as administrative units (Fig.,\ref{fig:region-size-dist}). 
Building on this robust empirical regularity and the theory of economic agglomeration, we develop a reduced-form forecasting model to predict the future geographic distribution of population at the 1-km grid level in Japan.

%
%
%
%
%
%


We focus on Japan, a country that has experienced the longest sustained period of population decline in recent history and is projected to undergo even more rapid shrinkage in the coming decades (\citeauthor{NIPSSR-2023}, hereafter NIPSSR, \citeyear{NIPSSR-2023}). 
Japan also offers an ideal context for examining the effects of decreasing distance frictions—a defining trend of recent decades—as technological advances in automated driving, logistics, and virtual reality are anticipated to further reduce the barriers of distance. These changes are expected to significantly influence both the size and spatial distribution of cities.

Over the past half century, Japan has undergone a significant but gradual reduction in transport costs, driven by the development and expansion of nationwide high-speed rail and expressway networks. 
Since the inauguration of the Tokaido Shinkansen (bullet train) in 1964 and the opening of the Meishin and Tomei Expressways in the 1960s (see Fig.\,\ref{fig:tcost}), connecting the largest two cities, Tokyo and Osaka, these networks have steadily grown to connect all major regions of the country, fundamentally transforming intercity accessibility.
Today, more than half of Japan’s population can reach Tokyo within a single day’s journey.%
\footnote{Japan’s core economic regions (Fig.\,\ref{fig:cities-2020}), from Fukuoka in the west to Sapporo in the east—are within the range of 1,000 kilometers from Tokyo. In this distance range, land transport remains dominant, with road and rail accounting for over 85\% of passenger flows  (MLIT, \citeyear{MLIT-2010}).
}\ 
As a result, changes in urban agglomeration patterns over the past fifty years can be seen as a response to the expansion of high-speed rail and expressways, reflecting the comparative statics of declining transport costs. In parallel, the widespread diffusion of the Internet has further reduced communication costs nationwide.

Population decline essentially shifts the power-law distribution of city sizes downward, leading to the disappearance of a greater number of smaller cities in the future. In contrast, the effects of declining transport costs are more complex. According to economic agglomeration theory, lower transport costs promote the concentration of population in larger and fewer cities \citep[][]{Akamatsu-et-al-DP2024}. This is because improved connectivity intensifies competition, making it difficult for small cities with limited local markets to survive, and their populations are subsequently absorbed by larger urban centers. As a result, larger cities that are geographically more distant from each other—and thus less likely to compete directly—are more likely to persist \citep[][]{Fujita-Krugman-Mori-EER1999, Tabuchi-Thisse-JUE2011, Hsu-EJ2012, Mori-et-al-DP2023}. This increased concentration in larger cities alters the power-law coefficient of the city size distribution, such that the logarithm of city population size declines more steeply with respect to the logarithm of city size rank.

The cities that persist, however, are likely to experience a flattening of their internal population distribution, characterized by declining population densities in and around the central business district (CBD). For example, as remote work becomes more prevalent, workers who once commuted to the CBD five days a week may now do so only two or three days, spending the remainder of their workweek at home. This shift incentivizes households to seek larger, more affordable homes in suburban areas, even at the cost of a longer commute. A similar logic applies to firms: as face-to-face transactions become less essential, the locational advantages of the CBD diminish, prompting businesses to consider relocating to suburban environments \citep{Akamatsu-et-al-DP2024,Osawa-Akamatsu-JET2020}.%
\footnote{The evolution of cities is shaped by a variety of factors beyond the reduction of transport and communication costs. For example, regional differences in age composition can lead to divergent migration patterns across areas \citep[e.g.,][]{Oizumi-et-al-PO2022}. Nevertheless, economic agglomeration theory remains a powerful framework for understanding urban change, as it can simultaneously account for the observed national-level concentration in larger cities, the flattening of population distributions within individual cities, and the persistence of the power law in city size distributions over the study period \citep[e.g.,][]{Akamatsu-et-al-DP2024,Mori-et-al-DP2023}.}\ 



Our forecasting model has three geographic levels: the 1 km grid cell level, the city level, and the country level. To predict the population growth and decline of each grid cell, we ensemble three standard time-series models and their extensions, which account for population growth in the neighboring cells.
We use the Grid Square Statistics of the Population Census of Japan, which provides us with the accurate population distribution over 1-km grid cells in Japan every five years between 1970 and 2020.
Each time series model is estimated to fit the evolution of the population size of each grid cell. All six models are estimated and ensembled to predict the future population growth of a given grid cell. 

Given the population distribution across grid cells, we identify cities as contiguous clusters of 1-km grid cells that each have a minimum population density of 1,000 persons per square kilometer and together comprise a total population of at least 10,000. This flexible definition allows city boundaries to evolve over time, expanding or contracting in response to demographic changes and overall population decline, as cells are added or removed based on whether they meet the density threshold. As a result, the geographic extent of each city adapts dynamically to shifting local and national conditions.%
\footnote{The projection results are not sensitive to the choice of specific thresholds for population density and minimum city size, unless they are set too low.
Although we follow \cite{Mori-et-al-PNAS2020}, our particular choice of thresholds results in cities accounting for 80\% of the total population in the base year 2020, which seems to be a reasonable benchmark case.
This simple bottom-up definition of a city based on population agglomeration allows us to easily identify the future distribution of cities and has been shown to exhibit a persistent power law for the city size distribution in a country and its recursively nested subregions in six countries, China, France, Germany, India, Japan, and the United States \citep[][]{Mori-et-al-PNAS2020}.
In related research, \cite{Rozenfeld-et-al-PNAS2008} has shown that a similar definition of cities as population agglomerations accounts for the power law for city size distribution when the grid population exhibits a random growth process.}\ 
We employ three standard time series models—mirroring those used at the grid cell level—to capture the evolution of each city's population.

At the country level, we impose three key conditions in our model. First, we take the official projection of the total population as given. Second, we assume that urbanization will continue in line with the trend observed between 1970 and 2020, during which the urban population share increased from 66\% to 80\%, and is projected to reach 90\% by 2120.\footnote{Specifically, the urban population share increased from 66\% in 1970 to 80\% in 2020, and is expected to reach 90\% in 2120.} Third, we maintain that the city size distribution will continue to approximately follow a power law, meaning that each city's population growth at a given time depends on its relative population size ranking. Assuming that this distribution will become increasingly skewed toward larger cities, as has been the case over the past half century, our power-law-based model predicts relatively faster population growth for larger cities compared to smaller ones. By combining this country-level adjustment with city-specific growth factors, which reflect, for example, differences in age and gender composition, we ensemble the power-law and time-series models to generate projections for each city's future population size. 

Using the models estimated from the 1970–2020 data, we project the geographic distribution of population across grid cells in five-year increments and re-identify cities at each step. At every projection interval, we ensure consistency among the projected total population, the urban population share, the population size of each city, and the population of each grid cell. Once this consistency is achieved, we proceed to the next five-year period. By repeating this process iteratively, we generate projections of the spatial distribution of population across grid cells through to 2120.

Our model predicts a substantial contraction of Japan’s city system, with each city's trajectory explicitly projected. The number of cities in each size class is expected to decline by more than one-third over the next 50 years and by more than half within 100 years. This consolidation is implied by the underlying theory and is driven both by population decline and by diminishing distance friction, which will disproportionately enhance the competitiveness of larger cities while accelerating the disappearance of smaller ones, leaving rural regions increasingly isolated. Furthermore, as population aging progresses more rapidly in eastern Japan, the population centroid is expected to shift westward. By identifying sustainable urban cores capable of anchoring regional economies, our model provides a framework for policymakers to manage population decline while maintaining resilience through optimized infrastructure and resource allocation focused on these key urban centers.

There is a substantial body of literature on gridded population projections that downscale national-level forecasts under various scenarios, such as the Shared Socioeconomic Pathways (SSPs) \citep{SSP-GEC2017}. Historical spatial patterns of gridded populations have been extrapolated for the projection. For example, gravity-based models have been employed to allocate populations across grid cells \citep{ONeil-et-al-CC2014, Jones-ONeil-ERL2016}, while other studies have incorporated multiple auxiliary variables—such as land cover, road length, and distance to cities—into spatial allocation models \citep{McKee-et-al-PNAS2015}, or machine learning techniques like random forests \citep{Stevens-et-al-PO2015, Wan-et-al-CGIS2022}. Spatial econometric models have also been developed to capture interactions among cities and urbanization potential \citep{Murakami-Yamagata-S2019}, and stochastic growth models have been used to allocate national populations to grid cells based on socioeconomic variables such as natural endowments, industrial structure, and political centrality \citep{Nam-Reilly-US2013}. In Japan, the NIPSSR has produced short-term municipal-level projections by downscaling official national forecasts, taking into account recent trends in natural and social population change as well as the age and gender structure of each municipality (NIPSSR, \citeyear{NIPSSR-Region-2023} \citeyear{NIPSSR-2023}).

Our approach differs from previous studies in two important respects. First, our projections are grounded in explicit theoretical implications regarding future changes in both the distribution of population across cities and within individual cities. Second, we incorporate the power law for city size distribution as a persistent empirical pattern expected to hold in the future, while allowing the power law coefficient to evolve in line with trends observed over recent decades.

The remainder of the paper is organized as follows. Section \ref{sec:theory} documents the strong correspondence between the observed evolution of cities and the implications of economic agglomeration theory, which forms the basis of our forecasting model. Section \ref{sec:data} describes the data sources, while Section \ref{sec:model} and Section \ref{sec:procedure} present the model and the projection methodology, respectively. In Section \ref{sec:results}, we report our projections of city growth and decline through 2120. Section \ref{sec:implications} discusses the policy implications of our findings, particularly in terms of infrastructure planning and regional economic resilience. Finally, Section \ref{sec:conclusion} concludes the paper and suggests directions for future research.

\section{Facts and theory about the evolution of cities in Japan}
\label{sec:theory}

In this section, we interpret the geographic distribution of the population and its changes over the past half century through the lens of economic agglomeration theory.
Our main data for estimating the model are the population counts in 30''$\times$45'' grids (about 1 km by 1 km) obtained from the Grid Square Statistics of the Population Census of Japan by MIC for every five years from 1970 to 2020.

\begin{figure}[htbp!]
    \centering
    \begin{minipage}[c]{\textwidth}
     \centering

    \captionsetup{width=\linewidth}
    
    \includegraphics[width=.7\textwidth]{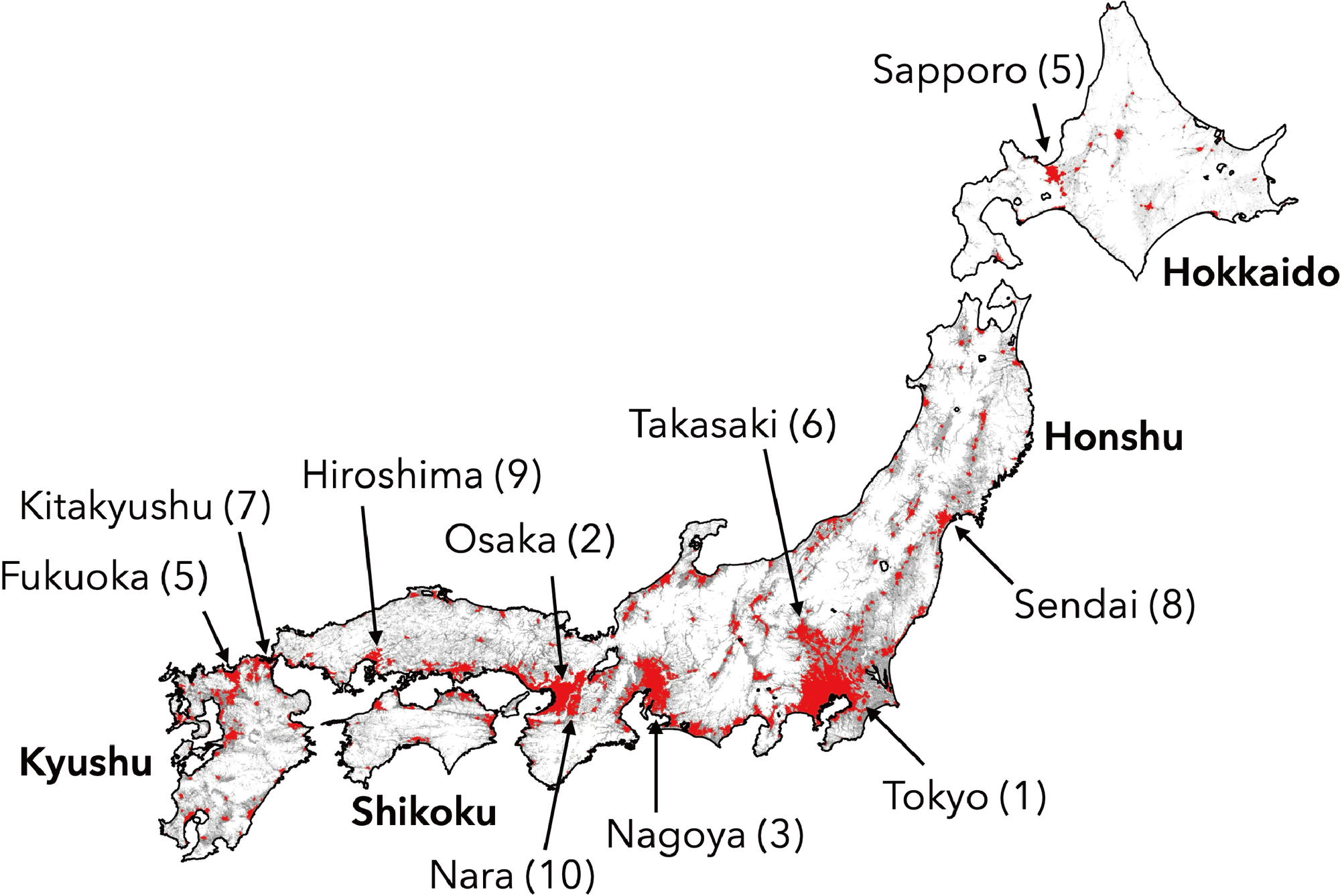}
    \bigskip
    
    \caption{Cities as urban agglomerations in 2020}
    \caption*{\footnotesize\textit{Note}: Each disjoint red area is a city as an urban agglomeration in Japan in 2020. To simplify the analysis, we focus on the part of Japan connected by road to Honshu (the main island) and Hokkaido (the northern island). The 10 largest cities in 2020 are listed, with the numbers in parentheses indicating their ranking by population size.}
    \label{fig:cities-2020}
    \end{minipage}
 
    \vspace{1cm}

    \begin{minipage}[c]{\textwidth}
    
    \centering
    \captionsetup{width=\linewidth}
    
    \includegraphics[width=\textwidth]{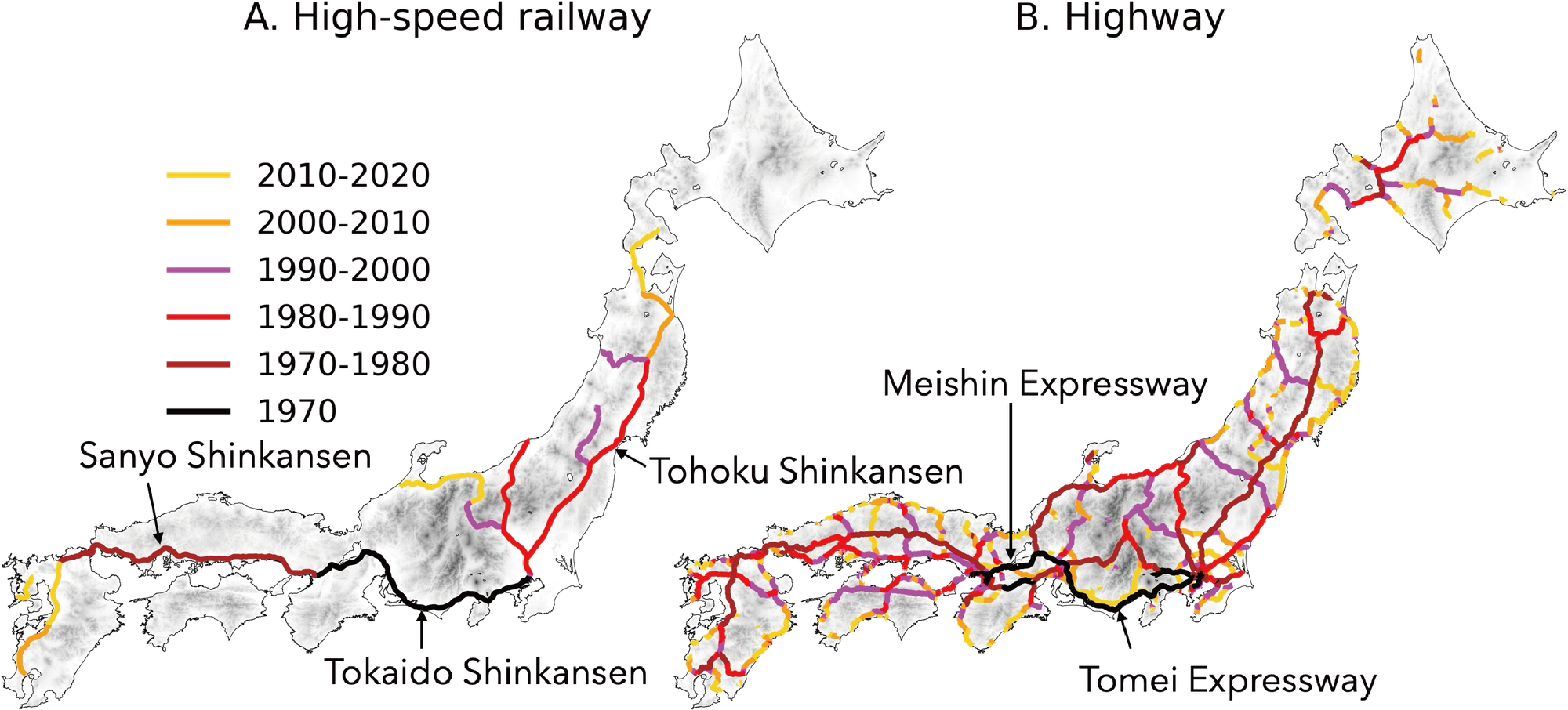}
    \caption{The development of high-speed transportation networks in Japan between 1970 and 2020.}
    \caption*{\footnotesize\textit{Note}: The shapefiles of the transport networks are obtained from the National Land Numerical Information Download Service by the Ministry of Land, Infrastructure, Transport and Tourism of Japan (\url{https://nlftp.mlit.go.jp/}).}
    \label{fig:tcost}
    \end{minipage}

\end{figure}

Throughout the paper, we focus on the part of Japanese archepelago which are connected by roads to either Honshu island or Hokkaido island (the coverage is shown in Fig.\,\ref{fig:cities-2020}), which consists of 365,470 grid cells. 
Each city is identified as an \textit{Urban Agglomeration} (\textit{UA}) defined by a spatially contiguous set of grid cells with a population density of at least 1,000 per 1 $\mathrm{km}^2$ and a total population of at least 10,000. 
The red areas in Fig.\,\ref{fig:cities-2020} indicate the 431 UAs in 2020 identified using this definition. These UAs account for 80\% of Japan's population while occupying 6\% of the total land area.
The urban share of the population has continued to increase and the location of the population is essentially dominated by cities.


\begin{figure}[hbtp!]
    \centering
    \includegraphics[width=\textwidth]{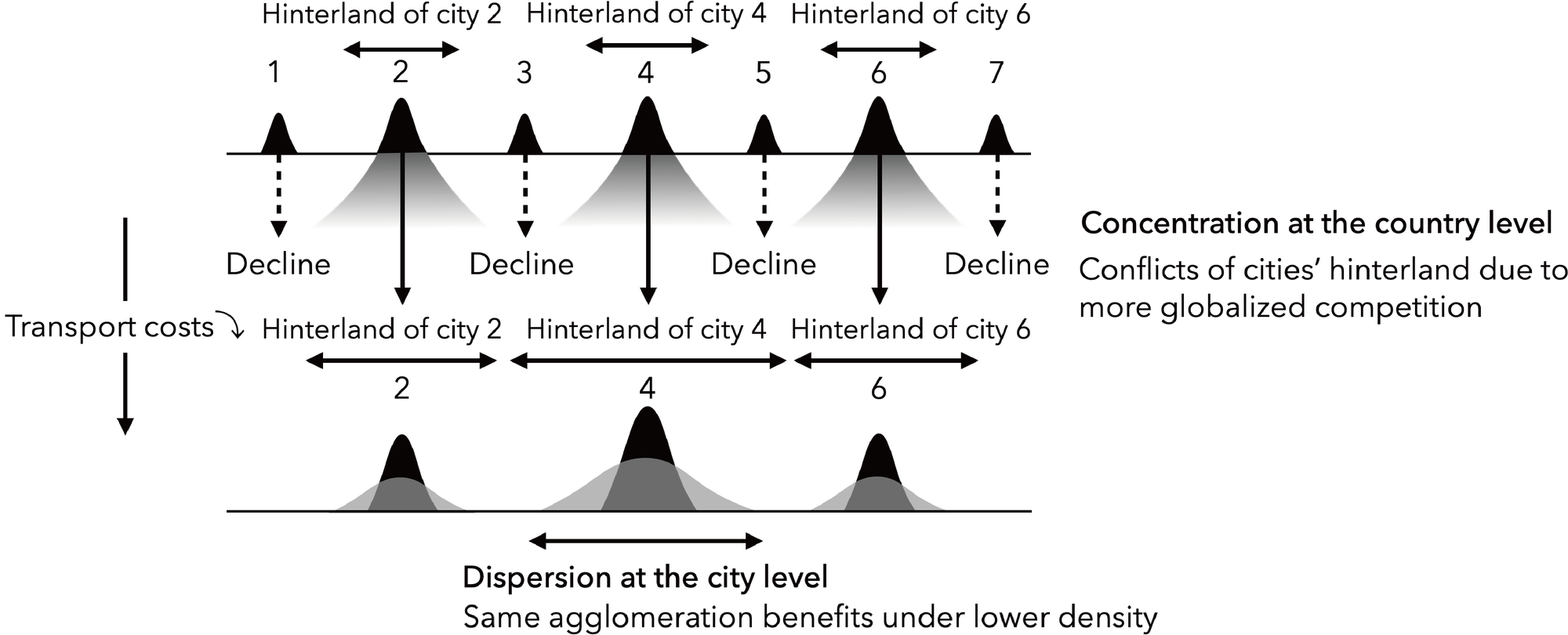}
    \smallskip
    
    \caption{Cities' response to the reduced transport costs}
    \label{fig:theory}
    \caption*{\footnotesize\textit{Note}: The comparative static results illustrated are based on \citeauthor{Akamatsu-et-al-DP2024} (\citeyear{Akamatsu-et-al-DP2024}, Fig.\,7) and \citeauthor{Osawa-Akamatsu-JET2020}(\citeyear{Osawa-Akamatsu-JET2020}, Fig.\,E7).}
\end{figure}

Japan's population has been growing since the beginning of time, except for short-term declines due to wars and natural disasters, and reached 104 million in 1970 at the beginning of our study period.
However, after reaching its peak of 128 million in 2010 (among the cenusus years), it began to decline and reached 126 million in 2020.
Thus, the 1970-2020 period includes this turning point.

It was also during this period that Japan's high-speed railways and highways were developed almost from scratch into full-fledged nationwide networks.
Construction of Japan's high-speed railways and highways began in time for the 1964 Tokyo Olympics.
The total length of high-speed railways increased from 515 km in 1964 to 3,067 km in 2020, and the total length of highways increased from 216 km to 3,067 km over the same period, significantly reducing transport costs in Japan (Fig.\,\ref{fig:tcost}).
Moreover, this expansion of the network occurred gradually (see Appendix \ref{app:network}).
Meanwhile, the Internet was introduced in the 1990s and became ubiquitous throughout the country in the 2000s and beyond, significantly reducing communication costs.

\cite{Akamatsu-et-al-DP2024} have shown that a general class of microeconomic models of economic agglomeration in the literature exhibit the response to the reductions in transport costs shown in Figure \ref{fig:theory}.%
\footnote{Here, we exclude the large literature on the systems-of-cities models \citep[see, e.g.,][]{Duranton-Puga-HB2014}, since they cannot address the geographic distribution of cities and has little to say about the location of cities as they abstract from inter-city space.}
On the one hand, at the country level, competition becomes more global, and the market areas of firms in neighboring cities begin to overlap.
As a result, the economic hinterlands of these cities begin to interfere with each other, and the cities with smaller home markets are squeezed out.
The surviving cities are farther apart, and grow in size by attracting population from the declining cities.
At the city scale, on the other hand, both firms and workers have less incentive to concentrate near the city center and disperse in search of cheaper space.%
\footnote{Think about a situation where a worker who used to commute to work five days a week now commutes two or three days a week and does the rest of their work remotely. Even if their commute time increases, they will have the motivation to move to the suburbs and rent or buy a larger home more cheaply. The same is true for companies, and if there is no longer a need to conduct all transactions face-to-face, the benefits of being in the city center will decrease.}
As a result, the spatial distribution of population in cities becomes flatter and more sprawling.%
\footnote{While \cite{Akamatsu-et-al-DP2024} consider ``regional models'' without commuting so that workers work where they live. Similar comparative statics results are obtained by a standard urban model with commuting studied by \cite{Osawa-Akamatsu-JET2020}.}

\begin{figure}[htbp!]
    \centering
    \begin{minipage}[c]{\textwidth}
    \centering
    
    \captionsetup{width=\linewidth}
    
    \includegraphics[width=.6\textwidth]{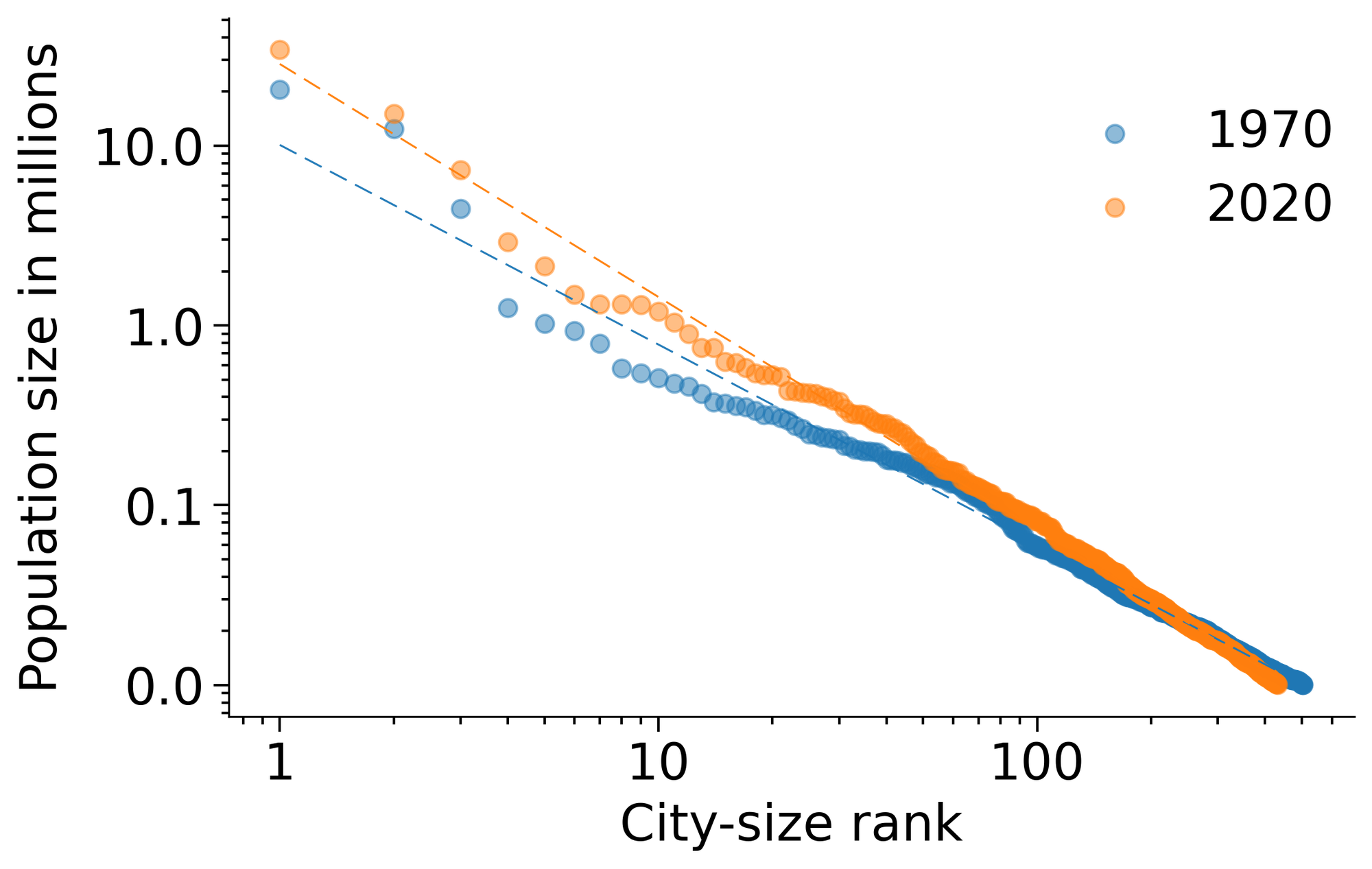}
    \caption{Concentration at the country level}
    \caption*{\footnotesize\textit{Note}: City size distributions of Japan in 1970 and 2020. The dashed lines are the fitted lines obtained from the log-linear regression of population size on the rankings of cities in terms of population size.}
    \label{fig:global-concentration}
    \end{minipage}
    
    \vspace{1cm}
    
    \begin{minipage}[c]{\textwidth}
    \centering
    \captionsetup{width=\linewidth}
    
    \includegraphics[width=.7\textwidth]{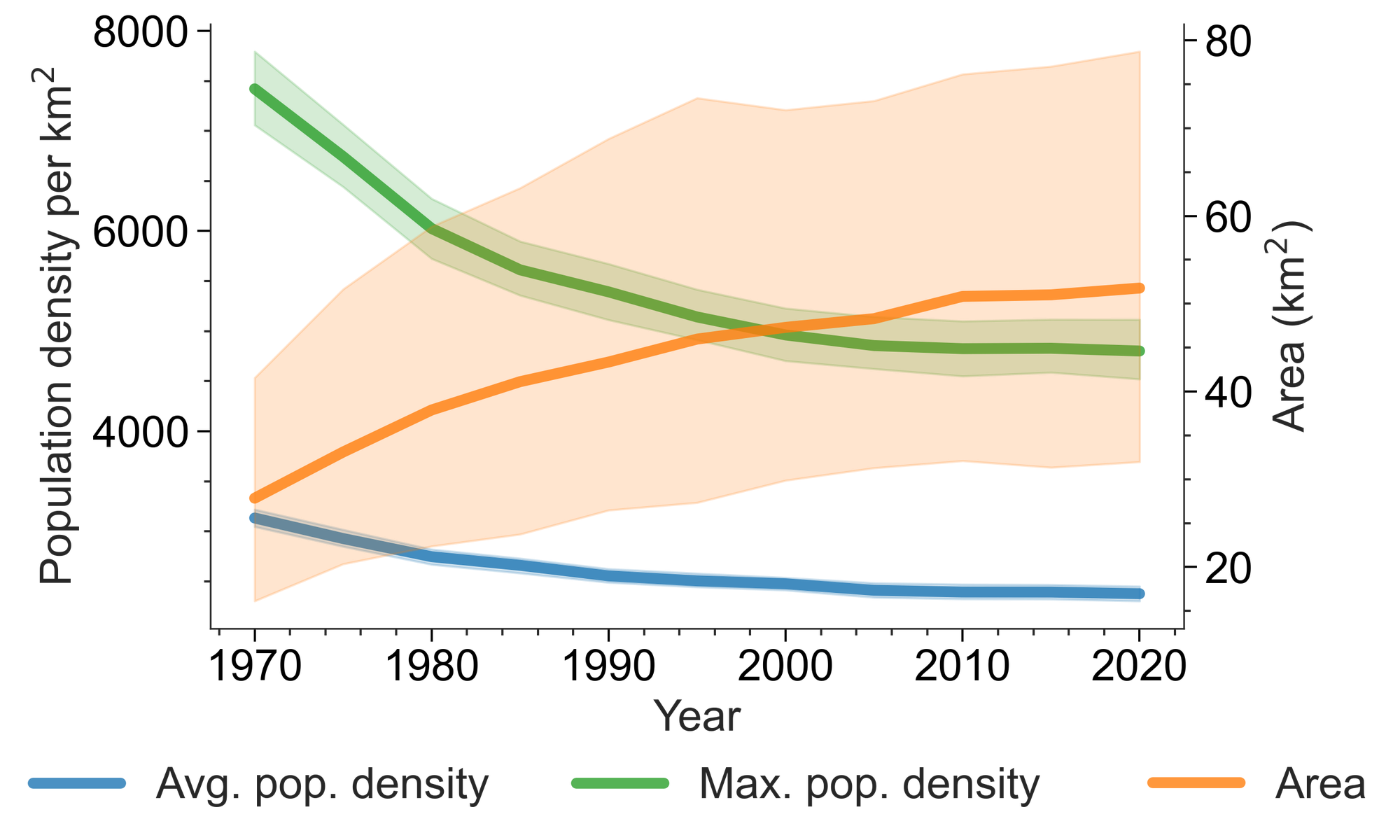}
    \caption{Dispersion at the city level}
    \caption*{\footnotesize\textit{Note}: The blue and green lines show the means of average and maximum population densities of a city and the orange line the mean total area of a city for each year indicated along the horizontal axis, while the shaded area indicates the range covering 95\% of the values for individual cities.}
    \label{fig:local-dispersion}
    \end{minipage}

    \vspace{1.5cm}
    
    \begin{minipage}[c]{\textwidth}
    \centering
    \includegraphics[width=.9\textwidth]{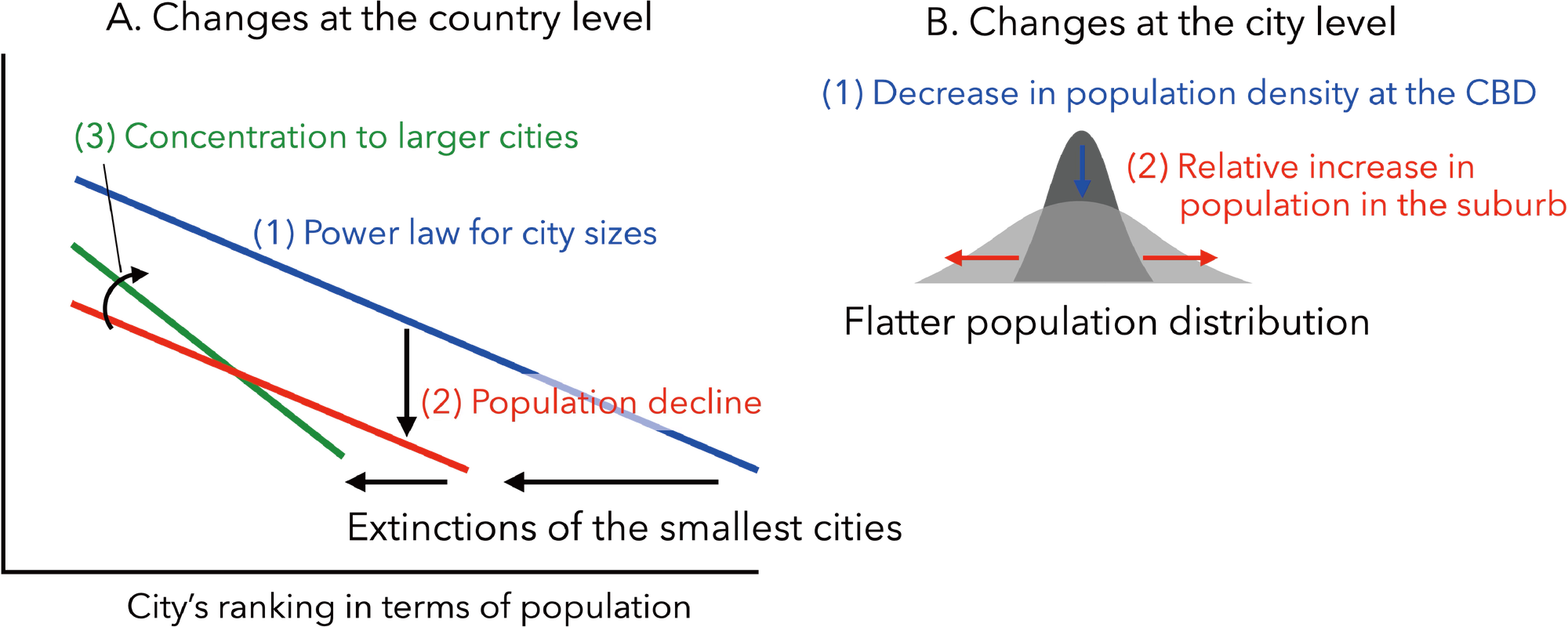}
    \medskip
    \caption{Expected changes at the country and city levels}
    \label{fig:model-behavior}
    \end{minipage}

\end{figure}

Figs.\,\ref{fig:global-concentration} and \ref{fig:local-dispersion} summarize what actually happened to cities in Japan during the period 1970--2020.
At the country level, the number of cities was 504 in 1970. It peaked at 511 in 1975, at the end of Japan's high-growth era, and has declined steadily since then to 431 in 2020.
Meanwhile, the distribution of city size has become clearly skewed toward larger cities (Fig.\,\ref{fig:global-concentration}), with larger cities becoming larger and smaller cities becoming smaller.%
\footnote{The decline in population at the city level results not only in a decline in employment in the city (the intensive margin), but also in a decline in industrial diversity, which means a decline in the range of job types available (the extensive margin).
Smaller cities, in particular, experience a disproportionate reduction in the breadth of their industrial base as their population declines. For detailed changes in industrial diversity across cities between 2015 and 2020, see \cref{app:industrial-diversity}.}\ 

Within a city, the average and maximum population densities continued to decrease, and the area occupied by a city continued to increase.
Specifically, the average and maximum population densities decreased by an average of 24\% and 35\%, respectively, and the area occupied by a city increased by an average of 86\% (Fig.\,\ref{fig:local-dispersion}), while the total population of Japan increased by 22\%.
Thus, cities have clearly flattened out.%
\footnote{The rapid decline in population density in urban areas from the 1970s to the 1990s is thought to reflect also the development of public transport networks such as subways and buses, as well as the proliferation of private cars. In particular, the number of private cars in Japan increased from about 16 million in 1970 to 37 million in 1980 and 57 million in 1990 \citep[][]{Yagi-Managi-TP2016}. }$^,$%
\footnote{Fig.\,\ref{fig:employment-dispersion} in Appendix \ref{app:employment} shows that the employment distribution within a city has also flattened out in the 1975--2015 period.}

In the coming decades, technological advances in automated driving and logistics as well as virtual reality could lead to an even more dramatic reduction in distance friction.
In addition, population decline will shift the distribution of city sizes downward, eliminating more of the smallest cities in the future.
An important validation of our forecasting model is the qualitative consistency of its future projections with theory (Fig.\,\ref{fig:model-behavior}).
That is, (i) the distribution of city sizes will remain roughly power-law shaped, while (ii) the distribution will shift downward as a result of population decline,%
\footnote{When agglomeration economies are the main force explaining a city's population size, population decline typically does not lead to a proportional decline of all cities, but to a greater concentration in larger cities, since it is the size, rather than the share, of a concentration that matters for the level of agglomeration economies \citep[e.g.,][]{Fujita-Krugman-RSUE1995,Fujita-Krugman-Mori-EER1999}.}\ 
and (iii) it will become more skewed toward larger cities, and the geographic distribution of population within a city will flatten out as a result of decreasing distance frictions.

\section{The basic setup and data preparation}
\label{sec:data}
Our goal is to predict the sustainability of individual cities in a future of rapid population decline. Even in Japan, a country at the forefront of population decline, it has only become apparent since the 2010 census. 
Therefore, it is not appropriate to use long time series data going back in time to predict the rise and fall of cities over the next 50 to 100 years, as it is likely to underestimate the effect of the recent trend of population decline as well as the decrease in distance frictions.%
\footnote{As we will see in Section \ref{sec:model}, even with our short time series data for model estimation, we had to give more weight to recent years to reflect the recent trend of population decline on future city growth.
Moreover, the geographic population distribution data for the entire country prior to 1970 are only available at the municipal level and thus have a much different resolution than the post-1970 grid data.}\ 
Our main data for estimating the model are the population counts in the 1-km grid cells obtained from the Grid Square Statistics of the Population Census of Japan for every five years from 1970 to 2020.%
\footnote{The 1970 data is based on 20\% subsample of population.}\ 

\begin{figure}[htbp!]
    \centering
    \begin{minipage}[c]{\textwidth}
    \centering
    \captionsetup{width=\linewidth}
    
    \includegraphics[width=.65\textwidth]{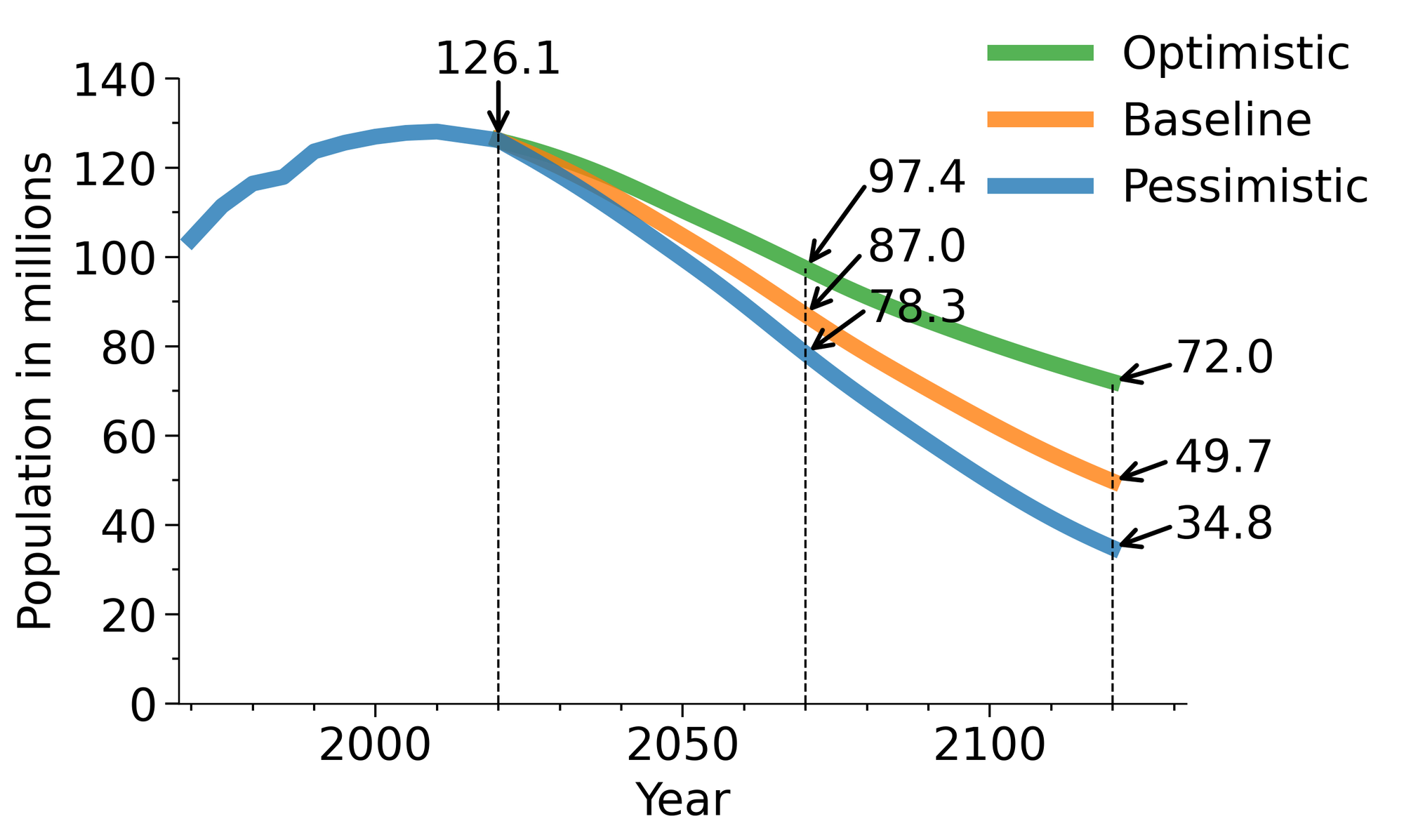}
    \caption{Projections of total population in Japan}
    \caption*{\footnotesize\textit{Note}: The population of Japan for 1970--2020 is the observed value. Population projections for 2021 to 2120 were obtained from \cite{NIPSSR-2023}. The ``baseline'' projection assumes that the fertility rate (1.33) and mortality rate in 2020 will continue in the future, while the pessimistic and optimistic projections assume fertility rates of 1.13 and 1.64, respectively (as well as higher and lower mortality rates). 
    The total population in our study area (the area connected by roads to either Honshu or Hokkaido as of 2020, shown in Fig.\,\ref{fig:cities-2020}) is 124 million in 2020, and its projections in 2120 are 34 million, 49 million, and 71 million under the pessimistic, baseline, and optimistic projections, respectively.}
    \label{fig:total-pop}
    \end{minipage}
    
    \vspace{1cm}

    \begin{minipage}[c]{\textwidth}
    \centering
    \captionsetup{width=\linewidth}
    \includegraphics[width=.6\textwidth]{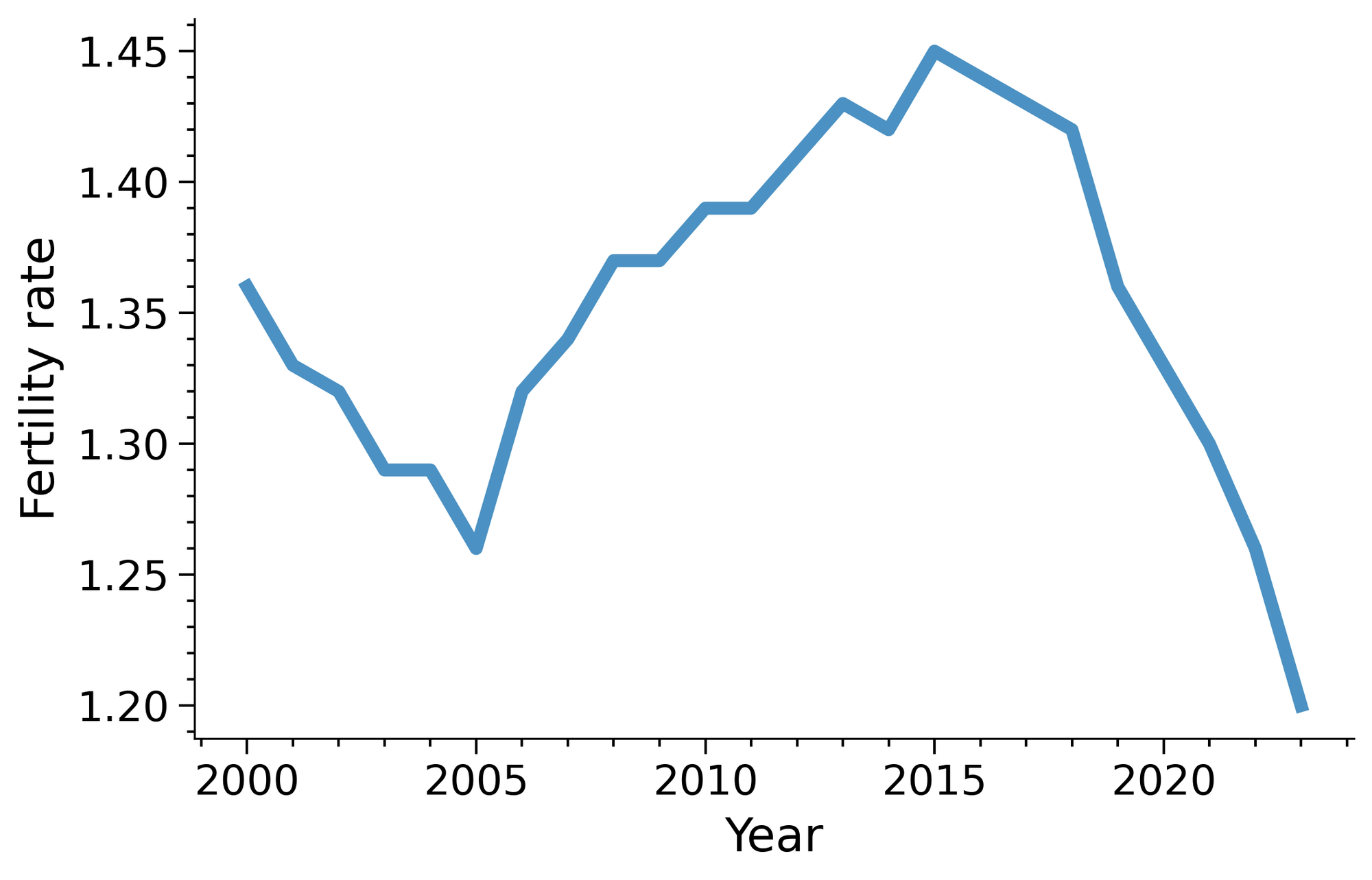}
    \caption{Fertility rates in Japan in 2000--2023}
    \caption*{\footnotesize\textit{Note}: For the years 2000--2022, data are available from \href{https://www.mhlw.go.jp/toukei/list/81-1.html}{The Vital Statistics}, Ministry of Health, Labour and Welfare (MHLW) of Japan. 
    The most recent data for 2023 can be found in \href{https://www.mhlw.go.jp/toukei/saikin/hw/jinkou/geppo/nengai23/dl/gaikyouR5.pdf}{Annual Report on Monthly Vital Statistics}, MHLW of Japan.}
    \label{fig:fertility}
    \end{minipage}
    

\end{figure}

To project the geographic distribution of the population in Japan in the future, we take as given the official projection of the total population of Japan provided by NIPSSR \citeyearpar{NIPSSR-2023} for the years 2025 to 2120 (Fig.\,\ref{fig:total-pop}). 
Their projections are based on three scenarios that differ essentially in the assumptions for the fertility and mortality rates.
The rates in the ``intermediate'' or ``baseline'' scenario are based on their values in 2020, whereas the ``pessimistic'' scenario assumes a lower fertility rate and a higher mortality rate, and the ``optimistic'' scenario assumes the opposite.

\begin{figure}[htbp!]
    \centering
    \includegraphics[width=.6\textwidth]{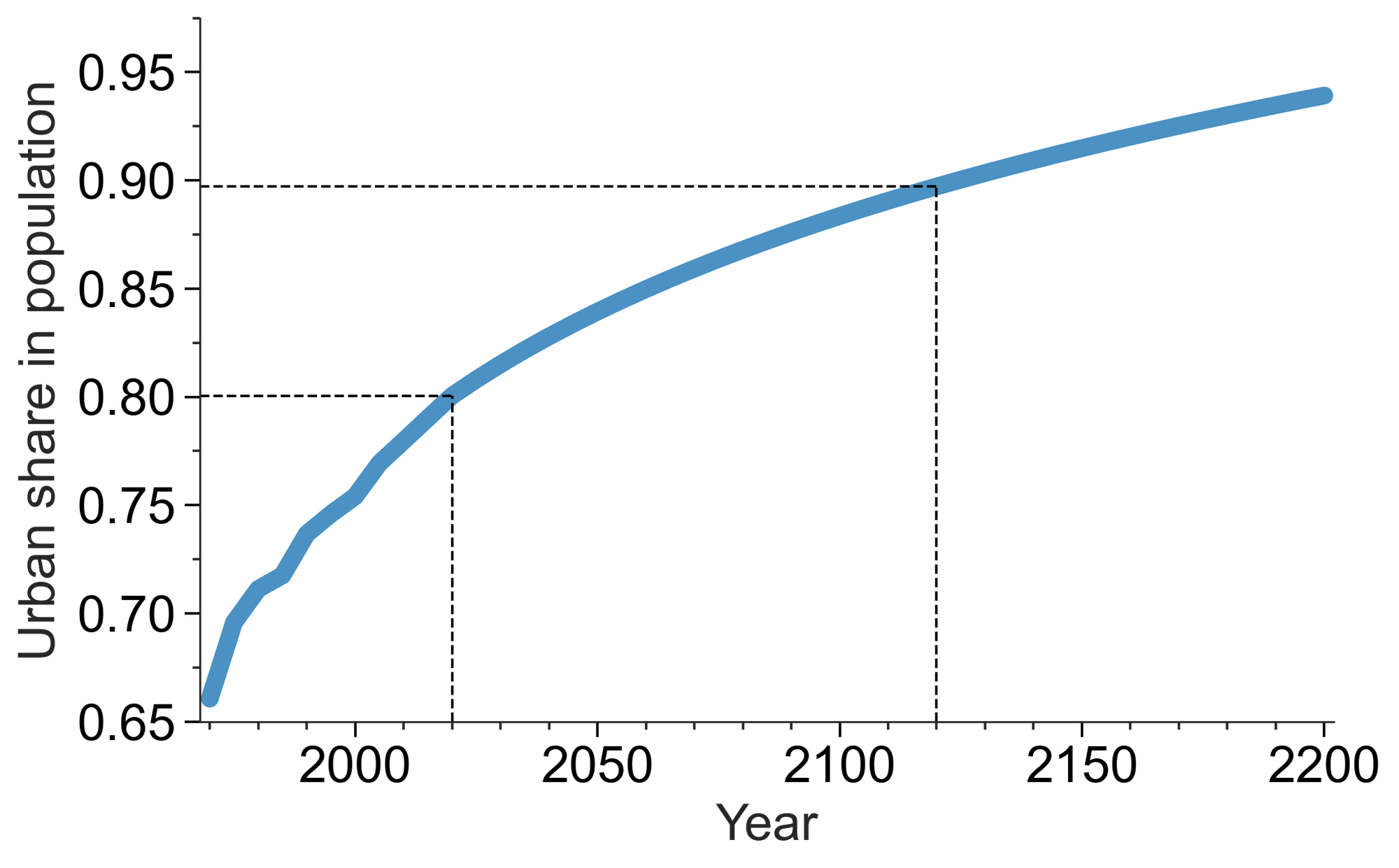}
    \caption{Projection of urbanization rate in Japan}
    \caption*{\footnotesize\textit{Note}: The urbanization rate is defined as the urban share of the total population. Observations in 1970--2020 are extrapolated to 2120 by assuming that the urban share increases log-linearly over time. The urbanization rates in 2020 and 2120 are 0.80 and 0.90, respectively.}
    \label{fig:urbanization}
\end{figure}
It should be noted that NIPSSR's projection of Japan's population is already optimistic because they fix the fertility rate in each scenario. 
In particular, their baseline scenario assumes that the fertility rate will be 1.33, the level in 2020, for the entire future.
But, as noted in the introduction, it has already fallen to 1.20 in 2023 and is expected to fall further (Fig.\,\ref{fig:fertility}).
Thus, the pessimistic scenario, assuming the fertility rate of 1.13, is probably the most realistic of the three shown in Fig.\,\ref{fig:total-pop}.%
\footnote{The optimistic scenario assumes the fertility rate of 1.64.}

We also assume that the urbanization trend of the last 50 years will continue into the future. Since the urban share of the population has a simple monotonic trend in the past, we extrapolate it by assuming a log-linear function of time (Fig.\,\ref{fig:urbanization}).

Let $t \in \{1, \ldots, 31\}$ represent the time index of the $t$-th year in the period 1970--2120 by 5-year intervals (i.e., $t=\frac{year-1965}{5}$). 
We identify cities in the past for $t=1,\ldots,11$ (i.e., in $1970, 1975,\ldots,2020$).
For the future, the set of cities is updated at every $t=12,\ldots,31$.
Thus, at each $t\geq 12$, new cities may form, multiple cities may merge, and a city may split, disappear, and reappear.
We track each city over time based on the overlap of its spatial area at different time points, assigning it a unique identifier throughout the study period (see Appendix \ref{app:ua} for details of the city detection).

\section{A forecasting model of city growth and decline}
\label{sec:model}
Our forecasting model has three levels: the country level, the city level, and the grid-cell level.
At the country level, the evolution of the total population and the total urban population, shown in Figs.\,\ref{fig:total-pop} and \ref{fig:urbanization}, respectively, are taken as given.
In addition, the evolution of the power law coefficient for the city size distribution estimated based on the 1970--2020 data is taken as given.
In particular, the city-size distribution is expected to become more skewed toward larger cities in the future, since the trend of falling transport and communication costs is expected to continue as it has over the past half century.
At the city and grid-cell levels, time series models are used to account for the city- and grid-specific growth factors. 
We also account for the population growth in nearby grid cells.
Below is a detailed description of the specifications at each level.

\subsection{The country level model\label{sec:country-level}}
Let $\mathbf{G}$ represent the set of grid cells, and $\mathbf{U}_t$ the set of cities (UAs) in time $t$. (Hereafter, the cardinality of a set $\mathbf{X}$ is expressed by its italic character $X\equiv |\mathbf{X}|$.)
Each $u\in\mathbf{U}_t$ consists of a contiguous set of grid cells, $\mathbf{G}_{u,t}\subseteq \mathbf{G}$, satisfying the population density condition, $p_{i,t}\geq \text{1,000}$ for $i\in \mathbf{G}_{u,t}$, and the population size condition, $P_{u,t} \equiv \sum_{i\in\mathbf{G}_{u,t}} p_{i,t} \geq \text{10,000}$.
Let $P_{u,t}$ represent the population of city $u$ that has the $r_{u,t}$-th largest population at time $t$.

The city size distribution is assumed to roughly follow the power law, which is empirically expressed as
\begin{align}
  \log P_{u,t} = A_t + B_t \log (r_{u,t}-0.5)+\epsilon_{u,t}^\textit{PL}, 
  \hspace{0.5cm} \epsilon_{u,t}^\textit{PL} \sim N(0, \sigma_\textit{PL}^2),\label{eq:PL}
\end{align} 
where ``$-0.5$'' is a bias correction \citep[][]{Gabaix-Ibragimov-JBES2011}, and $\sigma_{\textit{PL},t}^2$ is a variance parameter.
The parameters $A_t$ and $B_t$ represent the population scale and the power coefficient, respectively. Their estimates $\hat{A}_t$, $\hat{B}_t$ and variances $V[\hat{A}_t],V[\hat{B}_t]$ are evaluated by applying the ordinary least squares (OLS) method for each $t \in \{1,\cdots,11\}$ independently.

As discussed in Section \ref{sec:data}, both the population scale $A_t$ and the power coefficient $B_t$ have changed largely monotonically in recent decades and are expected to change monotonically with time in the future, we estimate their evolution by the following log-linear models.
\begin{align}
  A_t &= a^{A}_0 + a^{A}_1 \log t + \epsilon_{t}^{A}, 
  \hspace{0.5cm} \epsilon_{t}^{A} \sim N(0, \sigma_A^2/t^2),\label{eq:PL_A} \\
  B_t &= a^{B}_0 + a^{B}_1 \log t+ \epsilon_{t}^{B}, 
  \hspace{0.5cm} \epsilon_{t}^{B} \sim N(0, \sigma_B^2/t^2),\label{eq:PL_B}
\end{align}
where the error variance is divided by $t^2$ to estimate the parameters emphasizing more recent samples in $t\in\{1,\cdots,11\}$.%
\footnote{Japan's population growth rate continued to decline between 1970 and 2020. But it finally became negative in the last two census years, 2015 and 2020. 
It is obvious that this declining trend will continue even under the rather optimistic future projection of the NIPSSR shown in Fig.\,\ref{fig:total-pop} and from the trend of the declining fertility rate shown in Fig.\,\ref{fig:fertility}. Thus, we believe that this weight adjustment is reasonable to obtain a realistic projection of the future population size of individual cities.}\ 
The parameters $a^{A}_0, a^{A}_1, a^{B}_0, a^{B}_1,\sigma_A^2$, and $\sigma_B^2$ are estimated by applying the weighted least squares estimation for eqs.\,\eqref{eq:PL_A} and \eqref{eq:PL_B} that replaces $A_t$ and $B_t$ with their estimates $\hat{A}_t$ and $\hat{B}_t$, respectively.

Given the evolution of the coefficients of the power-law distribution, \eqref{eq:PL_A} and \eqref{eq:PL_B}, we predict the population $P^{PL}_{u,t+1}$ of city $u\in\mathbf{U}_{t}$ at time $t+1$, that is consistent with the power law for city size distribution, using the expected population given past populations.
Its expected value, $E[P^{PL}_{u,t+1}|\hat{P}_{u,t}]\equiv\hat{P}^{PL}_{u,t+1}$, is obtained by substituting \eqref{eq:PL_A} and \eqref{eq:PL_B} into \eqref{eq:PL}, and differencing eq.\,\eqref{eq:PL} and with respect to $t$:

\paragraph*{PL: the power law model}
\begin{equation}
  \hat{P}^{PL}_{u,t+1} 
 =\hat{P}_{u,t} + \Delta \hat{P}^{PL}_{u,t}
  \approx\left(1+\frac{a_1^{A} + a_1^{B}log(r_{u,t}-0.5)}{t} \right)\hat{P}_{u,t}. \label{eq:condexp}
\end{equation} 
where $\Delta \hat{P}^{PL}_{u,t}=\hat{P}^{PL}_{u,t+1}-\hat{P}^{PL}_{u,t}$.

The value $\hat{P}_{u,t}$ is a given projection of the population size of city $u$ at time $t$.
At the future time $t+1$, it is given by the model ensembled from the city-level time series \eqref{eq:ensemble_city} (introduced below) and the power law model \eqref{eq:condexp} at the previous time $t$.
Note that our PL model \eqref{eq:condexp} allows for deviations from the power law, reflecting city-specific growth factors captured by the time-series model.
%
%
\bigskip

\subsection{City-level model}\label{sec:city-level}
We use standard time series models to capture the city-specific growth process, in particular the inertia that the trend of population size and area of each city do not change suddenly in response to changes in basic parameters such as total population and transport costs.
For example, the second largest city Osaka has been stagnant since 2000, unlike other top cities, deviating from the power law.

Given the short time series data available to estimate the models, we use the following relatively simple three time-series models.%
\footnote{In a more ideal situation, relevant time-series models include the autoregressive integrated moving average (ARIMA) model \citep[][]{Box-et-al-Book2015}, the state-space model \citep[SSM; see][]{Durbin-Koopman-Book2012}, and recurrent neural networks \citep[RNN; see][]{Graves-DP2013}.
However, the observed time series data are typically too short for our forecasting model to apply SSM and RNN. 
Even for the ARIMA model, which is simpler than SSM and RNN, our preliminary analysis showed that the projected population often takes extremely large or small values given our long projection period of a century.}


\paragraph*{ARI1: the ARIMA(1,1,0) model}
 \begin{align}
        \Delta P_{u,t+1}=\rho_u \Delta P_{u,t}+\epsilon_{u,t}^{\textit{ARI}1},\quad 
        \epsilon_{u,t}^{\textit{ARI}1}\sim N(0,\sigma^2_{u,\textit{ARI}1})\label{eq:AR1_city}
    \end{align}
for $u\in\mathbf{U}_t$, where $\Delta P_{u,t+1}\equiv P_{u,t+1} - P_{u,t}$. The auto-correlation coefficient $\rho_u$ takes a positive value in the presence of positive temporal correlation while the opposite is true for negative correlation. 
We used a difference auto-regressive process because differentiation eliminates non-stationary trends such as depopulation in recent years. To estimate the difference in population growth pattern by city, parameters are estimated by city using a maximum likelihood method.

\paragraph*{ARI2: the ARIMA(2,1,0) model}
    \begin{align}
        \Delta P_{u,t+1}=
        \rho_{1,u} \Delta P_{u,t}+
        \rho_{2,u} \Delta P_{u,t-1}+
        \epsilon_{u,t}^{\textit{ARI}2},\hspace{0.5cm} 
        \epsilon_{u,t}^{\textit{ARI}2}\sim N(0,\sigma^2_{u,\textit{ARI}2}).\label{eq:AR2_city}
    \end{align}
This model considers second order time series using parameters $\rho_{1,u}$ and $\rho_{2,u}$. Because of the limited time periods, we do not consider higher order dependence. In addition, since the moving average process only explains short-term dependence, we ignore it (e.g., the second-order moving average process only affects the second period ahead).
    
\paragraph*{LL: Log-linear-in-time model \citep[][]{Smith-et-al-Book2005}}
    \begin{align}
    P_{u,t+1} = a_{0,u}^\textit{LL} + a_{1,u}^\textit{LL} \log (t+1) + \epsilon_{u,t}^\textit{LL},\quad \epsilon_{u,t}^\textit{LL} \sim N\left(0, \frac{\sigma^2_{u,\textit{LL}}}{t^2}\right),\label{eq:LL_city}     
    \end{align}
where $a_{0,u}^\textit{LL}$ and $a_{1,u}^\textit{LL}$ are regression coefficients estimated by minimizing the (weighted) squared error under the constraint that the predictive function path through the actual population in 2020 (see Appendix \ref{app:const-ls}).%

It is a common exercise in the literature of the population projection to be constrained to match the actual and projected population sizes of the target region at the latest time in the learning data (NIPSSR, \citeyear{NIPSSR-2023,NIPSSR-Region-2023}; Jones and O'Neill, \citeyear{Jones-ONeil-ERL2016}; Wang, Meng, and Long, \citeyear{Wang-Meng-Long-SD2022}). 
In our case, we applied this constraint to the LL model which most directly captures the inatia of city size.
The model is estimated to minimize the least squares error weighted by $t^2$, with a greater weight in time $t$ or just before, in order to reflect the trend of declining population in the recent past.


We average the projections by the models, $\mathbf{M}\equiv \{\textit{ARI}1,\textit{ARI}2,\textit{LL}\}$, using the inverse-variance weighting method, a model averaging method known to minimize the predictive variance \citep[e.g.,][]{Hartung-Knapp-Sinha-Book2008}. The predictive mean and variance of the synthesized time-series model are given as follows.
\paragraph*{City-level TS: City-level time series model}
\begin{equation}
\hat{P}^{TS}_{u,t}=\frac{\Sigma_{m\in \mathbf{M}} w_{u,t}^m \hat{P}_{u,t}^m}{\Sigma_{m\in\mathbf{M}} w_{u,t}^m}, \hspace{0.5cm}
V[\hat{P}^{TS}_{u,t}]=\frac{1}{\Sigma_{m\in\mathbf{M}} w_{u,t}^m},
    \label{eq:ensemble_city}
\end{equation}
where $w_{u,t}^m=\frac{1}{V[\hat{P}^m_{u,t}]}$ and $\hat{P}^m_{u,t}$ is the population size of city $u$ at time $t$ projected by model $m$. The variance $V[\hat{P}^m_{u,t}]$ for each model is evaluated by the conventional procedures for these models (see Appendix \ref{app:variance-city}).

Finally, the future population size of each city is projected by ensembling the power-law model \eqref{eq:condexp} and the time-series model \eqref{eq:ensemble_city}. 
The city-level populations $\hat{P}^{PL}_{u,t+1}$ and $\hat{P}^{TS}_{u,t+1}$ are averaged by the inverse-variance weighting.
The predictive mean and variance yield 

\paragraph*{Weighted average of the city-level models}
\begin{align}
    \hat{P}_{u,t+1} &=\frac{w^{PL}_{u,t+1}\hat{P}^{PL}_{u,t+1}+w^{TS}_{u,t+1}\hat{P}^{TS}_{u,t+1}}{w^{PL}_{u,t+1}+w^{TS}_{u,t+1}},\label{eq:PL_TS_avg}\\
    V[\hat{P}_{u,t+1}]&= \frac{1}{w^{PL}_{u,t+1}+w^{TS}_{u,t+1}},\label{eq:V_PL_TS_avg}
\end{align}
where $w^m_{u,t+1}=1/V[\hat{P}^m_{u,t+1}]$, $m=TS$ and $PL$.
$V[\hat{P}^{TS}_{u,t+1}]$ is evaluated using eq.\,\eqref{eq:ensemble_city}, and $V[\hat{P}^{PL}_{u,t+1}]$ is evaluated by the bootstrap resampling explained in Appendix \ref{app:variance-pl}. 

Specifically, for the initial time, $t=11$ (year 2020), we assume that $\hat{P}_{u,t}$ of city $u$ obeys our time-series model so that $V[\hat{P}_{u,t}]$ is evaluated by eq.\,\eqref{eq:ensemble_city}.
This simplifying assumption has an operational advantage.
Since $\hat{P}^{PL}_{u,t+1}$ is evaluated by multiplying $\hat{P}_{u,t}$ and a deterministic variable $\left(1+\frac{a_1^{A} + a_1^{B}log(r_{u,t}-0.5)}{t} \right)$, $\hat{P}^{PL}_{u,t+1}$ follows a Gaussian distribution for $t=11$. 
The same holds recursively for $t=12$ onward, since $\hat{P}_{u,t+1}$ is given by an average of the Gaussian time series model projection and $\hat{P}^{PL}_{u,t+1}$, eq.\,\eqref{eq:PL_TS_avg}. 

\subsection{Grid-level models\label{sec:grid-level}}
At the grid level, we consider the same set of time-series models adopted for the city level projection.

\paragraph*{ARI1: the ARIMA(1,1,0) model}
 \begin{align}
        \Delta p_{i,t+1}=\rho_i \Delta p_{i,t}+\epsilon_{i,t}^{\textit{ARI}1},\quad 
        \epsilon_{i,t}^{\textit{ARI}1}\sim N(0,\sigma^2_{i,\textit{ARI}1})\label{eq:AR1_grid}
    \end{align}
for $i\in\mathbf{G}$, where $\Delta p_{i,t+1}\equiv p_{i,t+1} - p_{i,t}$, $\rho_i$ is an auto-correlation coefficient. 

\paragraph*{ARI2: the ARIMA(2,1,0) model}
    \begin{align}
        \Delta p_{i,t+1}=
        \rho_{1,i} \Delta p_{i,t}+
        \rho_{2,i} \Delta p_{i,t-1}+
        \epsilon_{i,t}^{\textit{ARI}2},\hspace{0.5cm} 
        \epsilon_{i,t}^{\textit{ARI}2}\sim N(0,\sigma^2_{i,\textit{ARI}2}),\label{eq:AR2_grid}
    \end{align}
where $\rho_{1,i}$ and $\rho_{2,i}$ are auto-correlation parameters.
        
\paragraph*{LL: Log-linear-in-time model}
    \begin{align}
    p_{i,t+1} = a_{i,0}^\textit{LL} + a_{i,1}^\textit{LL} \log(t+1) + \epsilon_{i,t}^\textit{LL},\quad \epsilon_{i,t}^\textit{LL} \sim N\left(0, \frac{\sigma^2_{i,\textit{LL}}}{t^2}\right).\label{eq:LL_grid}     
    \end{align}
Similar to the city-level LL, the regression coefficients are estimated by minimizing the (weighted) squared error under the constraint that the predictive function path through the actual population in 2020 (see Appendix \ref{app:const-ls}).

\subsubsection*{Models for the neighboring populations}
To capture the influence from the population growth pattern in the neighboring areas, the same three time-series models are applied to the average population $q_{i,t}$ in the neighboring (NB) grid cells that share border or point with the $i$-th grid cell:%
\footnote{The 8NN of a grid $i$ includes the edge- and point-adjacent neighbors of grid $i$, excluding grid $i$ itself. While each 8NN of grid $i$ has at most 8 grid cells, it may have fewer grid cells if, for example, grid $i$ is located at the coast.
Alternatively, we considered a broader neighbor size, 15NN and 24NN. But, the results change only marginally. Thus, we adopt the 8NN which includes only the adjacent neighbor cells.}$^,$%
\footnote{The regression coefficients for the LL model are estimated by minimizing the (weighted) squared error under the constraint that the predictive function path through the actual population in 2020 (see Appendix \ref{app:const-ls}).}\ 
\medskip

\begin{itemize}
    \item \textit{ARI}1 model: The ARI1 model for $q_{i,t}$
    \item \textit{ARI}2 model: The ARI2 model for $q_{i,t}$
    \item $LL_N$ model: The LL model for $q_{i,t}$
\end{itemize}

The above three models, which are estimated by using the NB population data up to 2020, are applied to project future population of the NBs. The projected population $\hat{q}_{i,t}$ is then used to predict the gridded population $p_{i,t}$ by substituting $\hat{q}_{i,t}$ into  
        \begin{equation}        
        p_{i,t}=b_i \hat{q}_{i,t}+\epsilon_{i,t}, \hspace{0.5cm} \epsilon_{i,t} \sim N(0,s^2 ).\label{eq:8nn}
    \end{equation}
The parameter $b_i$ is estimated a priori by regressing the gridded population $p_{i,t}$ in 2020 on the average NB population $q_{i,t}$.%

All the time-series models $\mathbf{M}\equiv \{\textit{ARI}1,\textit{ARI}2,\textit{LL},\textit{ARI}1_N,\textit{ARI}2_N, \textit{LL}_N\}$ are estimated by grid cell. To reduce uncertainty, their projections are averaged by the inverse-variance weighting method. The averaged predictive mean and variance are given as follows.
\begin{equation}
    \hat{p}^{TS}_{i,t}=\frac{\Sigma_{m\in \mathbf{M}} w_{i,t}^m \hat{p}_{i,t}^m}{\Sigma_{m\in\mathbf{M}} w_{i,t}^m}, \hspace{0.5cm}
 V[\hat{p}_{i,t}^{TS}]=\frac{1}{\Sigma_{m\in\mathbf{M}} w_{i,t}^m},
 \label{eq:ensemble_grid}
\end{equation}
where $\hat{p}_{i,t}^m$ is the grid population projected by model $m$, $V[\hat{p}_{i,t}^m]$ is the variance, and $w_{i,t}^m=\frac{1}{V[\hat{p}_{i,t}^m]}$. 
For $\textit{ARI}1,\textit{ARI}2,\textit{LL}$, $V[\hat{p}^m_{u,t}]$ is evaluated similarly to the case of the city-level time-series model explained in Appendix \ref{app:variance-city}.
The computation of the variances for $\textit{ARI}1_N,\textit{ARI}2_N, \textit{LL}_N$ are explained in Appendix \ref{app:variance-grid}.

\section{Projection procedure\label{sec:procedure}}
Let $P_t$ be the total population, $P^U_t$ the projected total urban population, and $P^R_t \equiv P_t-P^U_t$ the total rural population in time $t$.
Denote by $\mathbf{G}^U_t$ the set of grid cells in cities in time $t$, and the set of those outside cities (rural area) by $\mathbf{G}^R_t\equiv \mathbf{G}\backslash\mathbf{G}^U_t$.

We use the gridded population data for the period 1970--2020 (every 5 years) from the Population Census of Japan to estimate the time series models \cref{eq:AR1_grid,eq:AR2_grid,eq:LL_grid,eq:8nn,eq:ensemble_grid} for each grid cell,  \cref{eq:AR1_city,eq:AR2_city,eq:LL_city} for each city, and the power law model \cref{eq:PL,eq:PL_A,eq:PL_B} at the country level.
Using these estimated models together with the given national and urban/rural population in each time $t$, we predict the grid and city population for $t = 12,\ldots,31$ (years $2025, 2030, \ldots,2120$), while updating the city set at each $t=12,13,\ldots,31$.

Specifically, we follow the steps below from $t=11$ (year 2020):

\noindent {\bf Step 0 (Parameter estimation).}  For each city existing in year 2020, ARI1, ARI2, LL are estimated using the population time series up to 2020. For each grid, the grid-level models ARI1, ARI2, LL, $\mathrm{ARI1_N, ARI2_N, LL_N}$ are estimated in the same manner. In addition, the parameters $a^A_1, a^B_1$ in the PL model are estimated using the city population data between 1970 and 2020 (see Section \ref{sec:country-level}).

\noindent {\bf Step 1 (Grid-cell level projection).} 
The population in each grid cell at time $t+1$ is predicted using the estimated time-series models ARI1, ARI2, LL, $\mathrm{ARI1_N, ARI2_N,}$ and $\mathrm{LL_N}$. The predicted populations are averaged using eq.\,\eqref{eq:ensemble_city} to obtain the synthesized population projection $\hat{p}^{TS}_{i,t+1}$ (see Section \ref{sec:grid-level}). 

\noindent {\bf Step 2 (City-level projection).} For each city for which at least one of ARI1, ARI2, LL could be estimated in {\bf Step 0}, the population at time $t+1$ is predicted using the estimated models. 
The predicted populations are averaged using eq.\,\eqref{eq:ensemble_grid} to obtain the synthesized projection $\hat{P}^{TS}_{u,t+1}$ (see Section \ref{sec:city-level}). 
In the averaging, zero weight $w_{u,t+1}^m=0$ is given to models that could not be estimated in {\bf Step 0}. 
For cities for which all time series models could not be estimated because they did not exist in the training period before 2020, or because the cities are newly formed (in {\bf Step 7}), $V[\hat{P}^{TS}_{u,t+1}]=\infty$ is assumed.%
\footnote{A typical case is that city $u$ is newly identified in time $t$ due to the split of an existing city. In this case, we set $\hat{P}_{u,t+1} = \hat{P}^{PL}_{u,t}$.}


\noindent {\bf Step 3 (Power law-based population projection).} Independent of {\bf Steps 1--3}, the population $\hat{P}^{PL}_{u,t+1}$ is evaluated by substituting $\hat{P}_{u,t}$ (determined in {\bf Step 4} in time $t-1$), $\hat{a}_1^{A}, \hat{a}_1^{B}$ into the power law-based projection model (\ref{eq:condexp}) (see Section \ref{sec:country-level}).

\noindent {\bf Step 4 (Model ensemble).} The city-level populations $\hat{P}^{PL}_{u,t+1}$  and $\hat{P}^{TS}_{u,t+1}$ are averaged by the inverse-variance weighting. The predictive mean and variance are given by eqs. \eqref{eq:PL_TS_avg} and \eqref{eq:V_PL_TS_avg}, respectively.

\noindent {\bf Step 5. (Rescale grid population to fit city population).} For each city $u\in \mathbf{U}_t$, the population in each grid cell in $\mathbf{G}^U_t$ is rescaled proportionally to match the projected population $\hat{P}_{u,t+1}$.

\noindent {\bf Step 6 (Urban-rural population adjustment).} 
We rescale the city populations and the rural population to match the assumed total urban and rural populations, respectively. The grid populations are rescaled in each city so that their aggregated values match the rescaled city populations. The gridded populations in the rural area are also rescaled in the same manner.%
\footnote{The population gap near the border between urban and rural areas can be adjusted by spatial smoothing (see Appendix \ref{app:smoothing}). However, we decided not to apply it because it typically over-smoothes and over-removes small cities.
While smoothing becomes important when cities are growing and expanding, it is less of an issue when cities are shrinking.}

\noindent {\bf Step 7 (City detection).} We detect cities based on the city definition (Appendix \ref{app:ua}). Denote the updated city set by $\mathbf{U}_{t+1}$.%
\footnote{We allow for a small inconsistency with the assumed urban share of the total population at this step. }
If a city $u$ is newly identified or was once merged with another city but revived in time $t$, its population size $\hat{P}_{u,t}$ is given as the sum of the populations in the grid cells $i\in\mathbf{G}_u$.

\bigskip

We repeat Steps 1--7, for $t=11,\ldots,31$, i.e., for years 2020 through 2115 to obtain the projections of grid and city populations for years 2025 through 2120.

%

For a long-run prediction, the most important model validation is to generate predictions that are consistent with the qualitative implications (comparative statics) of economic agglomeration theory (Fig.\,\ref{fig:theory}). 
Namely, we expect cities to generally shrink in both population and spatial size as the national population declines rapidly.
With lower distance frictions, we expect that a smaller number of larger cities will attract a larger share of the population, and that these cities will tend to be farther apart.
Within each city, on the other hand, there will be a flatter geographic distribution of population.

\cref{app:performance,app:bias-correction,app:proj-annual} present model validation exercises.
\cref{app:performance} discusses the role of each specific model at the grid, city and country level adjustments in correctly predicting the growth of individual cities.
\cref{app:bias-correction} provides a justification for the intercept restrictions applied to the LL models.
\cref{app:proj-annual} shows that our projection is robust even when our five-yearly data are replaced by the annual municipal population data during the period of population decline after 2010.

\section{Results}
\label{sec:results}
In this section, we present the results of our predictions regarding city growth and decline based on Japan’s official population projections. As noted in the Introduction, the optimistic scenario appears overly optimistic; therefore, our analyses focus  on the more realistic baseline and pessimistic scenarios.
Section \ref{sec:concentration} examines the anticipated concentration of population in larger cities nationwide as the national population declines. We further demonstrate that this concentration trend is not limited to the national scale, but is also evident within each of Japan’s subregions, highlighting the spatial fractal structure of the country’s urban system.
Section \ref{sec:dispersion} explores the projected patterns of suburbanization that individual surviving cities are expected to experience.
\subsection{Concentration at the country and regional levels}
\label{sec:concentration}
Figure \ref{fig:city-count} shows the projected number of surviving cities under each scenario.
In the baseline scenario, cities decrease from 431 in 2020 to 230 by 2120.
In the pessimistic scenario, only 195 cities are expected to remain by 2120, less than half of the current total.

\subsubsection*{Concentration at the country level}
The trend of concentration at the country level is captured by the key parameters of the PL model, which are estimated as $\hat{a}_1^A=0.424$ and $\hat{a}_1^B=-0.0822$. In the PL model \eqref{eq:condexp}, these values indicate that the higher ranking cities are expected to grow faster than lower ranking ones so that the city size distribution will be more skewed towards larger cities in the future.

\begin{figure}[htbp!]
    \centering
    \begin{minipage}[c]{\textwidth}
    \centering
    \captionsetup{width=\linewidth}
    \includegraphics[width=0.6\textwidth]{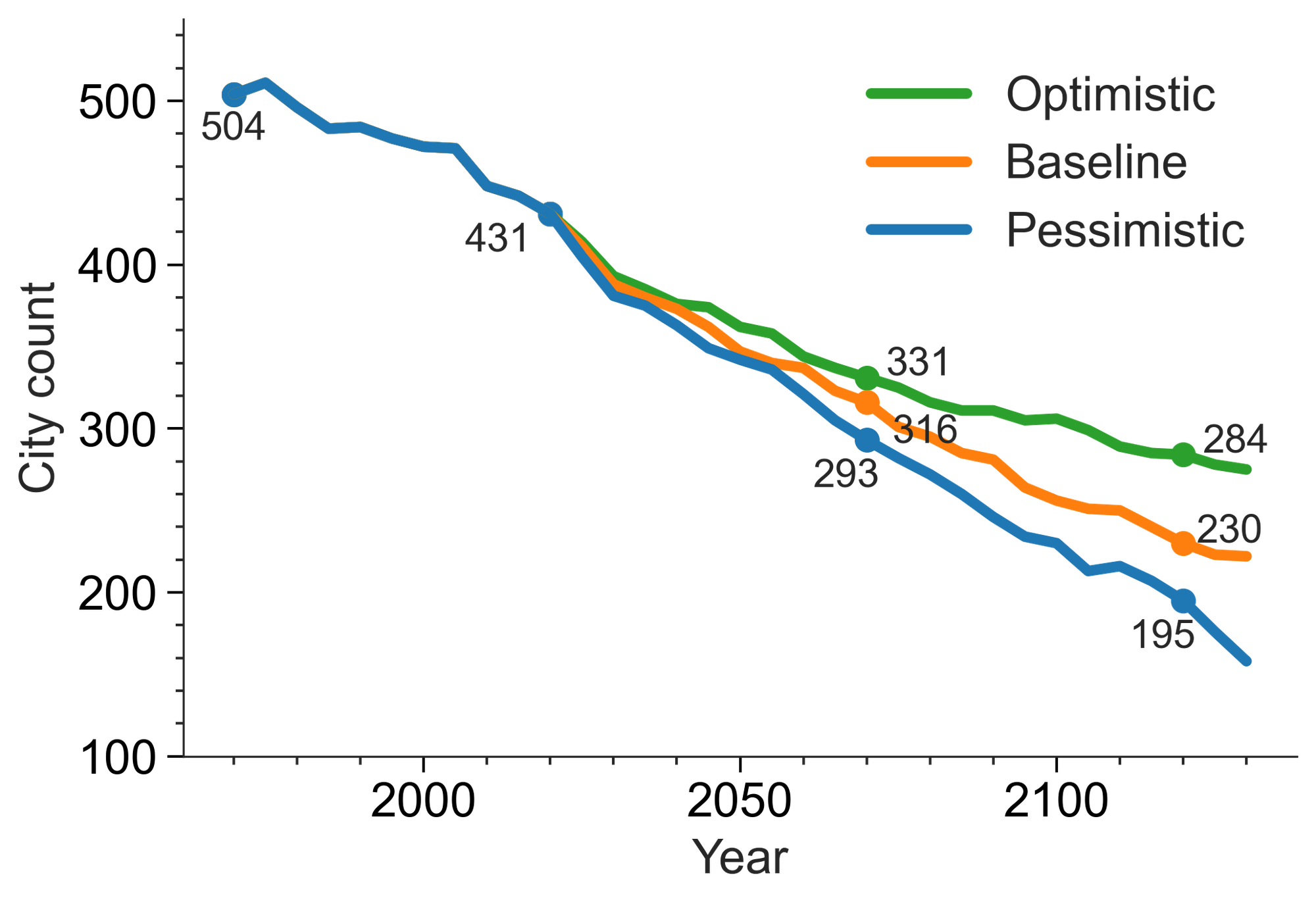}
    \bigskip
    
    \caption{City counts in 1970--2120 under the three scenarios of population decline in Japan}
    \caption*{\footnotesize\textit{Note}: The city counts in 1970--2020 are realized values, and those in 2025-2120 are our projections under three scenarios of population decline in Japan.}
    \label{fig:city-count}
    \end{minipage}
    
    \vspace{1cm}
    
    \begin{minipage}[c]{\textwidth}
    \centering
    \captionsetup{width=\linewidth}
    \includegraphics[width=\textwidth]{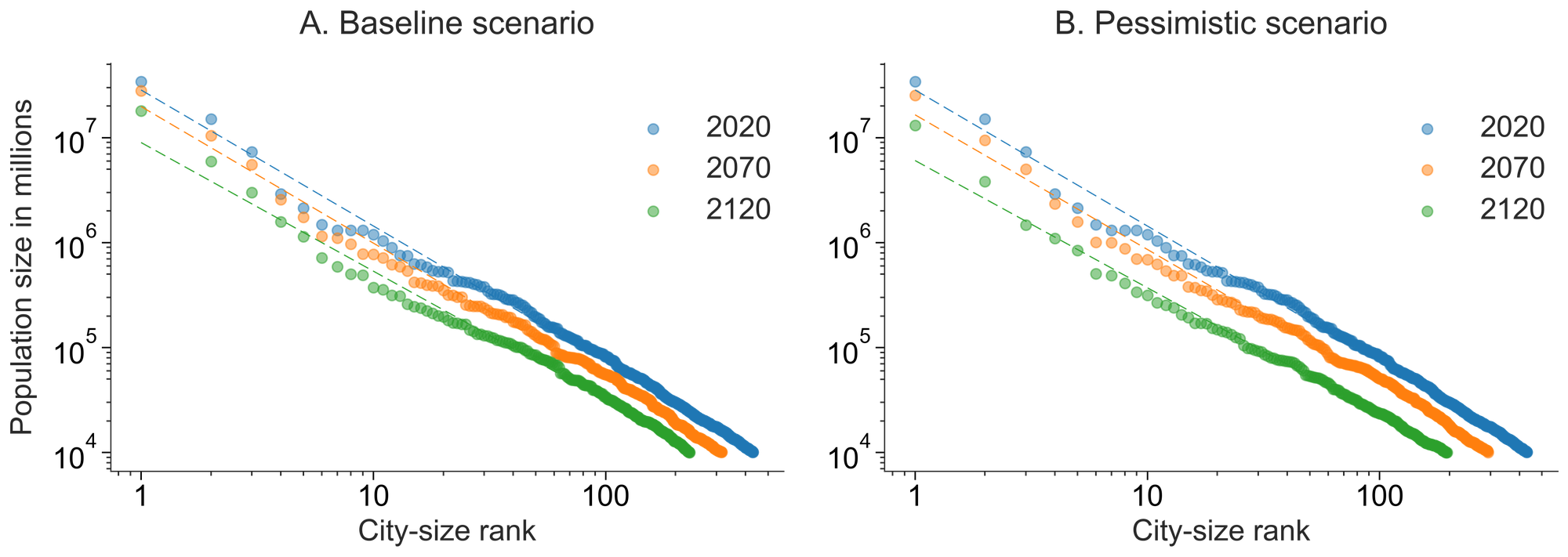}
    \smallskip
    \caption{Predicted city-size distributions under the baseline and pessimistic scenarios}
    \caption*{\footnotesize\textit{Note}: The dashed lines are the fitted regression lines. The largest Tokyo's share in urban population is predicted to increase from 34\% in 2020 to 38\% in 2070 and to 41\% in 2120 under the baseline scenario, and to 39\% in 2070 and to 44\% in 2120 under the the pessimistic scenarios.}
    \label{fig:size-dist}
    \end{minipage} 
    
    \vspace{1cm}
    
    \begin{minipage}[c]{\textwidth}
    \centering
    \captionsetup{width=\linewidth}
    
    \includegraphics[width=\textwidth]{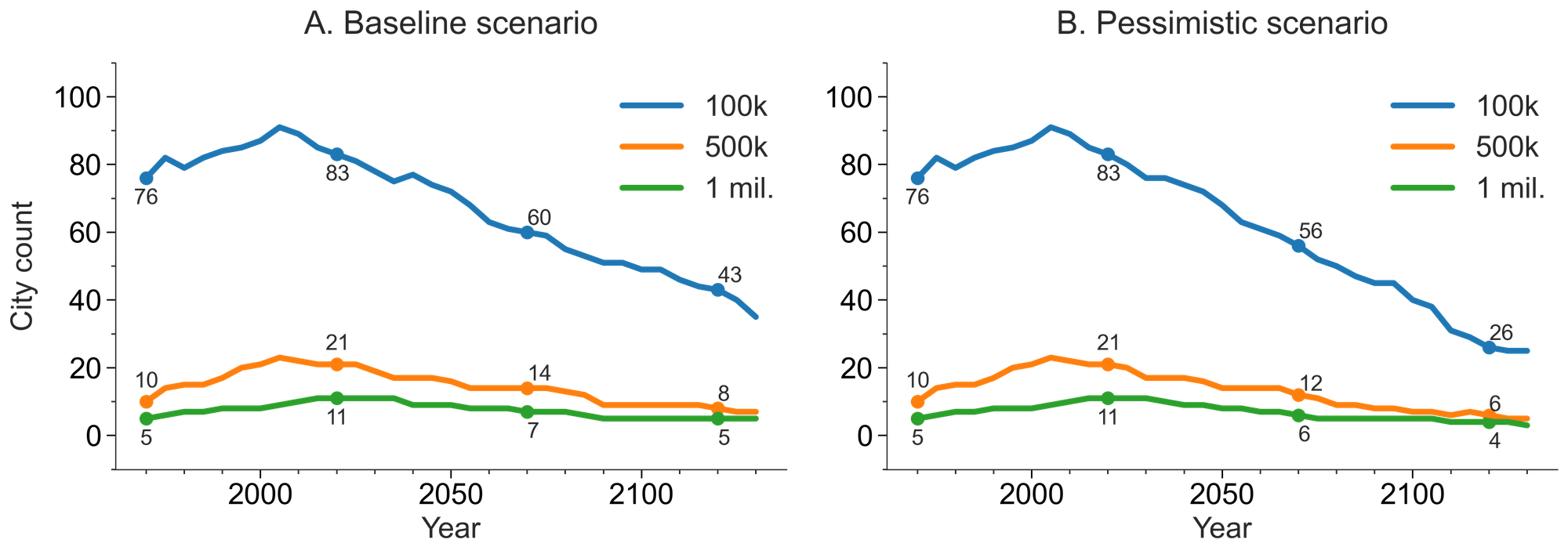}
    \caption{Numbers of cities of at least 100, 500 thausand and 1 million inhabitants}
    \label{fig:city-count-given-size}
    \end{minipage}
\end{figure}

Figure \ref{fig:size-dist} presents the observed city size distribution in 2020 and the projected distributions for 2070 and 2120 under the baseline (panel A) and pessimistic (panel B) scenarios.
While the overall slopes of the fitted regression lines remain largely unchanged, indicating little increase in skewness, Tokyo’s share of the urban population grows disproportionately.
Tokyo accounts for 34\% of the urban population in 2020; this is projected to rise to 38\% and 39\% in 2070, and to 41\% and 44\% in 2120 under the baseline and pessimistic scenarios, respectively
Thus, Tokyo’s dominance intensifies, especially under the pessimistic scenario.
From 2020 to 2070, this represents an increase of 4–5 percentage points—equivalent to 2.9–3.3 million people shifting from rural cities and regions to Tokyo.

Figure \ref{fig:city-count-given-size}A and B illustrate the projected changes in the number of cities with populations of at least 100,000, 500,000, and 1 million under the baseline and pessimistic scenarios, respectively.
As the total population declines, the number of cities with at least 100,000 inhabitants decreases more rapidly than that of larger cities.
This trend is especially pronounced in the pessimistic scenario, where the disappearance of smaller cities is even more marked.

\begin{figure}[htbp!]
    \centering
    \begin{minipage}[c]{\textwidth}
    \centering
    \captionsetup{width=\linewidth}
    
    \includegraphics[width=\textwidth]{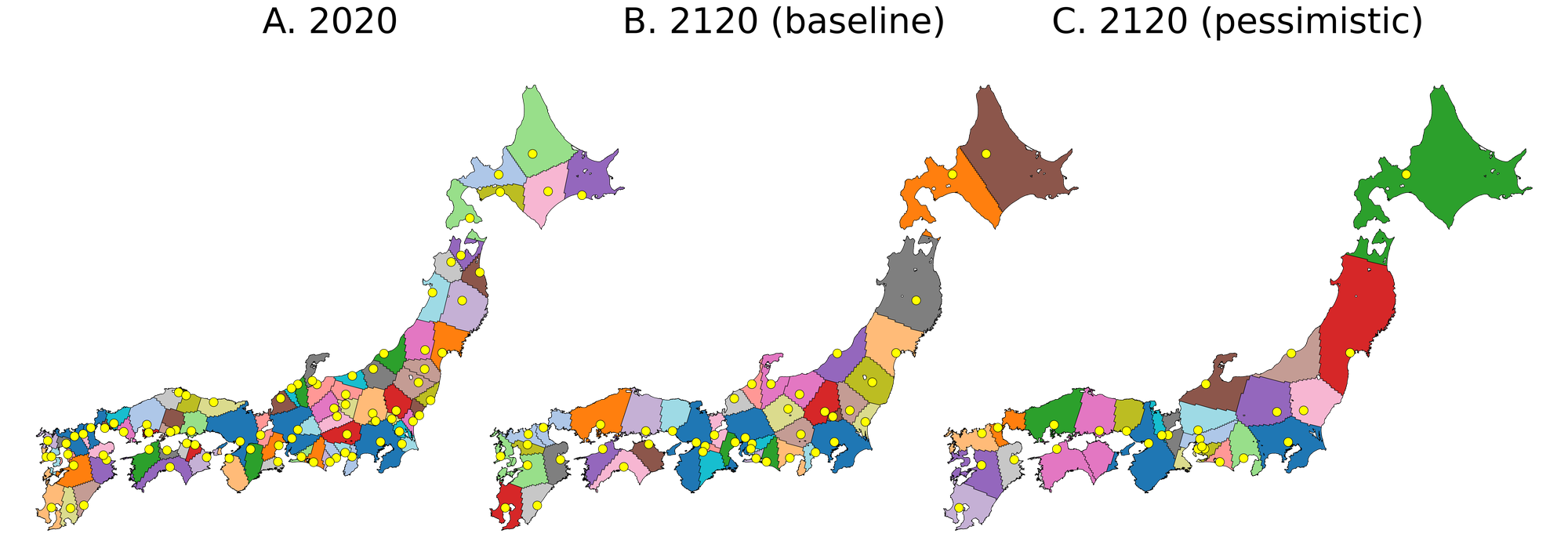}
    \caption{Locations of cities with at least 100,000 inhabitants in 2020 and 2120}
    \caption*{\footnotesize\textit{Note}: Locations (yellow circles) of cities with at least 100,000 inhabitants in 2020 (panel A), in 2120 in the baseline scenario (panel B) and in the pessimistic scenario (panel C). 
    There are 83 such cities in 2020, and 43 and 26 in 2120 in the baseline and pessimistic scenarios, respectively.
    The colored cells represent the Voronoi partition of the country with respect to each of these cities, where each grid cell is assigned to the closest one of the cities with at least 100,000 inhabitants. To obtain the partition, the distance is calculated as a linear distance.}
    \label{fig:100k}
    \end{minipage}
    
    \vspace{1cm}
    
    \begin{minipage}[c]{\textwidth}
    \centering
    \captionsetup{width=\linewidth}
    
    \includegraphics[width=\textwidth]{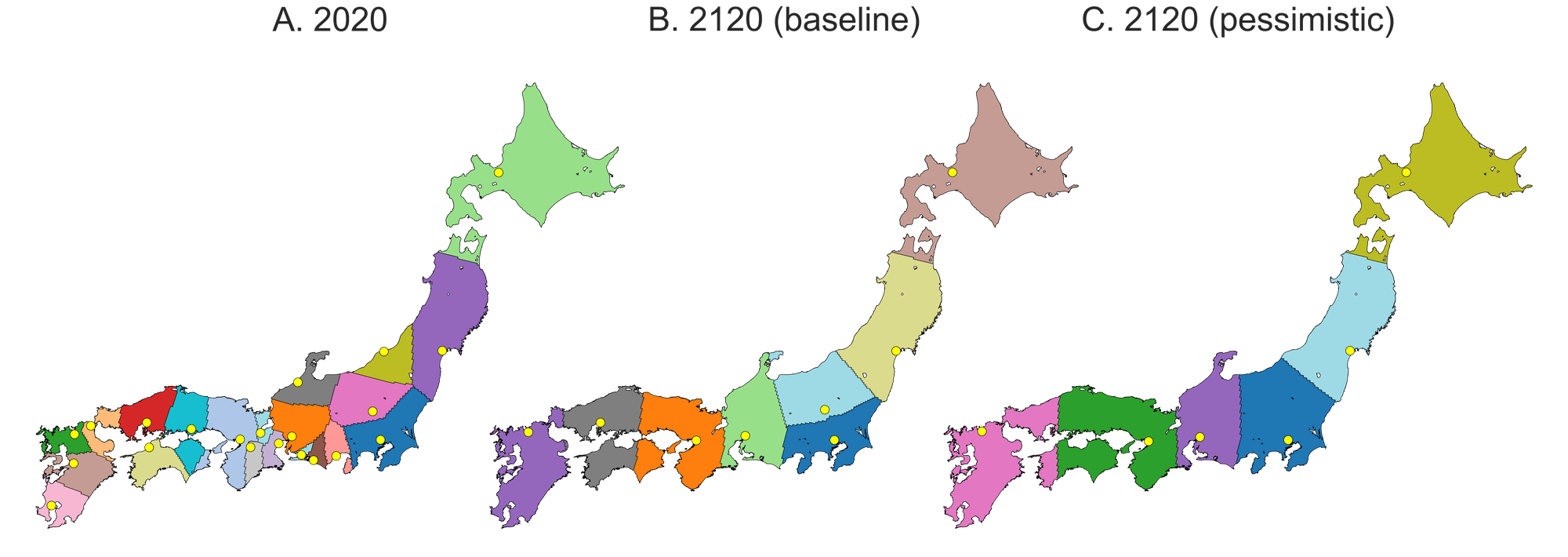}
    \caption{Locations of cities with at least 500,000 inhabitants in 2020 and 2120}
    \caption*{\footnotesize\textit{Note}: Locations (yellow circles) of cities of at least 500,000 inhabitants in 2020 (panel A), in 2120 in the baseline scenario (panel B), and in the pessimistic scenario (panel C). 
    There are 21 such cities in 2020, and 8 and 6 in 2120 in the baseline and pessimistic scenarios, respectively.
    The colored cells represent the Voronoi partition of the country with respect to each of these cities. In obtaining the partition, the distance is calculated as the straight-line distance.}
    \label{fig:500k}
    \end{minipage}
    

\end{figure}

Figure \ref{fig:100k} depicts the changing spatial distribution of cities with at least 100,000 inhabitants (yellow circles) from 2020 to 2120, with colored cells indicating the Voronoi partition based on linear distance. In 2020 (panel A), 83 such cities are spread across the country, but by 2120, the remaining cities are largely concentrated in today’s major urban regions and along key transportation corridors (see Fig.,\ref{fig:tcost}), in both the baseline (panel B) and pessimistic (panel C) scenarios. The pessimistic scenario, in particular, reveals a sharper decline in eastern Honshu and Hokkaido, areas already affected by population aging and decline, resulting in more cities falling below the 100,000 population threshold. These region-specific dynamics are effectively captured by city- and grid-level time-series models.

\begin{figure}[htbp!]
    \centering
    \captionsetup{width=\linewidth}
    \includegraphics[width=\textwidth]{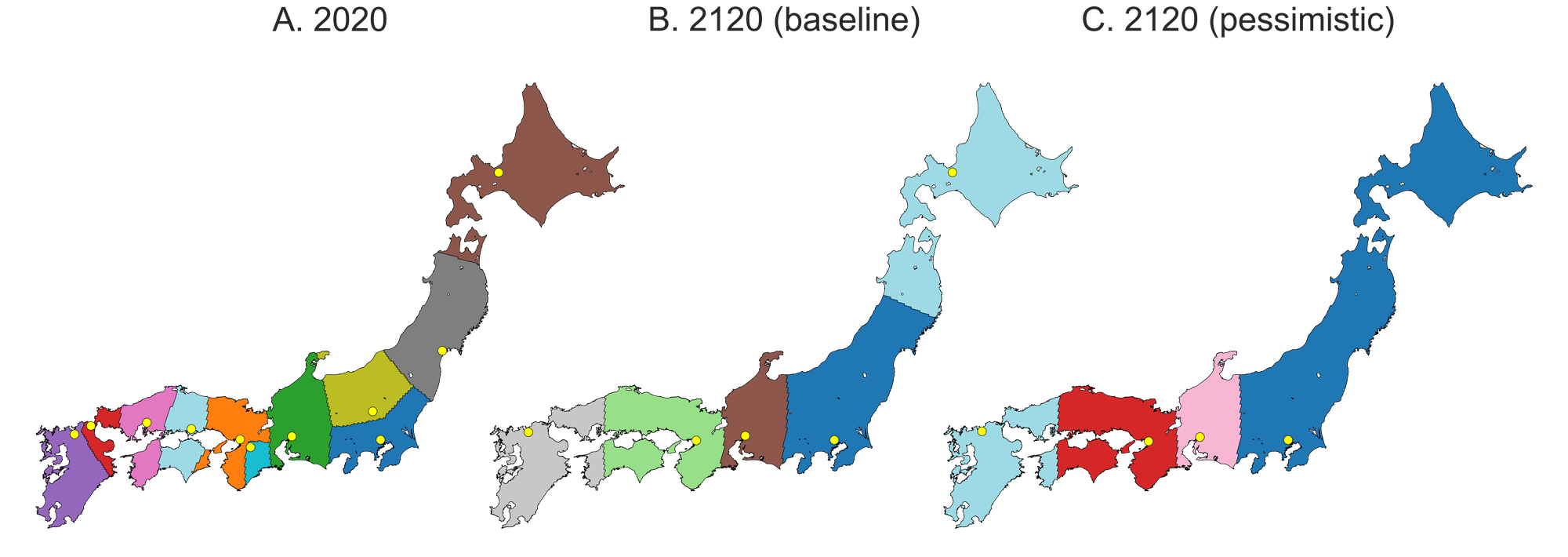}
    \caption{Locations of cities with at least 1 million inhabitants in 2020 and 2120}
    \caption*{\footnotesize\textit{Note}: Locations (yellow circles) of cities of at least 1 million inhabitants in 2020 (panel A), in 2120 in the baseline scenario (panel B), and in the pessimistic scenario (panel C). 
    There are 11 such cities in 2020, and 5 and 4 in 2120 in the baseline and pessimistic scenarios, respectively.
    The colored cells represent the Voronoi partition of the country with respect to each of these cities. In obtaining the partition, the distance is calculated as the straight-line distance.}
    \label{fig:1mil}
 \end{figure}

The enlargement of the Voronoi areas for surviving cities indicates that cities of a given size will be spaced farther apart, reflecting the spatial fractal structure of Japan’s urban system. This increased separation is consistent with economic theories of agglomeration, which suggest that reduced distance frictions can lead to greater dispersion among cities of similar size \citep[][]{Akamatsu-et-al-DP2024}. Similar trends are observed for cities with populations of at least 500,000 and 1 million (Figs.,\ref{fig:500k} and \ref{fig:1mil}), where their numbers are projected to decline by more than half in the baseline scenario and by about two-thirds in the pessimistic scenario over the next century, resulting in a more dispersed geographic distribution within these larger city size classes.



%

\subsubsection*{Concentration at the regional levels} 
\begin{figure}[htbp!]
    \centering
    \captionsetup{width=\linewidth}
    
    \includegraphics[width=.7\textwidth]{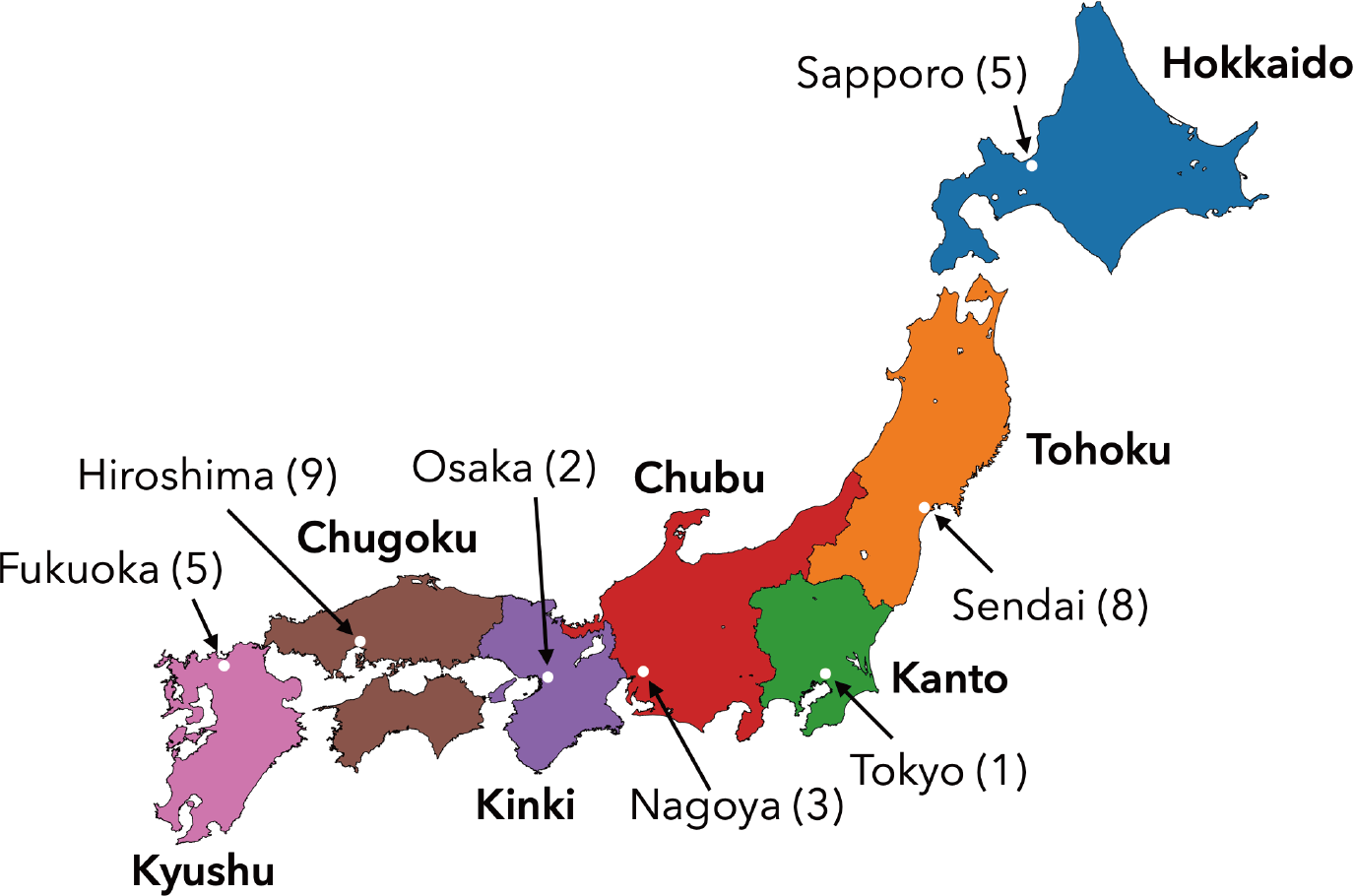}
    \caption{Seven regional divisions with their largest cities}
    \caption*{\footnotesize\textit{Note}: The location of the largest city in each region is indicated by a white circle, with the number in parentheses being the city's population size ranking in the country.}
    \label{fig:7regions}
\end{figure}

\begin{figure}[htbp!]
    \centering

    
    \begin{minipage}[c]{\textwidth}
    \centering
    \captionsetup{width=\linewidth}
    
    \includegraphics[width=.8\textwidth]{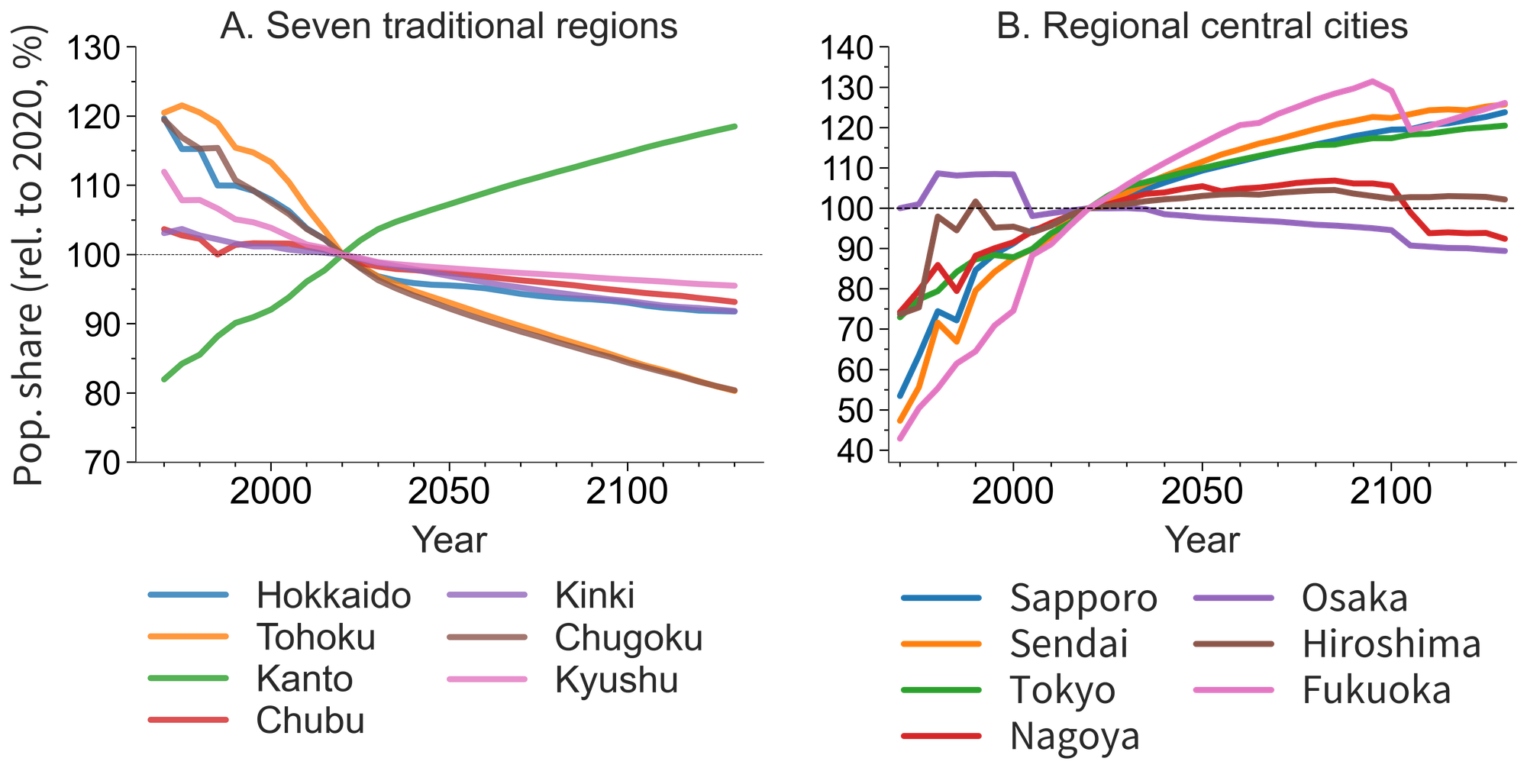}
    \caption{Concetration in the country and regional levels (baseline)}
    \caption*{\footnotesize\textit{Note}: The national population is assumed to decrease according to the baseline scenario. (A) Shares of the seven regional divisions in the national population, relative to their shares in 2020. (B) Shares of the largest cities in the seven regions in the national population, relative to their shares in 2020.}
    \label{fig:7regions-growth-intermediate}
    \end{minipage}

    \vspace{1cm}

    \begin{minipage}[c]{\textwidth}
    \centering
    \captionsetup{width=\linewidth}
    \includegraphics[width=.8\textwidth]{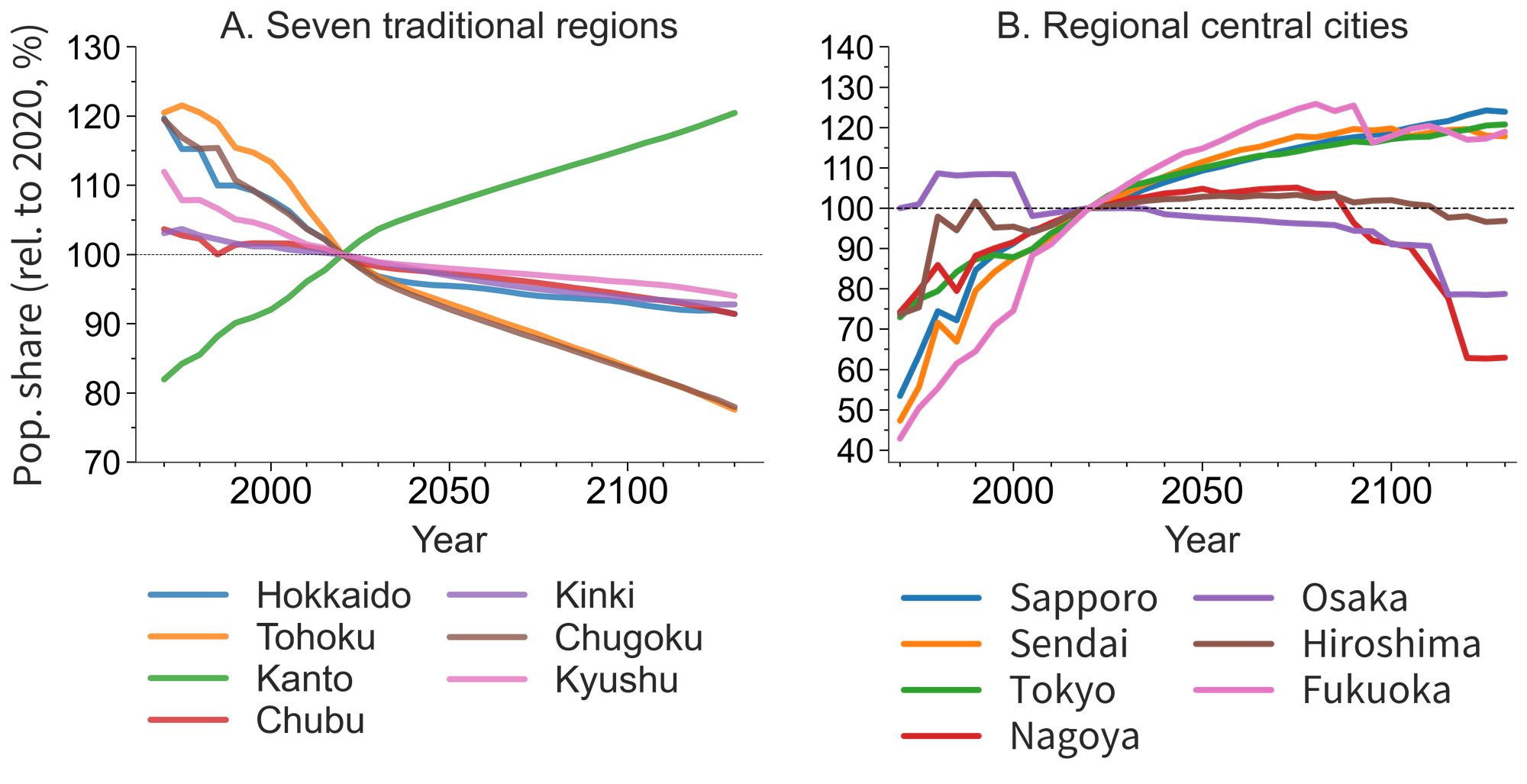}
    \caption{Concetration in the country and regional levels (pessimistic)}
    \caption*{\footnotesize\textit{Note}: The national population is assumed to decrease according to the pessimistic scenario. (A) Shares of the seven regional divisions in the national population, relative to their shares in 2020. (B) Shares of the largest cities in the seven regions in the national population, relative to their shares in 2020.}
    \label{fig:7regions-growth-pessimistic}
    \end{minipage}

\end{figure}

Population concentration in Japan is projected to intensify not only at the national level but also within its subregions. Figure \ref{fig:7regions} shows the traditional seven regional divisions and the locations of their largest cities. According to Figure \ref{fig:7regions-growth-intermediate}A, under the baseline scenario, the Kanto region—which includes Tokyo—will capture an increasingly disproportionate share of the population, while all other regions are expected to lose population share. This trend suggests that the monopolar structure centered on Tokyo will become even more pronounced in the future.

At the subregional scale, Figure \ref{fig:7regions-growth-intermediate}B reveals a similar pattern: the largest city in each region, with the exception of Osaka in the Kinki region, is projected to increase its share of the regional population, mirroring the concentration observed at the national level. This indicates that the monopolar structure is replicated within subregions as well.

Osaka, the largest city in the Kinki region and Japan’s second largest city, has stagnated since 2000, coinciding with the expansion of the Nozomi Shinkansen, which significantly shortened travel times and greatly increased the frequency of trains between Tokyo and Osaka.%
\footnote{The travel time between Tokyo and Osaka was 2 hours and 50 minutes in 1991, before the introduction of the Nozomi, and was reduced to 2 hours and 30 minutes in 2000 after the start of the Nozomi. 
More importantly, the frequency of the Nozomi trains has improved from 2 round trips per day when it was introduced in 1992 to 15-20 round trips in 2000 and more than 100 round trips in 2020.}\ 
The improved connectivity has diminished Osaka’s competitiveness as a regional center, as Tokyo’s economic influence now extends more easily into western Japan. This phenomenon, known in economic geography as the "agglomeration shadow," occurs when enhanced transport links allow the dominant city to outcompete secondary centers by expanding its market reach \citep[e.g.,][]{Fujita-Krugman-RSUE1995}.

Nagoya and Hiroshima, the third and ninth largest cities, also show relatively lower concentration, likely due to their location on the Nozomi line and improved access to Tokyo. 
Nagoya, situated between Tokyo and Osaka, faces particularly intense competition from Tokyo, but its resilience compared to Osaka may be attributed to its strong manufacturing base, as it has recently been Japan’s leading manufacturing exporter \citep[see, e.g.,][]{Mori-Wrona-RSUE2024}. 
Nevertheless, both Nagoya and Hiroshima are projected to decline relative to their 2020 levels, with Nagoya’s decline accelerating after 2095. 
This is partly due to Nagoya's low population density, a result of its car-oriented urban planning with a wide street grid, which makes the city more vulnerable to population decline as density thresholds are more easily violated.

Overall, these findings suggest that Japan’s urban hierarchy will become increasingly centralized around Tokyo, with similar monopolar patterns emerging within subregions. Improvements in transportation infrastructure and differences in urban form are key factors shaping the resilience or vulnerability of individual cities in the face of ongoing demographic change.

Under the pessimistic scenario, where population decline accelerates, the concentration toward the largest cities intensifies at both national and regional levels (Fig.,\ref{fig:7regions-growth-pessimistic}). By 2070 and 2120, the Kanto region’s population share rises by 10\% and 19\%, respectively, under this scenario—slightly higher than the 10\% and 17\% increases projected in the baseline. Notably, fewer large cities maintain population share growth in the pessimistic scenario compared to the baseline (Fig.,\ref{fig:7regions-growth-pessimistic}B), as the shrinking total population must be distributed across a diminishing number of urban centers. This dynamic reinforces the dominance of Tokyo and select regional hubs, while accelerating the decline of smaller cities.

\subsection{Dispersion at the city level}
\label{sec:dispersion}
\begin{figure}[htbp!]
    \centering
    \begin{minipage}[c]{\textwidth}
    \centering
    \captionsetup{width=\linewidth}
    
    \includegraphics[width=.9\textwidth]{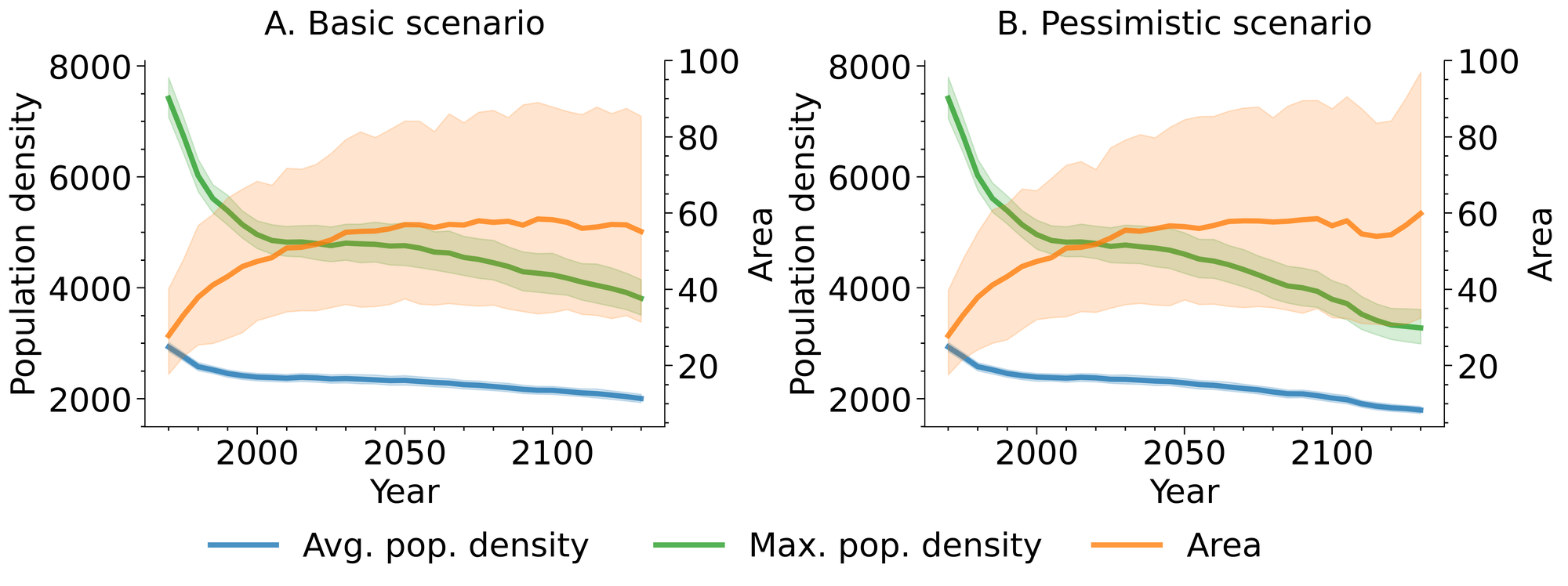}
    \caption{Predicted dispersion within a city}
    \caption*{\footnotesize\textit{Note}: (A) and (B) show the prediction of population density and area within individual cities under the baseline and pessimistic scenarios, respectively, of the NIPSSR's projection of Japan's population. 
    In each panel, the blue and green lines show the means of the average and maximum population densities within a city, and the orange line shows the average total area of a city for each year indicated along the horizontal axis, while the shaded area indicates the range covering 95\% of the values for individual cities.}
    \label{fig:dispersion}
    \end{minipage}
    
    \vspace{1cm}
    
    \begin{minipage}[c]{\textwidth}
    \centering
    \includegraphics[width=.9\textwidth]{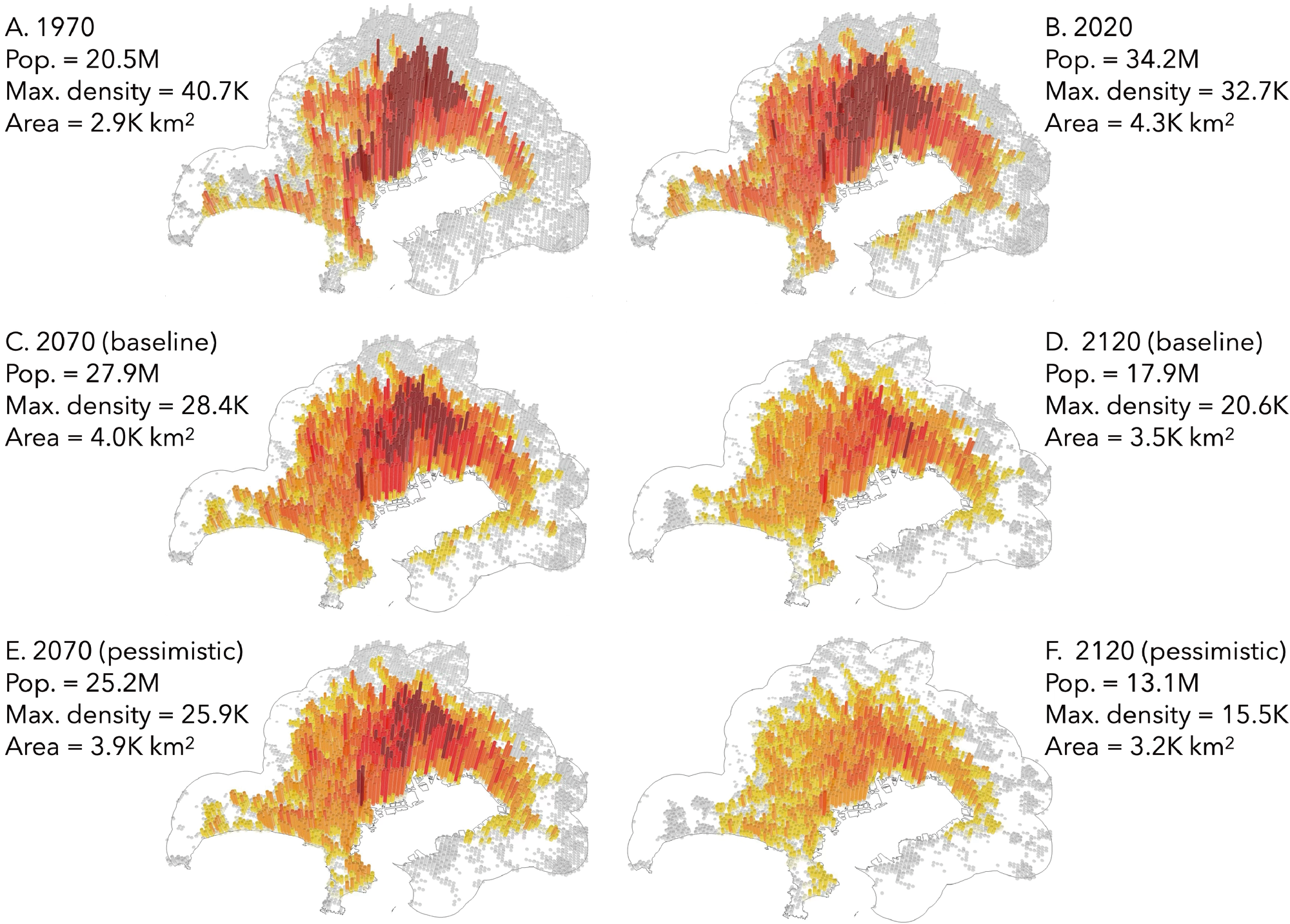}
    \caption{Geographic population distribution in Tokyo. The grid cells with bars colored from yellow to red indicate the coverage of Tokyo UA.}
    \caption*{\footnotesize\textit{Note}: (A) and (B) are the realized population distributions over the grid cells in Tokyo UA in 1970 and 2020, respectively. (C) and (D) 
 are the projected population distributions in 2070 and 2120, respectively, under the baseline scenario, while (E) and (F) are those under the pessimistic scenario. The warmer colors indicate larger populations. The darkest grid cells have at least 20,000 inhabitants. The other thresholds are 15,000, 10,000, 5,000, 2,000, and 1,000 inhabitants.}
    \label{fig:tokyo}
    \end{minipage}
 
\end{figure}

The evolution of the spatial distribution of population within each city projected into the future largely mirrors the trends observed over the past 50 years. \cref{fig:dispersion}A and B extend the results shown in \cref{fig:local-dispersion} to future years (2025–2120) under the baseline and pessimistic scenarios, respectively. In the baseline scenario, both the maximum and average population densities within cities are projected to decrease by 5\% from 2020 to 2070, and by 17\% and 13\%, respectively, by 2120. In the pessimistic scenario, the decline in urban density is, on average, twice as rapid: the maximum and average densities are expected to fall by 10\% and 8\% from 2020 to 2070, and by 31\% and 23\% by 2120, respectively.

Although the tendency for larger cities to attract population from across the country will partially mitigate their shrinkage, it will not be sufficient to maintain their population size and density. Notably, the distribution of city area sizes shows no significant shift after 2070, indicating that the flattening of the population distribution within cities is progressing faster than the overall decline in city population.

\begin{figure}[htbp!]
 \centering
 \includegraphics[width=.9\textwidth]{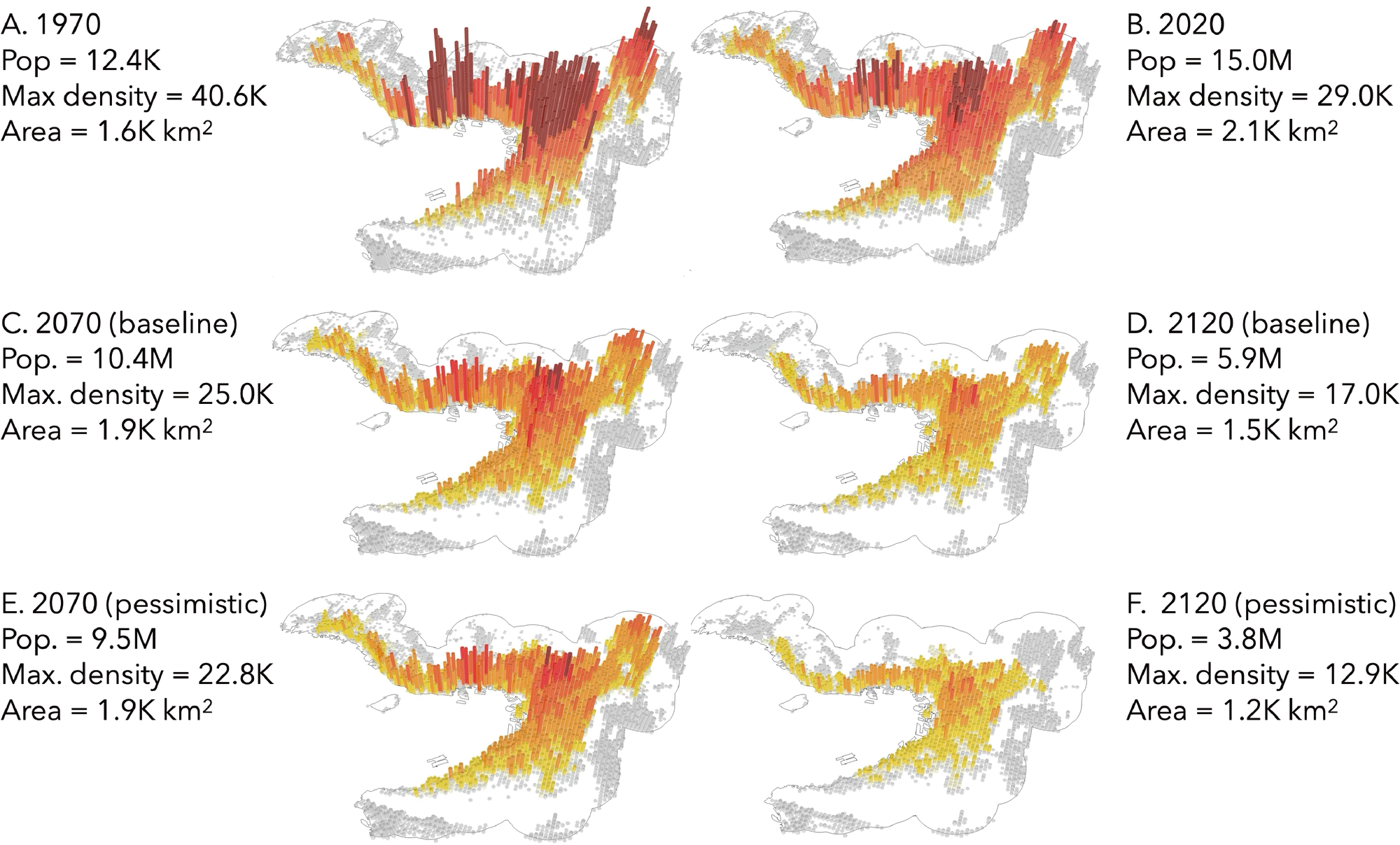}
 \caption{Geographic population distribution in Osaka. The grid cells with bars colored from yellow to red indicate the coverage of Osaka UA.}
 \caption*{\footnotesize\textit{Note}: (A) and (B) are the realized population distributions over the grid cells in Osaka UA in 1970 and 2020, respectively. (C) and (D) 
 are the projected population distributions in 2070 and 2120, respectively, under the baseline scenario, while (E) and (F) are those under the pessimistic scenario. The warmer colors indicate larger populations. The darkest grid cells have at least 20,000 inhabitants. The other thresholds are 15,000, 10,000, 5,000, 2,000, and 1,000 inhabitants.} 
 \label{fig:osaka}
\end{figure}

\begin{figure}[htbp!]
 \centering
 \includegraphics[width=.9\textwidth]{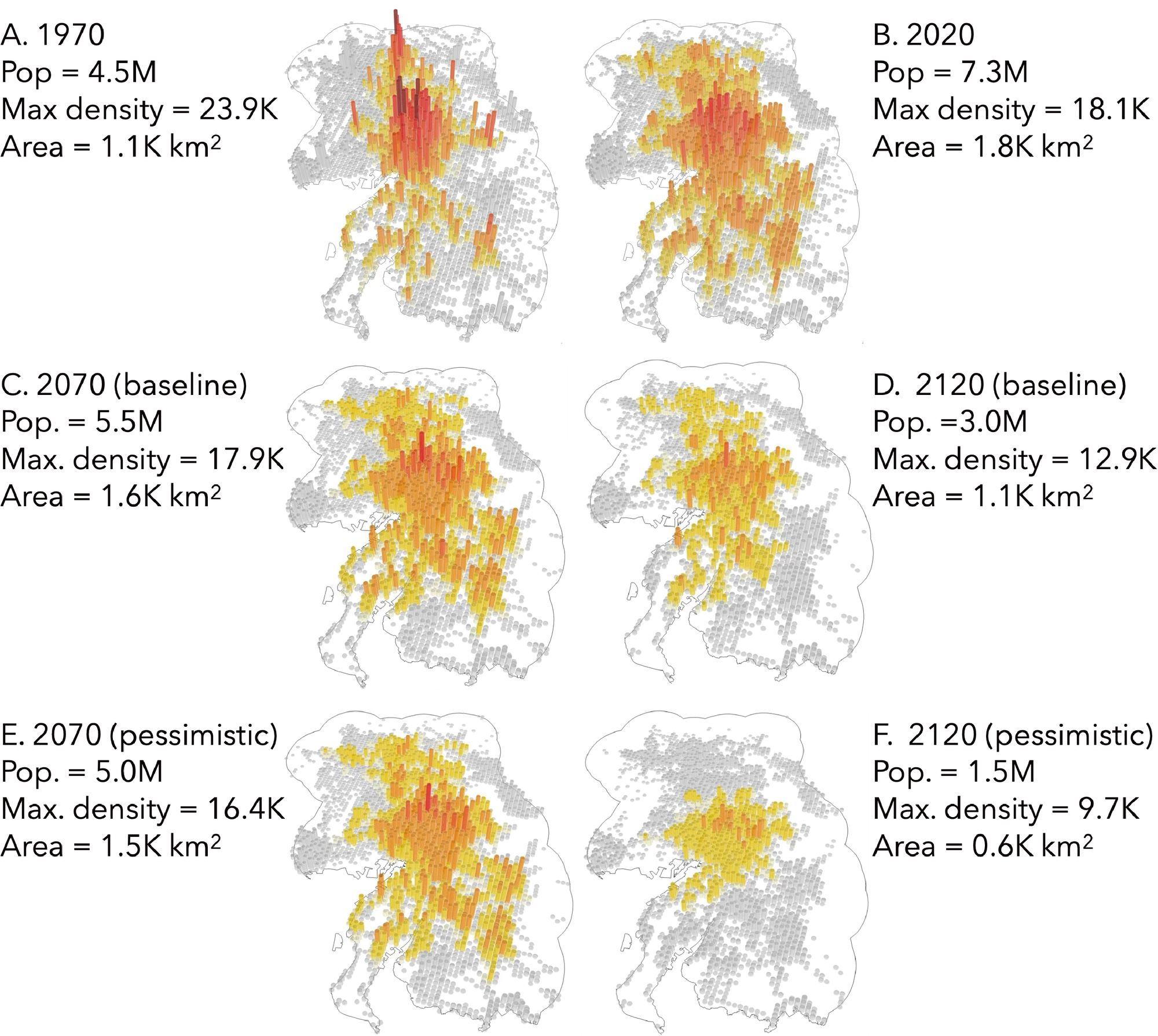}
 \caption{Geographic population distribution in Nagoya. The grid cells with bars colored from yellow to red indicate the coverage of Nagoya UA.}
 \caption*{\footnotesize\textit{Note}: (A) and (B) are the realized population distributions over the grid cells in Nagoya UA in 1970 and 2020, respectively. (C) and (D) 
 are the projected population distributions in 2070 and 2120, respectively, under the baseline scenario, while (E) and (F) are those under the pessimistic scenario. The warmer colors indicate larger populations. The darkest grid cells have at least 20,000 inhabitants. The other thresholds are 15,000, 10,000, 5,000, 2,000, and 1,000 inhabitants.} 
 \label{fig:nagoya}
\end{figure}

\begin{figure}[htbp!]
 \centering
 \includegraphics[width=.9\textwidth]{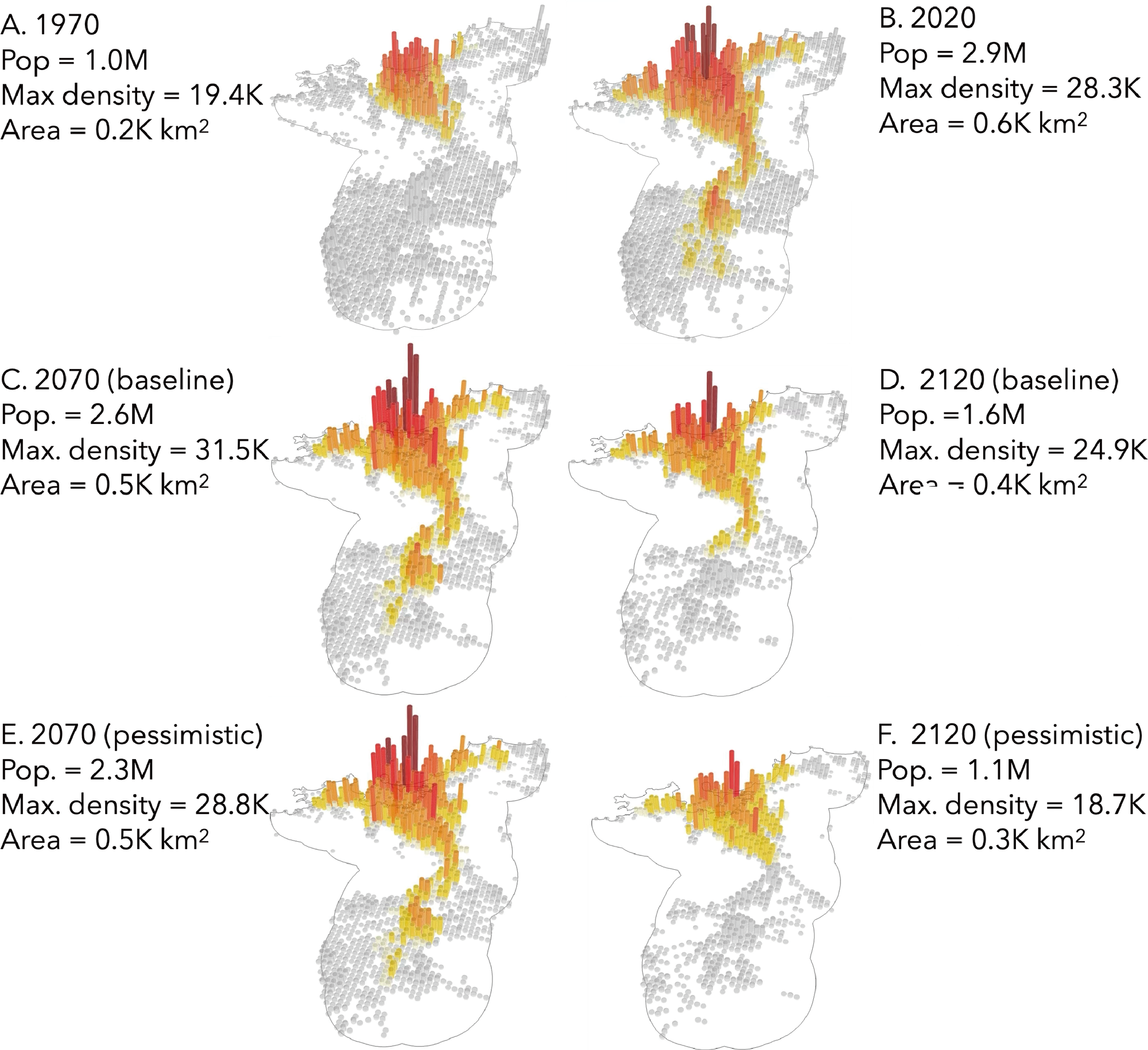}
 \caption{Geographic population distribution in Fukuoka. The grid cells with bars colored from yellow to red indicate the coverage of Fukuoka UA.}
 \caption*{\footnotesize\textit{Note}: (A) and (B) are the realized population distributions over the grid cells in Fukuoka UA in 1970 and 2020, respectively. (C) and (D) 
 are the projected population distributions in 2070 and 2120, respectively, under the baseline scenario, while (E) and (F) are those under the pessimistic scenario. The warmer colors indicate larger populations. The darkest grid cells have at least 20,000 inhabitants. The other thresholds are 15,000, 10,000, 5,000, 2,000, and 1,000 inhabitants.} 
 \label{fig:fukuoka}
\end{figure}

We take a closer look at Japan’s four largest cities. In \cref{fig:tokyo,fig:osaka,fig:nagoya,fig:fukuoka}, panels A and B show the population distributions in Tokyo, Osaka, Nagoya, and Fukuoka in 1970 and 2020, respectively. Panels C and D present our projections for 2070 and 2120 under the baseline scenario, while panels E and F provide the corresponding projections under the pessimistic scenario.

\subsubsection*{Tokyo -- The largest city}
Tokyo’s population grew by 67\%, from 13.7 million in 1970 to 34.2 million in 2020, far outpacing the national growth rate of 21\% over the same period. 
This increase alone exceeds the entire population of Osaka, Japan’s second largest city, in 1970. 
However, Tokyo’s internal population distribution has become noticeably flatter: the 100th, 99th, and 95th percentile values for population density by grid cell have declined by 21\% (from 41,000 to 33,000), 11\% (from 31,000 to 27,000), and 8\% (from 22,000 to 21,000), respectively, indicating that the most densely populated areas have seen the steepest declines.

Looking ahead, Tokyo’s population is projected to decrease by 18\% by 2070 and 48\% by 2120 in the baseline scenario, falling to 28 million and 18 million, respectively. This decline is slower than the projected national population decrease of 31\% by 2070 and 61\% by 2120. In the pessimistic scenario, Tokyo’s population is expected to fall by 25\% to 25 million by 2070 and by 62\% to 13 million by 2120, still less severe than the national declines of 38\% and 73\%, respectively. These trends suggest that Japan’s demographic and economic landscape will become increasingly polarized toward Tokyo.

At the same time, Tokyo’s internal population distribution is expected to flatten further. The city’s maximum population density is projected to decrease by 13\% to 28,000 per km$^2$ by 2070 and by 37\% to 21,000 per km$^2$ by 2120 in the baseline scenario, while the city’s area contracts only modestly by 6\% to 4,000 km$^2$ by 2070 and 19\% to 3,500 km$^2$ by 2120. In the pessimistic scenario, maximum density is projected to drop by 21\% to 26,000 per km$^2$ by 2070 and by 53\% to 15,000 per km$^2$ by 2120, with the city’s area decreasing by 9\% to 3,900 km$^2$ and 26\% to 3,200 km$^2$, respectively. This indicates that the decline in density will outpace the reduction in Tokyo’s spatial extent, further flattening the city’s population distribution.

\subsubsection*{Osaka -- The second largest city}

Osaka’s population grew by 22\%, from 12 million to 15 million between 1970 and 2020, closely tracking the national growth rate over the same period. Despite this growth, Osaka experienced a notable flattening of its internal population distribution: the 100th, 99th, and 95th percentile values of population density by grid cell declined by 28\% (from 41,000 to 29,000), 30\% (from 31,000 to 22,000), and 25\% (from 23,000 to 17,000), respectively. This indicates that density has decreased most sharply in the city’s densest areas.

Looking ahead, Osaka’s population is projected to decline by 30\% to 10 million by 2070 and by 61\% to 6 million by 2120 in the baseline scenario. Under the pessimistic scenario, the declines are even steeper: 37\% to 9 million by 2070 and 74\% to 4 million by 2120. These trends suggest that Osaka is gradually losing its centrality relative to Tokyo.

Mirroring Tokyo, Osaka’s internal population distribution is expected to flatten further in the coming decades. In the baseline scenario, the city’s maximum population density is projected to decrease by 14\% to 25,000 per km$^2$ by 2070 and by 41\% to 17,000 per km$^2$ by 2120, while its area contracts modestly by 10\% to 1,900 km$^2$ by 2070 and 27\% to 1,500 km$^2$ by 2120. 
In the pessimistic scenario, the maximum density is expected to fall by 21\% to 23,000 per km$^2$ by 2070, and by 56\% to 13,000 per km$^2$ by 2120, while its area decreases by 9.5\% to 1,900 km$^2$ by 2070 and by 43\% to 1,200 km$^2$ by 2120. These projections highlight that the decline in population density will outpace the reduction in Osaka’s spatial extent, further accelerating the city’s internal flattening.

\subsubsection*{Nagoya -- The third largest city}

Nagoya’s population grew by 67\%, from 4.5 million to 7.3 million between 1970 and 2020, yet, like Tokyo and Osaka, the city experienced a significant flattening of its internal population distribution. During this period, the 100th, 99th, and 95th percentile values of population density by grid cell declined by 25\% (from 24,000 to 18,000), 31\% (from 19,000 to 13,000), and 28\% (from 13,000 to 10,000), respectively. Notably, Nagoya’s population density in 2020 is considerably lower than that of other major cities such as Tokyo and Osaka, a result of its distinctive urban planning that favors lower density. This characteristic makes Nagoya especially vulnerable to rapid spatial contraction as its population declines, since it is more likely to fall below the density threshold required for inclusion as an urban area.

Projections indicate that Nagoya’s population will decrease by 23\% to 5.5 million by 2070 and by 59\% to 3.0 million by 2120 in the baseline scenario. In the pessimistic scenario, the declines are even steeper—32\% to 5.0 million by 2070 and 77\% to 1.5 million by 2120. 

Although Nagoya's future decline in maximum population density is modest compared to the two larger cities, its spatial extent is projected to shrink significantly: by 15\% to 1,600 km$^2$ by 2070 and by
42\% to 1,400 km$^2$ by 2120 in the baseline scenario, and by 18\% to 1,500 km$^2$ and by 69\% to just 600 km$^2$ in the pessimistic scenario. This pronounced decrease in both population and area underscores how Nagoya's relatively low density accelerates the city's shrinkage as demographic decline intensifies.

\subsubsection*{Fukuoka -- The fourth largest city}

Fukuoka’s demographic trajectory diverges markedly from Japan’s three largest cities. Between 1970 and 2020, its population surged by 184\%, from 1.0 million to 2.9 million, driven by the 1975 extension of the Sanyo Shinkansen to Fukuoka and the 1982 expansion of the Tohoku Shinkansen to Sendai, which enhanced its connectivity to Tokyo. This infrastructure positioned Fukuoka within a one-day travel radius of half of Japan’s population, fueling concentrated growth around its urban core. Unlike Tokyo, Osaka, and Nagoya, Fukuoka saw increases in its 100th, 99th, and 95th percentile population density values by 46\% (19,000 to 28,000), 25\% (17,000 to 20,000), and 9\% (13,000 to 14,000), respectively, reflecting a centralized growth pattern.

However, our projections suggest diminishing locational advantages as distance frictions decrease nationwide. Under the baseline scenario, Fukuoka’s population is expected to decline by 12\% to 2.6 million by 2070 and by 46\% to 1.6 million by 2120. The pessimistic scenario forecasts steeper drops of 20\% to 2.3 million and 62\% to 1.1 million over the same periods. 
Interestingly, despite overall population loss, Fukuoka’s maximum density is projected to rise temporarily by 11\% (3,200/km$^2$) and 2\% (2,900/km$^2$) by 2070 in baseline and pessimistic scenarios, respectively, before declining to 2,500 per km$^2$ (-12\%) and 1,900 per km$^2$ (-34\%) by 2120. This contrasts with Osaka and Nagoya, where density declines are more immediate and severe. By 2120, Fukuoka’s urban core density is projected to surpass both cities, though it cannot escape Japan’s broader demographic collapse. The city’s trajectory underscores how even vibrant regional hubs face inevitable shrinkage and flattening under nationwide population decline.

While other surviving cities exhibit distinct demographic patterns, they generally follow a trajectory of population decline and flattening internal population distributions. These trends are further illustrated in Appendix \ref{app:pop-dist}, which details historical and projected population distributions across Japan’s grid cells under both baseline and pessimistic scenarios.

\section{Policy implications}
\label{sec:implications}

There is a considerable divergence between our predictions for the future of Japan's cities and regions and those envisioned by national policy and private urban redevelopment.
On the one hand, Japan's regional policies have repeatedly implemented short-term measures aimed at curbing continued population concentration in major cities, particularly Tokyo.
On the other hand, large cities where people from all over the country are concentrated have managed to maintain population growth or avoid significant population decline even as Japan's overall population has declined over the past 15 years. 
From now on, these major cities will be competing for their own survival, fighting over the limited shrinking total population.
Below, we warn of the future that may result from accepting this divergence and suggest a redirection of regional policy and private incentives.


\subsection{Public policies in rural regions\label{sec:public-policies}}
To promote and sustain local economies, the Japanese government is currently pursuing two major policy initiatives: the “Compact City”  and “Chiho Sosei” (Regional Revitalization) policies. The former seeks to concentrate urban functions and residential areas closer to city centers, thereby enhancing efficiency and livability. The latter encompasses a broad set of measures aimed at maintaining and increasing local populations, revitalizing regional economies, and addressing demographic challenges through job creation, support for in-migration, and the development of local resources.

\begin{figure}[htbp!]
    \centering
    \captionsetup{width=\linewidth}
    \includegraphics[width=\textwidth]{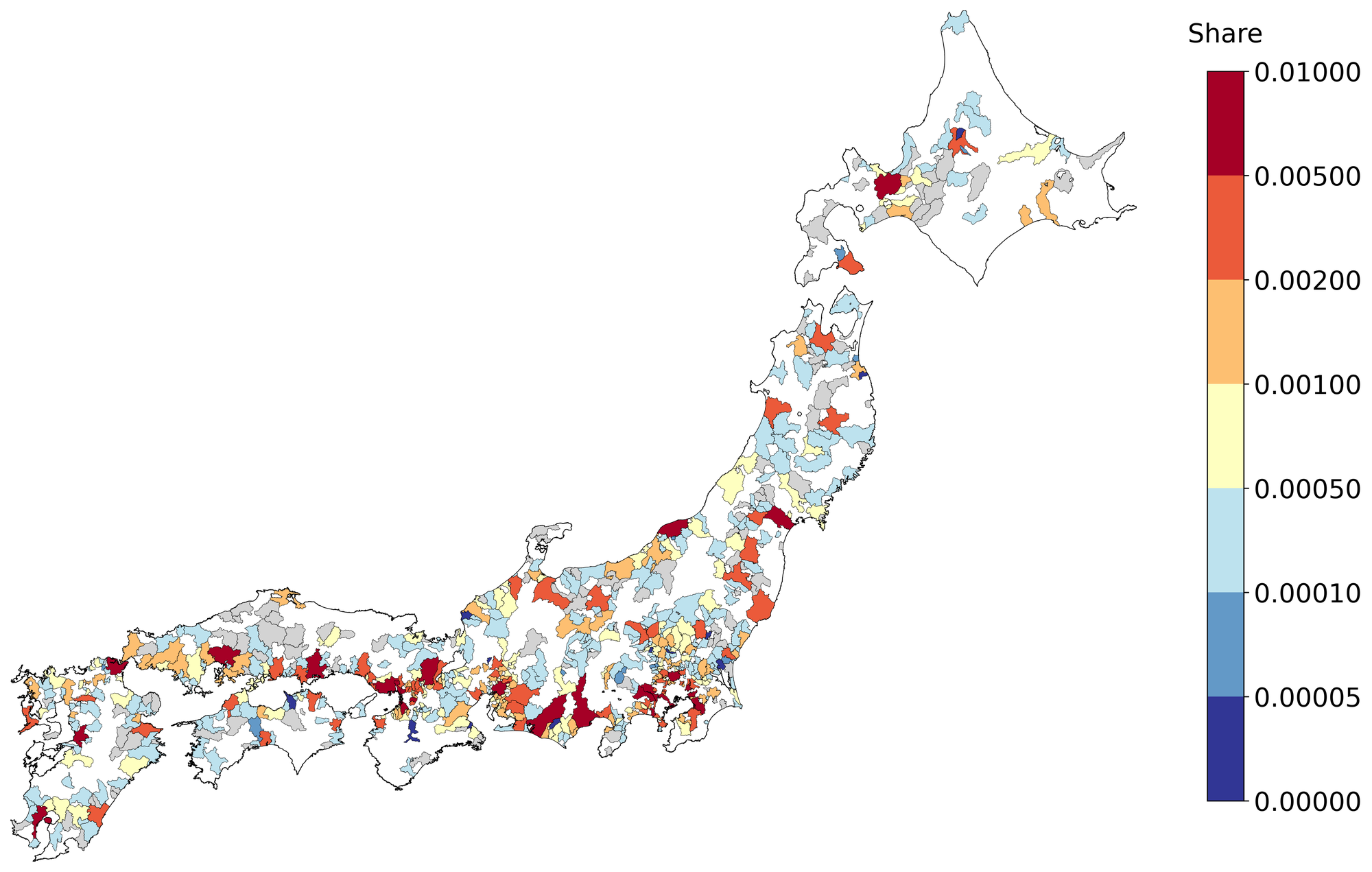}
    \bigskip
    
    \caption{Urban population in the muunicipalities pursuing compact city policies in 2020}
    \caption*{\footnotesize\textit{Note}: Out of a total of 703 municipalities with a compact city policy, our study area includes 694.
    In 2020, 566 of them have a positive urban population, accounting for 73\% of the total urban population, while the rest of the 128 gray municipalities have zero urban population.}
    \label{fig:compact-city-2020}
\end{figure}

According to the MLIT of Japan, 703 of Japan's 1,718 municipalities are currently working on the compact city project.%
\footnote{It is mentioned in \href{https://www.mlit.go.jp/toshi/city_plan/content/001751720.pdf}{the website of MLIT}.}\ 
Fig.\,\ref{fig:compact-city-2020} shows 694 of these municipalities included in our study area in 2020, with warmer colors indicating higher urban population shares. 
Among them, 128 municipalities shown in gray have zero urban population.

\begin{figure}[p!]
    \centering    
    \begin{minipage}[c]{\textwidth}
    \centering
    \captionsetup{width=\linewidth}
    \includegraphics[width=\textwidth]{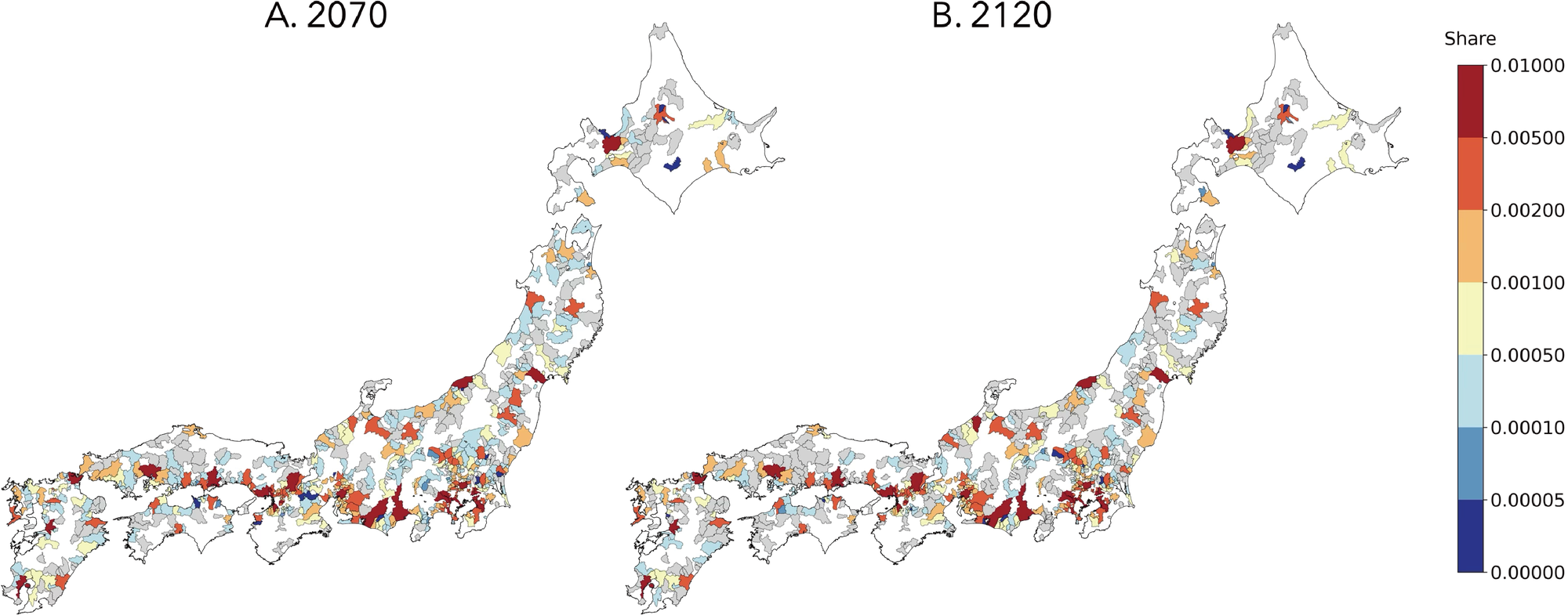}
    \smallskip
    \caption{Urban population in the municipalities pursuing compact city policy in 2070 and 2120 (baseline scenario)}
    \caption*{\footnotesize\textit{Note}: Out of a total of 703 municipalities with a compact city policy, our study area includes 694. 
    (A) In 2070, 484 of them have a positive urban population, accounting for 72\% of the total urban population, while the rest of the 210 gray municipalities have zero urban population.
    (B) In 2120, the number of municipalities with an urban population is projected to decrease to 386 accounting for 70\% share of urban population, while the number of municipalities without an urban population is projected to increase to 308.}
    \label{fig:compact-city-baseline}
    \end{minipage} 
    
    \vspace{1cm}
    
    \begin{minipage}[c]{\textwidth}
    \centering
    \captionsetup{width=\linewidth}
    
    \includegraphics[width=\textwidth]{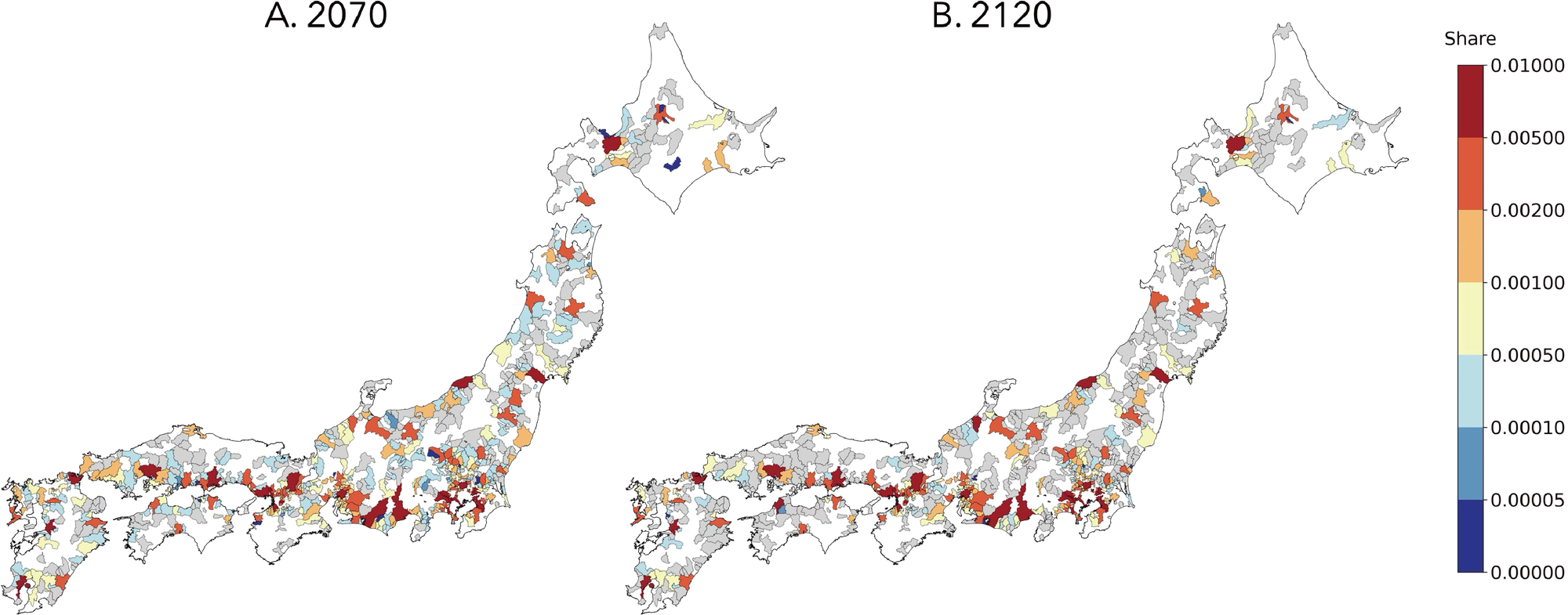}
    \caption{Urban population in the municipalities pursuing compact city policy in 2070 and 2120 (pessimistic scenario)}
    \label{fig:compact-city-pessimistic}
    \caption*{\footnotesize\textit{Note}: Out of a total of 703 municipalities with a compact city policy, our study area includes 694. 
    (A) In 2070, 464 of them have a positive urban population, accounting for 72\% of the total urban population, while the rest of the 230 gray municipalities have zero urban population.
    (B) In 2120, the number of municipalities with an urban population is projected to decrease to 315, accounting for 69\% share of urban population, while the number of municipalities without an urban population is projected to increase to 379.}    
    \end{minipage}
\end{figure}

Figs.\,\ref{fig:compact-city-baseline} and \ref{fig:compact-city-pessimistic} show the urban population shares of these municipalities in the future under the baseline and pessimistic scenarios, respectively.
In 2070 and 2120, the number of municipalities with zero urban population is projected to increase to 210 and 308, respectively, in the baseline scenario, while the corresponding numbers are 230 and 379 in the pessimistic scenario.
Therefore, even if these municipalities develop urban centers, about one third in 50 years and a half of them in 100 years from now will be depopulated and much of the infrastructure and housing developed will be abandoned.
What is needed is to take the predicted sustainable cities as the centers around which the regional economy is reorganized.

The MLIT considers \href{https://www.mlit.go.jp/policy/shingikai/content/001389683.pdf}{``regional living areas''} to be the basic regional unit for maintaining Japan's regions in the future.
A regional living area is defined as a compact regional unit reachable within 60-90 minutes by car that can autonomously maintain essential living and business infrastructure such as healthcare, welfare, transportation, and education, with a population threshold of 100,000.
Our prediction is helpful in identifying the \textit{focal cities}, the urban agglomeration of at least a certain population size, around which sustainable regional living areas are more systematically identified.
%
%
%

\begin{figure}[h!]
 \centering

     \begin{minipage}[c]{\textwidth}
     \includegraphics[width=\textwidth]{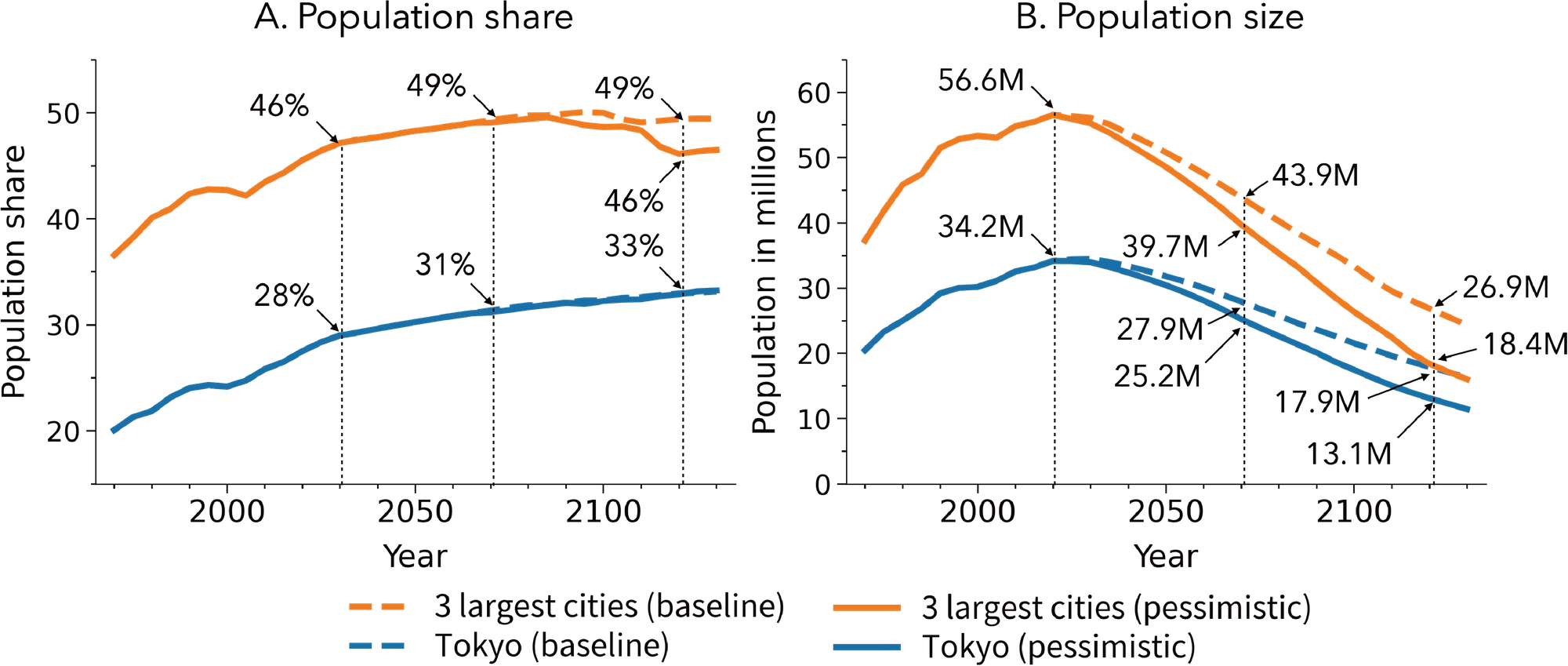}
     \caption{Population shares and sizes of the largest three cities and Tokyo.}
     \label{fig:large-city-growth}

    \end{minipage}
    \vspace{1cm}

     \begin{minipage}[c]{\textwidth}

 \centering
 \includegraphics[width=.6\textwidth]{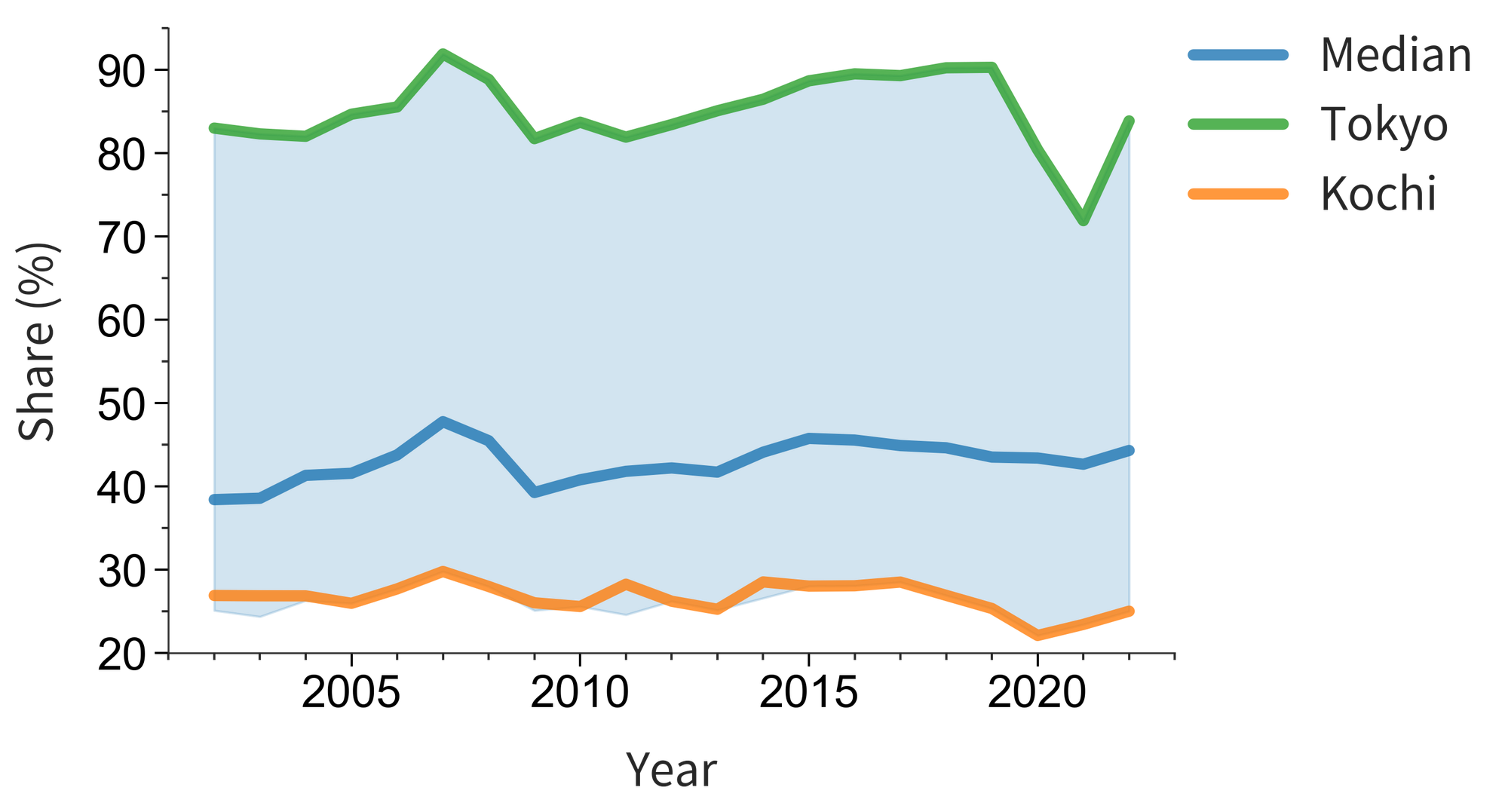}
 \caption{Share of own sources of revenue of prefectures in 2002--2022.}
 \caption*{\footnotesize\textit{Note}: Data source: \href{https://www.soumu.go.jp/iken/kessan\_jokyo\_1.html}{To-Do-Fu-Ken Kessan Jokyo Shirabe} (Prefectural Financial Settlement Status Survey) by MIC of Japan. The shaded band covers all the 47 prefectures.  Tokyo Prefecture has maintained the highest level of self-financing, while Kochi Prefecture has maintained the lowest level most of the time.}
 \label{fig:ownshare}
 \end{minipage}
\end{figure}

Ongoing regional revitalization does not seem promising either. By 2120, Japan's population will decline by 60--70\%, with Tokyo's share rising from 28\% in 2020 to 33\% in 2120 in both the baseline and pessimistic scenarios.
This 5\% increase in share translates into 2.7 million and 2.0 million people in 2120 in the baseline and pessimistic scenarios, respectively. As shown in Fig.\,\ref{fig:large-city-growth}A, Tokyo accounts for most of the share growth among the three largest cities.
This expected relative growth of Tokyo's population is comparable to the disappearance of some prefectures, as 10 out of 47 prefectures have a population of less than 1 million.
Our projections suggest that these most rural areas may not be able to maintain or increase their population, and their infrastructure and facilities invested in the name of regional revitalization may fall into disrepair.

Japan's regional economies are heavily dependent on government subsidies.
Fig.\,\ref{fig:ownshare} shows the own-source revenue shares of 47 prefectures over the period 2002-2022.
The median of the prefectures' own-source revenue shares is only about 40\% throughout the period. 
Not surprisingly, Tokyo Prefecture continues to have the highest level of self-financing.
For these prefectures, which are highly dependent on the government, \hyperlink{https://www.nga.gr.jp/english/}{the National Governors' Conference} made a \hyperlink{https://www.nga.gr.jp/conference/item/423c2cfaac05552d461fb490dbf64be2_2.pdf}{proposal} on August 2, 2024, stating that the reason why regional revitalization policies have not produced results at the macro level is that ``the government has failed to focus policies on the concentration of population in specific areas (Tokyo).  It is necessary to take measures to halt population decline in both rural and urban areas and to alleviate social decline in each region.'' 
But is it really ideal to ``stop population decline'' in all of Japan's rural areas?

The prefectures that have relied most heavily on government subsidies have tended to be rural prefectures that are off the main transport network and were slow to industrialize during the high-growth period from 1955 to the early 1970s.
Although less industrialized, they tend to be rich in agricultural and forestry resources, often along with fisheries resources.
For example, Kochi Prefecture in Shikoku, which currently has the lowest ratio (25\%) of self-financing to prefectural revenue, is a pioneer in Japan in the adoption of the Internet of Plants (IoP) and has a comparative advantage in primary industry, known for its high productivity in horticulture. %
Although Kochi possesses a comparative advantage in primary industries, its industrial structure appears to be heavily shaped by public expenditure. In 2020, primary industries accounted for less than 4\% of the prefecture’s total output, while tertiary industries comprised 78\%.%
\footnote{Data source: \href{https://www.esri.cao.go.jp/jp/sna/data/data\_list/kenmin/files/contents/main\_2021.html}{Kenmin Keizai Keisan} of the Cabinet Office of Japan.}\ 
The optimal population size of Kochi may turn out to be much smaller than the 690,000 residents recorded in 2020.
A similar assessment could apply to other rural prefectures, where current population levels may exceed what is economically sustainable.
Given these circumstances, it may be an opportune moment to reconsider the appropriate population size for each region, rather than simply aiming to maintain existing population levels.

Regional disparities will not only be between the countryside and big cities, but also between core regional cities and Tokyo.
Our prediction implicitly assumes that the reduction in distance frictions over the past 50 years will continue in the future. 
In the future, however, transport access between Tokyo and the rest of Japan will inevitably deteriorate under the current transport network, and improved access may be limited to a few large cities.
So far, interregional travel has depended on mass transportation such as high-speed rail and air travel. 
However, the current level of transport access realized by these modes cannot be maintained in the future because the mass demand cannot be sustained at the current level under the expected population decline.

Fig.\,\ref{fig:large-city-growth}B shows the evolution of the population in the three largest cities, Tokyo, Osaka, and Nagoya, and that of Tokyo alone in the baseline and pessimistic scenarios. 
While the three largest cities will continue to account for half of the national population, and Tokyo alone for more than 30\%, it is important to have a large population mass rather than a large population share in order to maintain the mass transit system.
In the pessimistic scenario, Tokyo in 2070 is much smaller in population than, for example, the current Seoul metropolitan area, and in 2120 is smaller than Osaka today.
In these futures, high-speed rail is unlikely to provide the current frequency of service in any direction from Tokyo.
Thus, regional centers such as Fukuoka, Sendai, and Hiroshima, for which access to Tokyo by bullet train is key to their existence, may face a much harsher future.

Currently, the bullet train lines (Hokuriku, Hokkaido, and Nishi Kyushu Shinkansen) are being extended by a total of about 400 kilometers to cover rural parts of Japan, based on the original development plan decided in 1973, while the opening of the Chuo Linear Shinkansen, which will be faster than the current Shinkansen, is scheduled to connect Tokyo and Osaka in 2037.
\footnote{The MLIT website explains the detailed plan for the development of Shinkansen lines (in Japanese).}\ 
However, further expansion of mass transit is unlikely to be sustainable in the coming decades due to the lack of scale of demand.
Again, more concentrated investments seem to be effective after selection and concentration based on the predicted spatial distribution of the population.

\subsection{Urban redevelopment\label{sec:urban-redevelopment}}
\begin{figure}[htbp!]
 \centering

 \includegraphics[width=.55\textwidth]{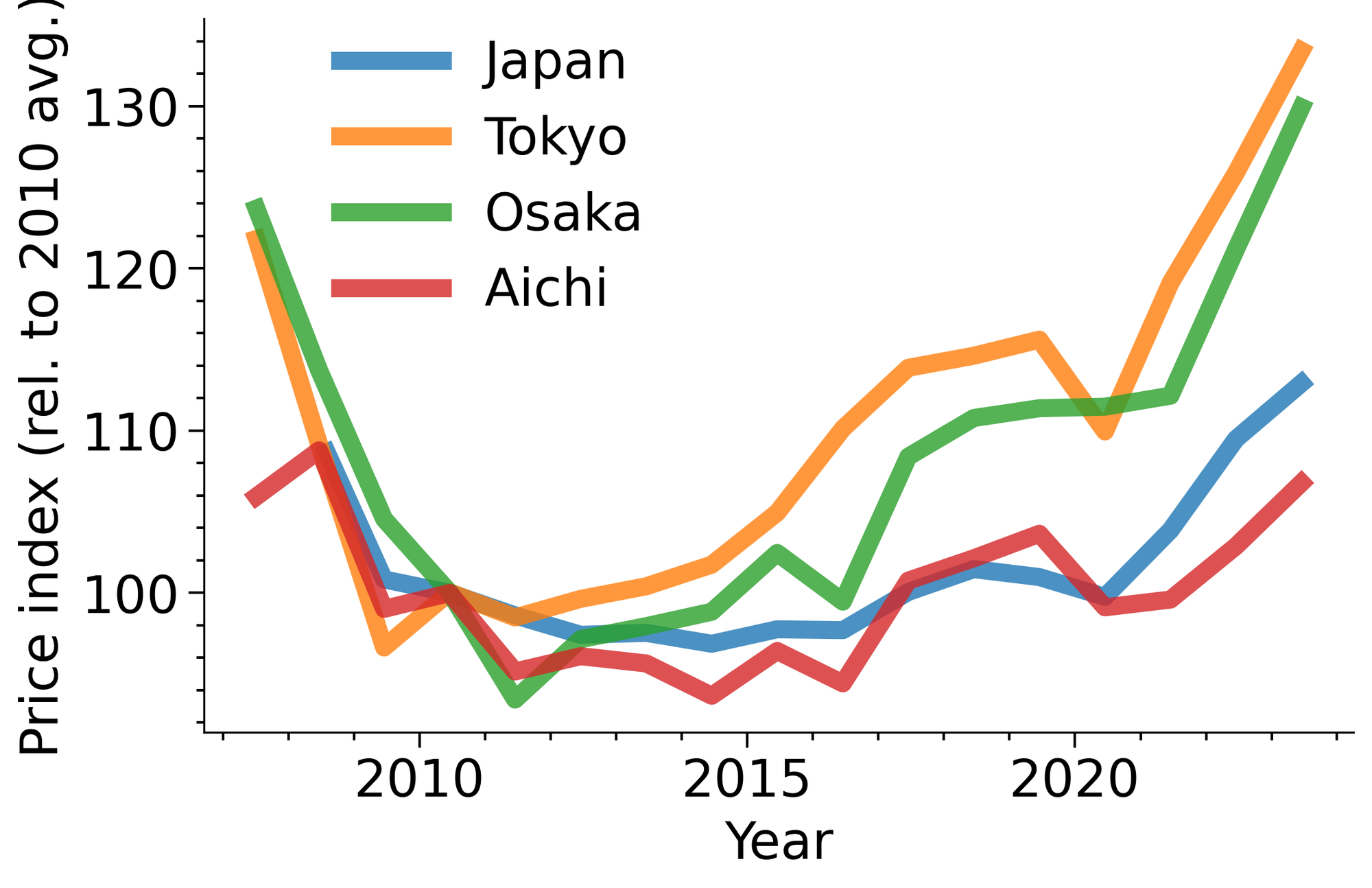}
 \caption{The annual average housing price indice in Japan, Tokyo Prefecture, Osaka Prefecture and Aichi Prefecture between 2008--2024.}
 \caption*{\footnotesize\textit{Note}: Data source: \href{https://www.mlit.go.jp/totikensangyo/totikensangyo\_tk5\_000085.html}{Monthly house price index} for the period from April 2008 to February 2024 by MILT. Tokyo, Osaka and Aichi prefectures overlap with the CBD of Tokyo, Osaka and Nagoya UAs.}
 \label{fig:housing-price-index}
 \end{figure}

In Japan, the redevelopment of urban centers with the construction of high-rise buildings, such as high-rise condominiums, has been actively pursued after the population decline became apparent in 2010, especially in major cities such as Tokyo and Osaka.
In 2023, 47 high-rise condos with 13,862 units have been built in Tokyo and the corresponding numbers are 30 and 8,497 in Osaka.%
\footnote{Data source: \href{https://www.mlit.go.jp/totikensangyo/totikensangyo\_tk5\_000085.html}{Real estate price index, MLIT, 2024}.}\ 
Reflecting this situation, the real estate price index published by MLIT for Tokyo and Osaka continued to rise until 2024 (except for 2020 due to the COVID-19 shock) (Fig.\,\ref{fig:housing-price-index}).

\begin{figure}[p]
 \centering

 \begin{minipage}[c]{\textwidth}
 \includegraphics[width=\textwidth]{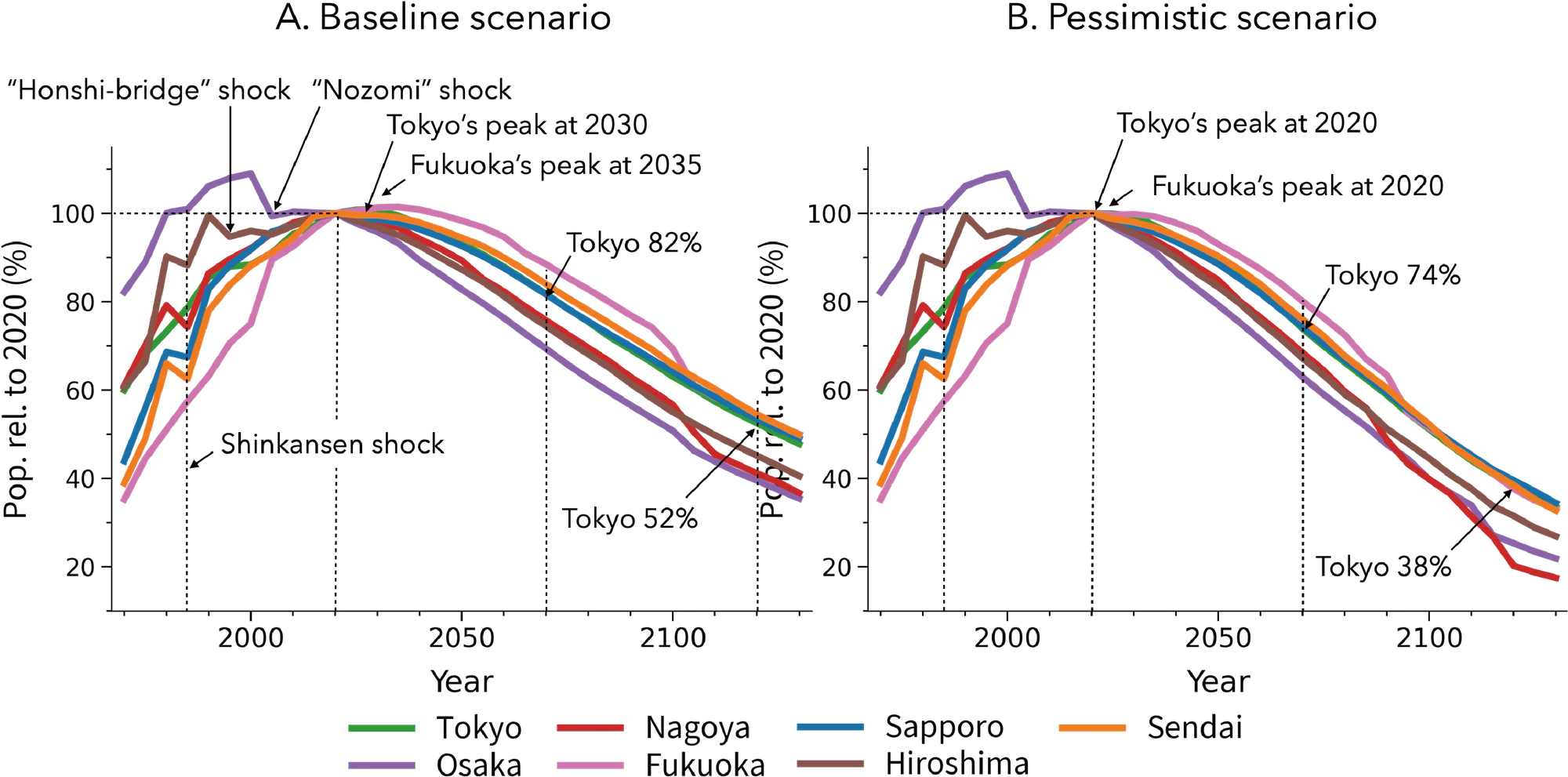}
 \caption{Evolution of population sizes of the largest cities in the seven regional divisions relative to their 2020 level}
 \label{fig:rel-size-seven-largest-cities}
 \end{minipage}

 \vspace{1.5cm}

 \begin{minipage}[c]{\textwidth}
 \includegraphics[width=\textwidth]{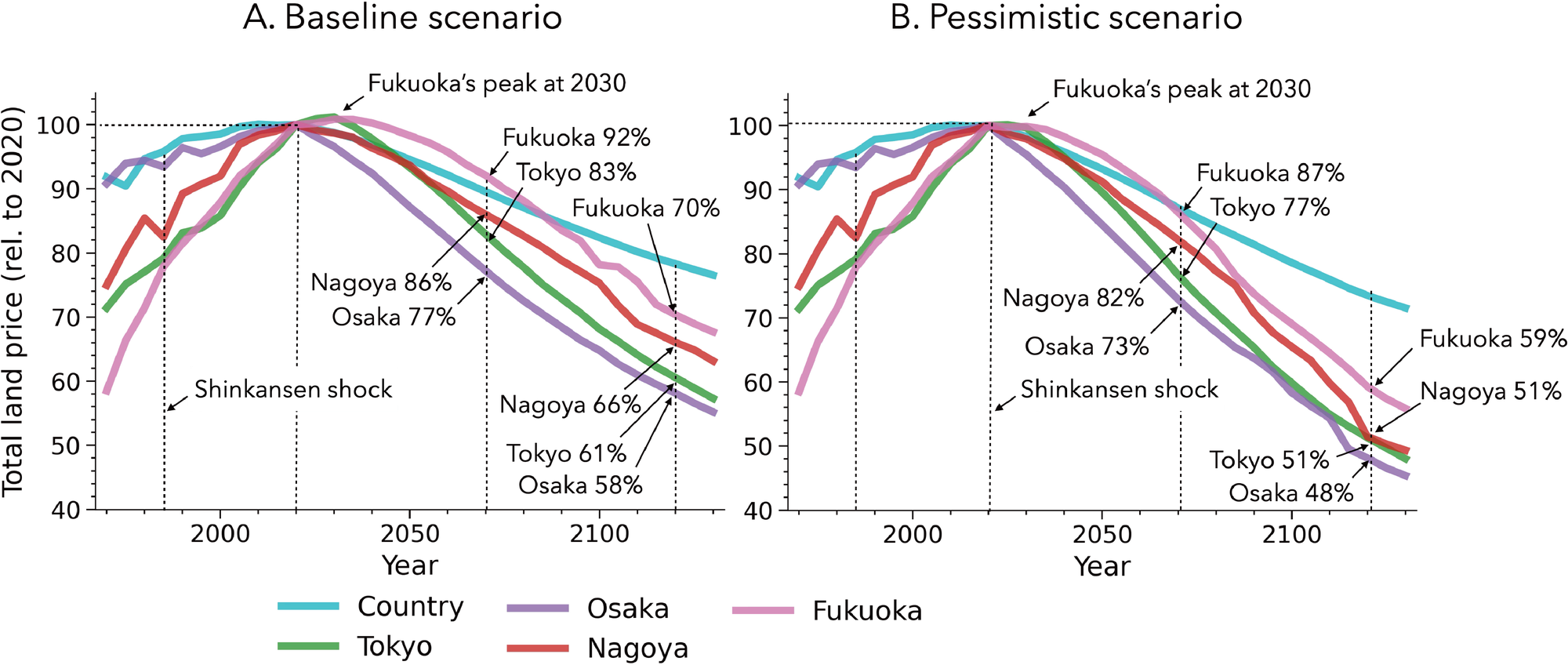}
 \caption{The predicted total residential land prices in 1970--2020 and their predicted values in 2025--2120 for the country, all cities, and the four largest cities, Tokyo, Osaka, Nagoya, and Fukuoka}
 \caption*{\footnotesize\textit{Note}: The land price at each grid cell under the projected population distribution over the grid cells by using the log-linear additive model \citep[][]{Wood-Book2017} using the population distribution over the grid cells, elevation and spatial coordinates. The model is estimated by using the land price data at the grid cell level in 1985--2020 (every 5 years) obtained from the Land Price Survey Data available from MILT (\url{https://nlftp.mlit.go.jp/ksj/gml/datalist/KsjTmplt-L01-2024.html}).
  } 
\label{fig:landprice}
\end{minipage}
\end{figure}

But with the rapid decline in population and the flattening of population distribution within cities, the rise in housing prices in these large cities is likely to be excessive.
Fig.\,\ref{fig:rel-size-seven-largest-cities} shows the population size of the largest city in each of the seven regional divisions (Fig.\,\ref{fig:7regions}) from 1970 to 2120, where the population size is measured relative to it's value in 2020.
In our projection, Tokyo will continue to attract population until 2030 in the baseline scenario (Fig.\,\ref{fig:rel-size-seven-largest-cities}A), but it will begin to decline immediately in 2025 in the pessimistic scenario (Fig.\,\ref{fig:rel-size-seven-largest-cities}B).
Tokyo's population is projected to fall to 82\% and 74\% of its 2020 level in 2070, and to 52\% and 38\% in 2120 in the baseline and pessimistic scenarios, respectively.
The fourth largest Fukuoka's growth may continue the longest, but only until 2035 in the baseline scenario, while other major cities are predicted to decline immediately.

Skyscrapers are said to reach the age of major renovation about 45 years after construction is completed \citep[e.g.,][Ch.2]{Sakaki-Book2020}. 
If the city's population is steadily declining, there will be no incentive to maintain a building through major renovations.
In the city center, where population density continues to decline, there will be fewer advantages to living in the upper floors of a sparsely populated building. If so, the demand for existing buildings is likely to fall.%
\footnote{Refer to Appendix \ref{app:tenjin} for the discussion on Fukuoka's urban redevelopment project, the Tenjin Big Bang.}\ 

Fig.\,\ref{fig:landprice}A and B show the projection of the total residential land price relative to their 2020 value in the baseline and pessimistic scenarios, respectively, in the country, and in each of the four largest cities.%
\footnote{In this projection, the city's polygon is fixed at those in 2020.
The spatial distribution of land price is estimated by the log-linear additive model \citep[][]{Wood-Book2017} using the projected population distribution, elevation and spatial coordinates (see details in Appendix \ref{app:land-price}).}\ 
The total value of land in cities declines faster than in the country (except for Fukuoka), responding to the flattening of the geographic distribution of population in each city.
While our model by construction does not reflect the recent surge in housing prices in large cities, which deviates substantially from the spatial pattern of population, we believe that our prediction is suggestive of a long-run spatial pattern of land prices.


The current building boom in Japan may be reminiscent of the devastation that Detroit experienced from the 1960s through its default in 2013. According to the Population Census of the US, the City of Detroit's population peaked at 1.8 million in 1950 and has continued to decline since then, falling to 640,000 by 2020. Many of Detroit's skyscrapers, built in the 1910s and 1920s, were abandoned in the 1980s and 1990s as public safety deteriorated, and remained abandoned for a decade or two before redevelopment began.
In the 2000s and beyond, some skyscrapers were repaired and restored, while others were turned into parking lots and parks.%
\footnote{The abandonment of Detroit's skyscrapers and its aftermath is detailed in \href{https://www.eherg.com/blog/2016/10/8/detroits-dirty-dozen-then-and-now}{an October 9, 2016 blog post by Eric Hergenreder}, along with the original February 16, 2004 article.}\ 

Once built, the skyscrapers (which are much taller than many of Detroit's skyscrapers at the time) will be very expensive to maintain, especially if the population continues to decline at the current rate.
What makes Japan's case different from Detroit's is that while Detroit's future may have been less clear when it developed its high-rise urban center, we know at this point that the population size and population density in the urban center of Japanese cities is likely to begin shrinking soon.

For large cities, one obvious advice from our projection is that, assuming population decline in the future and a decrease in population density in the city center, buildings in the city center should be lower rather than higher. 
This will reduce building maintenance costs, promote human interaction, and enhance, not diminish, the benefits of urban centers.
Furthermore, with proper zoning, lower density can make cities more resilient to disasters by eliminating the need to live in disaster-prone areas. 
In addition, as cities shrink and become less dense, there will be a need to consolidate transport networks. 
In the process, it is in principle possible to change the structure to one that is more compatible with automated driving and logistics.

\section{Conclusion}
\label{sec:conclusion}

The purpose of this research is to provide initial evidence of the impending problem of rapid population decline in order to stimulate discussion of countermeasures to sustain urban and regional economies in Japan.
Since population decline itself is a recent phenomenon, the data available for model estimation are limited, so the validity of our projection method needs to be closely examined and verified through ex-post evaluation.
Nevertheless, we believe it is important to quantify the possible future of the regional economy as systematically as possible at this point, because deciding how much infrastructure support each region needs requires long-term planning.
The projected gridded population makes it possible to predict the divergence between the scale of infrastructure and population in each region, and to assess the sustainability of existing infrastructure in those regions.%
\footnote{For example, the East Japan Railway Company, which serves the eastern half of the island of Honshu, including the Tokyo metropolitan area, faces a deficit of 75 billion yen in 2023.
The deficit is concentrated in the Tohoku region and is largely offset by profits from the Tokyo metropolitan area \citep[][]{JRE-News-2024}.
But such an operation is not going to be sustainable under the present population decline.}\ 

Three key conditions allowed us to make a systematic projection of the size and spatial distribution of Japanese cities.
First, accurate official gridded population data are available for the last half century.
Second, the cities defined in terms of population agglomeration in a bottom-up manner exhibit an approximate but persistent power law for the distribution of city size at any point in time.
Third, the evolution of cities defined in this way over the past half century is consistent with the comparative statics implied by economic agglomeration theory, particularly with respect to declining distance frictions.

The expected concentration of population in larger cities is clearly reflected in the change in the power-law coefficient that helped predict the future population size of cities, while the dispersion that occurs within individual cities is reflected in the flattening of the internal population distribution within a city.
The predicted concentration at the national and regional levels and dispersion at the city level suggest a rethinking of current regional and urban policies that seek to facilitate dispersion from larger to smaller cities at the national level and also facilitate concentration toward centers within a city.

Our model predicts a substantial shrinkage of the urban system in Japan.
The number of large cities with at least 1 million and 500,000 inhabitants will decrease by one-third to one-half in 50 years and by one-half to more than 70\% in 100 years, and they will be concentrated in Tokyo and west of Tokyo. 
The number of cities with at least 10,000 inhabitants, which will play the role of regional centers, will decrease by a third in 50 years and by 50\% to 70\% in 100 years, and they will be more widely dispersed, leaving the regions in between more isolated than before.

Shrinkage is not uniform in all regions, but is more severe in the rural areas, which will require us to make the corresponding adjustment in maintaining the infrastructure.
Specifically, one of the most important problems to be solved is how to transform the current fixed-capacity infrastructure into a scalable, self-sustaining system according to the community size in the shrinking regions.

In shrinking cities and regions, there will not be a large mass of transport demand, so maintaining the current mass transit system will not be viable.
A concrete vision of the extent of shrinkage of the regional economy facilitates the development of alternative, more appropriate technologies. 
For example, a recent development of flying cars suggests a solution as a scalable mode of transportation.%
\footnote{For example, the ASKA A5, from a company called ASKA in Moutain View, California, is the world's first flying car to begin the type certification process with the Federal Aviation Administration (FAA) in the United States in 2024. About the size of a sport utility vehicle in driving mode, it can take off and land vertically, eliminating the need for fixed infrastructure such as runways and rail networks.}\ 
As discussed in Section \ref{sec:public-policies}, Japan's rural areas have a strong comparative advantage in primary industries.
With a smaller more suitable population of these regions, there may be fewer sustainable regions as a place to live. 
For example, sustainable living places may be the cities with at least 100,000 inhabitants (Fig.\,\ref{fig:100k}). 
With such a new scalable mode of transportation, it may be possible to earn in the rural regions while living in the urban area.

Scalability is also key for other types of infrastructure.
For water and sewage, for example, containerized water purification systems have reached a practical level. 
Originally developed for use in developing countries, a containerized water purification infrastructure system is now being used in shrinking rural communities in Japan.%
\footnote{To cite two examples, an Indian company \href{https://drinkwellsystems.com/}{DrinkWell} provides automated water dispensers to communities affected by arsenic and fluorine.
Rather than delivering water through a water network, it is supplied by filtering the water in a reservoir.
A Japanese company, \href{https://wota.co.jp/en/}{WOTA}, produces a containerized water purification system that recycles and reuses nearly 100\% of domestic wastewater. By collecting and treating rainwater and domestic wastewater on site and then returning it to the same house, it is possible to provide all the water needed for domestic use in an independent environment without relying on existing water and sewer infrastructure.
}\

We close the paper by discussing three directions for future research.

\noindent{\bf Dissaster resilience} -- As discussed in Section \ref{sec:implications}, one of the benefits of declining population is to be able to avoid disaster-prone areas to reside, and to make cities more resilient to disasters.
As a simple extension of our projection, instead of assuming homogeneous grid cells as in this study, we can impose zoning so that the impedance to population growth depends on the disaster resilience of each zone.
Such a projection can help in planning future disaster-resilient cities.

\noindent{\bf Environmental load} -- 
Population dynamics influence urban expansion/shrinkage and transportation patterns. Therefore, our gridded populations would help refine projection of carbon emissions and energy consumptions, which are strongly influenced by these factors \citep[see,][]{yamagata2015comparison}. By integrating gridded population data with energy system models \citep[see,][]{chong2021calibrating}, future energy demand can be projected more accurately. Additionally, the population projection is valuable for modeling land-use changes, including deforestation, which has significant environmental impacts. Such applications can support policy making for a sustainable, low-carbon society. In other words, the projected populations enhance the precision and relevance of climate and energy policy assessments.

\noindent{\bf Provision of essential services} -- Finally, another direction is to relate the population size of a city to its industrial composition and to assess the sustainability of cities in terms of the availability of essential services.
If we focus on cities as industrial locations, their location pattern shows a strong \textit{hierarchical property} that an industry located in a smaller city is also located in larger cities, which is especially strong in the composition of tertiary industries \citep[e.g.,][]{Mori-et-al-JRS2008,Hsu-EJ2012,Davis-Dingel-JIE2020}.

This property allows us to identify the threshold population size of a city for a given set of industries to be able to operate in the city. In particular, by defining the sets of essential industries for modern life, such as emergency and maternity hospitals, high schools, funeral services, supermarkets, one can identify the \textit{essential population size} of a city that can provide the full set of such essential services.
Combined with our future projection of individual city sizes, this suggests a guideline for strategic reduction of the country's economic geography by identifying the future spatial distribution of sustainable cities. Accordingly, it will allow for the reorganisation of functional regions around more explicitly defined sustainable cities to ensure that the country can adapt to its declining population \citep[see preliminary analysis by][]{Mori-Ogawa-DP2024}.

\clearpage

\newpage

\bibliographystyle{aea}
\bibliography{main}

@misc{BRR,
	author = {{Ministry of Internal Affairs and Communications of Japan}},
	title = {The Basic Resident Registration},
	year = {2013--2020}}

@electronic{OECD-FDB2023,
	author = {OECD},
	institution = {OECD},
	title = {OECD Family Database},
	url = {https://www.oecd.org/content/dam/oecd/en/data/datasets/family-database/sf_2_1_fertility_rates.pdf},
	year = {2023}}

@unpublished{JRE-News-2024,
	author = {{East Japan Railway Company}},
	month = {10},
	note = {\url{https://www.jreast.co.jp/press/2024/20241029_ho01.pdf}},
	title = {JRE News},
	year = {2024}}

@article{chong2021calibrating,
	author = {Chong, Adrian and Gu, Yaonan and Jia, Hongyuan},
	journal = {Energy and Buildings},
	pages = {111533},
	title = {Calibrating building energy simulation models: A review of the basics to guide future work},
	volume = {253},
	year = {2021}}

@article{yamagata2015comparison,
	author = {Yamagata, Yoshiki and Murakami, Daisuke and Seya, Hajime},
	journal = {Applied Energy},
	pages = {255--262},
	title = {A comparison of grid-level residential electricity demand scenarios in Japan for 2050},
	volume = {158},
	year = {2015}}

@misc{ECensus-2009-2016,
	author = {{Ministry of Internal Affairs and Communications of Japan}},
	title = {Grid Square Statistics of Economic Census for Business Frame},
	year = {2009, 2016}}

@misc{ECensus-1975-2006,
	author = {{Ministry of Internal Affairs and Communications of Japan}},
	title = {Grid Square Statistics of Establishment and Enterprise Census},
	year = {1975, 1978, 1981, 1986, 1991, 1996, 2001, 2006}}

@article{Yagi-Managi-TP2016,
	author = {Yagi, Michiyuki and Shunsuke Managi},
	journal = {Transport Policy},
	pages = {37-53},
	title = {Demographic determinants of car ownership in Japan},
	volume = {50},
	year = {2016}}

@article{Fujinami-RF2020,
	author = {Fujinami, Takumi},
	journal = {Research Focus},
	month = {August},
	pages = {1-9},
	title = {The background to the sudden decline in births and future measures to tackle the falling birthrate (in Japanese)},
	volume = {2020-19},
	year = {2020}}

@article{Davis-Dingel-JIE2020,
	author = {Donald R. Davis and Jonathan I. Dingel},
	doi = {https://doi.org/10.1016/j.jinteco.2020.103291},
	issn = {0022-1996},
	journal = {Journal of International Economics},
	pages = {103291},
	title = {The Comparative Advantage of Cities},
	url = {http://www.sciencedirect.com/science/article/pii/S0022199620300106},
	volume = {123},
	year = {2020}}

@article{Mori-et-al-JRS2008,
	author = {Tomoya Mori and Koji Nishikimi and Tony E. Smith},
	journal = {Journal of Regional Science},
	number = {1},
	pages = {165-211},
	title = {{The Number-Average Size Rule: A New Empirical Relationship Between Industrial Location And City Size}},
	url = {https://ideas.repec.org/a/bla/jregsc/v48y2008i1p165-211.html},
	volume = {48},
	year = {2008}}

@unpublished{Mori-Ogawa-DP2024,
	author = {Mori, Tomoya and Miki Ogawa},
	note = {{RIETI Policy Discussion Paper 25-P-006}},
	title = {Sustainability of cities from the perspective of viable locations for essential industries under declining populations},
	year = {2024}}

@article{Oizumi-et-al-PO2022,
	author = {Oizumi, Ryo and Hisashi Inaba and Takenori Takada and Youichi Enatsu and Kensaku Kinjo},
	journal = {PLOS One},
	number = {9},
	pages = {e0273817},
	title = {Sensitivity analysis on the declining population in Japan: Effects of prefecture-specific fertility and interregional migration},
	volume = {17},
	year = {2022}}

@article{Wan-et-al-CGIS2022,
	author = {Wan, Heng and Jim Yoon and Vivek Srikrishnan and Brent Daniel and David Judi},
	journal = {Cartography and Geographic Information Science},
	number = {1},
	pages = {18--31},
	title = {Population downscaling using high-resolution, temporally-rich US property data},
	volume = {49},
	year = {2022}}

@article{Murakami-Yamagata-S2019,
	author = {Murakami, Daisuke and Yoshiki Yamagata},
	journal = {Sustainability},
	number = {7},
	pages = {2106},
	title = {Estimation of gridded population and GDP scenarios with spatially explicit statistical downscaling},
	volume = {11},
	year = {2019}}

@article{Wang-Meng-Long-SD2022,
	author = {Wang, Xinyu and Xiangfeng Meng and Ying Long},
	journal = {Scientific Data},
	number = {1},
	pages = {563},
	title = {Projecting 1 km-grid population distributions from 2020 to 2100 globally under shared socioeconomic pathways},
	volume = {9},
	year = {2022}}

@article{Stevens-et-al-PO2015,
	author = {Stevens, Forrest R. and Andrea E. Gaughan and Catherine Linard and Andrew J. Talem},
	journal = {PLOS One},
	number = {2},
	pages = {e0107042},
	title = {Disaggregating census data for population mapping using random forests with remotely-sensed and ancillary data},
	volume = {10},
	year = {2015}}

@article{McKee-et-al-PNAS2015,
	author = {McKee, Jacob J. and Amy N. Rose and Edward A. Bright and Timmy Huynh and Budhendra L. Bhaduri},
	journal = {Proceedings of the National Academy of Science of the United States of America},
	number = {5},
	pages = {1344--1349},
	title = {Locally adaptive, spatially explicit projection of US population for 2030 and 2050},
	volume = {112},
	year = {2015}}

@article{Osawa-Akamatsu-JET2020,
	author = {Osawa, Minoru and Akamatsu, Takashi},
	journal = {Journal of Economic Theory},
	pages = {105025},
	title = {Equilibrium refinement for a model of non-monocentric internal structures of cities: A potential game approach},
	volume = {187},
	year = {2020}}

@article{Fujita-Krugman-RSUE1995,
	author = {Fujita, Masahisa and Paul Krugman},
	journal = {Regional Science and Urban Economics},
	pages = {505-528},
	title = {When is the economy monocentric?: von Thu\"{u}nen and Chamberlin unified},
	volume = {25},
	year = {1995}}

@article{ONeil-et-al-CC2014,
	author = {O'Neill, Brian C. and Elmar Kriegler and Keywan Riahi and Kristie L. Ebi and Stephane Hallegatte and Timothy R. Carter and Ritu Mathur and Detlef P. van Vuuren},
	journal = {Climatic Change},
	pages = {387--400},
	title = {A new scenario framework for climate change research: the concept of shared socioeconomic pathways},
	volume = {122},
	year = {2014}}

@article{Rozenfeld-et-al-PNAS2008,
	author = {Rozenfeldt, Hern\'{a}n D. and Diego Rybski and Jos\'{e} S. {Andrade, Jr.}},
	journal = {Proceedings of the National Academy of Sciences of the United States of America},
	number = {48},
	pages = {18702--18707},
	title = {Lows of population growth},
	volume = {105},
	year = {2008}}

@article{SSP-GEC2017,
	author = {Riahi, Keywan and Detlef P. van Vuuren and Elmar Kriegler and Jae Edmonds and Brian C. O'Neill and Shinichiro Fujimori and Nico Bauer and Katherine V. Calvin and Rob Dellink and Oliver Fricko and Wolfgang Lutz and Alexander Popp and Jesus Crespo Cuaresma and Samir KC and Marian Leimbach and Leiwen Jiang and Tom Kram and Shilpa Rao and Johannes Emmerling and Kristie Ebi and Massimo Tavoni},
	journal = {Global Environmental Change},
	pages = {153-168},
	title = {The Shared Socioeconomic Pathways and their energy, land use, and greenhouse gas emissions implications: An overview},
	volume = {42},
	year = {2017}}

@article{Wood-JRSS2011,
	author = {Wood, Simon N.},
	journal = {Journal of the Royal Statistical Society Series B: Statistical Methodology},
	number = {1},
	pages = {3--36},
	title = {Fast stable restricted maximum likelihood and marginal likelihood estimation of semiparametric generalized linear models},
	volume = {73},
	year = {2001}}

@article{Nakajima-Takano-RSUE2023,
	author = {Nakajima, Kentaro and Keisuke Takano},
	journal = {Regional Science and Urban Economics},
	pages = {103955},
	title = {Estimating the effect of land use regulation on land price: At the kink point of building height limits in Fukuoka},
	volume = {103},
	year = {2023}}

@book{Hartung-Knapp-Sinha-Book2008,
	author = {Hartung, Joachim and Guido Knapp and Bimal K. Sinha},
	publisher = {John Wiley \& Sons},
	title = {Statistical Meta-Analysis with Applications},
	year = {2008}}

@article{Mori-Wrona-RSUE2024,
	author = {Mori, Tomoya and Jens Wrona},
	journal = {Regional Science and Urban Economics},
	pages = {104060},
	title = {Centrality bias in the inter-city trade},
	volume = {109},
	year = {2024}}

@article{Nam-Reilly-US2013,
	author = {Nam, Kyung-Min and John M. Reilly},
	journal = {Urban Studies},
	number = {1},
	pages = {208--225},
	title = {City size distribution as a function of socioeconomic conditions: An eclectic approach to downscalling global population},
	volume = {50},
	year = {2013}}

@article{Jones-ONeil-ERL2016,
	author = {Jones, Bryan and Brian C. O'Neill},
	journal = {Environmental Research Letters},
	pages = {084003},
	title = {Spatially explicit global population scenarios consistent with the Shared Socioeconomic Pathways},
	volume = {11},
	year = {2016}}

@article{Tabuchi-Thisse-JUE2011,
	author = {Tabuchi, Takatoshi and Jacques-Fran\c{c}ois Thisse},
	journal = {Journal of Urban Economics},
	month = {March},
	number = {2},
	pages = {240-252},
	title = {A new economic geography model of central places},
	volume = {69},
	year = {2011}}

@unpublished{MLIT-2010,
	author = {{The Ministry of Land, Infrastructure, Transport and Tourism of Japan}},
	note = {\url{https://www.mlit.go.jp/common/001005632.pdf}},
	title = {Actual State of Trunk Link Passenger Flow},
	url = {https://www.mlit.go.jp/common/001005632.pdf},
	year = {2010}}

@webpage{UN-2022,
	author = {{United Nations}},
	title = {World Population Prospects},
	url = {https://population.un.org/wpp/},
	year = {2022}}

@misc{MIC2008,
	author = {{Ministry of Internal Affairs and Communications of Japan}},
	title = {Estimated Population by Age (Single Year) and Sex for Japan},
	year = {2008}}

@webpage{MIC2024,
	author = {{Ministry of Internal Affairs and Communications of Japan}},
	title = {Key points on population, population dynamics and number of households based on the Basic Resident Register (in Japanese)},
	url = {https://www.soumu.go.jp/main_content/000892926.pdf},
	year = {2024}}

@book{Wood-Book2017,
	author = {Wood, Simon N},
	publisher = {chapman and hall/CRC},
	title = {Generalized additive models: an introduction with R},
	year = {2017}}

@book{Sakaki-Book2020,
	author = {Sakaki, Atsushi},
	publisher = {East Press},
	title = {Yokoso 2050 no Tokyo he: Ikinokoru hudosan, haikyoninaru hudosan},
	year = {2020}}

@article{Fujita-Krugman-Mori-EER1999,
	author = {Fujita, Masahisa and Paul Krugman and Tomoya Mori},
	journal = {European Economic Review},
	pages = {209-251},
	title = {On the evolution of hierarchical urban systems},
	volume = {43},
	year = {1999}}

@book{Smith-et-al-Book2005,
	author = {Smith, Stanley K and Tayman, Jeff and Swanson, David A},
	publisher = {Springer Dordrecht},
	series = {The Springer Methods and Population Analysis},
	title = {State and Local Population Projections: Methodology and Analysis},
	year = {2002}}

@unpublished{Graves-DP2013,
	author = {Graves, Alex},
	journal = {arXiv preprint arXiv:1308.0850},
	title = {Generating sequences with recurrent neural networks},
	year = {2013}}

@book{Durbin-Koopman-Book2012,
	author = {Durbin, James and Koopman, Siem Jan},
	publisher = {OUP Oxford},
	title = {Time series analysis by state space methods},
	volume = {38},
	year = {2012}}

@book{Box-et-al-Book2015,
	author = {Box, George EP and Jenkins, Gwilym M and Reinsel, Gregory C and Ljung, Greta M},
	publisher = {John Wiley \& Sons},
	title = {Time series analysis: forecasting and control},
	year = {2015}}

@article{Gabaix-Ibragimov-JBES2011,
	author = {Gabaix, Xavier and Ibragimov, Rustam},
	journal = {Journal of Business \& Economic Statistics},
	number = {1},
	pages = {24--39},
	publisher = {Taylor \& Francis},
	title = {Rank- 1/2: a simple way to improve the OLS estimation of tail exponents},
	volume = {29},
	year = {2011}}

@incollection{Duranton-Puga-HB2014,
	author = {Duranton, Gilles and Diego Puga},
	booktitle = {Handbook of Economic Growth},
	chapter = {5},
	editor = {Durlauf, Steven N. and Philippe Aghion},
	pages = {781--853},
	publisher = {North-Holland},
	title = {The growth of cities},
	volume = {2},
	year = {2014}}

@article{Mori-et-al-PNAS2020,
	author = {Mori, Tomoya and Tony E. Smith and Wen-Tai Hsu},
	journal = {Proceedings of the National Academy of Sciences of the United States of America},
	month = {March},
	number = {12},
	pages = {6469-6475},
	title = {Common power laws for cities and spatial fractal structures},
	volume = {117},
	year = {2020}}

@article{Hsu-EJ2012,
	author = {Hsu, Wen-Tai},
	journal = {Economic Journal},
	pages = {903-932},
	title = {Central place theory and city size distribution},
	volume = {122},
	year = {2012}}

@incollection{Gabaix-Ioannides-HB2004,
	author = {Gabaix, Xavier and Ioannides, Yannis M.},
	booktitle = {Handbook of Regional and Urban Economics},
	chapter = {53},
	editor = {Henderson, J. Vernon and Thisse, Jacques-Fran{\c{c}}ois},
	pages = {2341--2378},
	publisher = {Elsevier},
	title = {The evolution of city size distributions},
	volume = {4},
	year = {2004}}

@misc{NIPSSR-2023,
	author = {{National Institute of Population and Social Security Research}},
	title = {Population Projections for Japan: 2021 to 2070 (With long-range Population Projections: 2071 to 2120)},
	url = {https://www.ipss.go.jp/pp-zenkoku/e/zenkoku_e2023/pp_zenkoku2023e.asp},
	year = {2023}}

@techreport{NIPSSR-Region-2023,
	author = {{National Institute of Population and Social Security Research}},
	title = {Regional Population Statistics of Japan},
	url = {https://www.ipss.go.jp/pp-shicyoson/j/shicyoson23/t-page.asp},
	year = {2023}}

@unpublished{Mori-et-al-DP2023,
	author = {Mori, Tomoya and Takashi Akamatsu and Yuki Takayama and Minoru Osawa},
	note = {arXiv:2207.05346},
	title = {Origin of power laws and their spatial fractal structure for city-size distributions},
	year = {2023}}

@unpublished{Akamatsu-et-al-DP2024,
	author = {Akamatsu, Takashi and Tomoya Mori and Minoru Osawa and Yuki Takayama},
	note = {arXvi:1912.05113},
	title = {Spatial scale of agglomeration and dispersion: Number, spacing, and the spatial extent of cities},
	year = {2024}}

@misc{UN-2018,
	author = {{United Nations}},
	title = {The 2018 Revision of World Urbanization Prospects},
	url = {https://population.un.org/wpp/},
	year = {2018}}

@article{IHME-Lancet2024,
	author = {{Institute for Health Metrics and Evaluation}},
	journal = {Lancet},
	pages = {2057-2099},
	title = {Global fertility in 204 countries and territories, 1950--2021, with forecasts to 2100: A comprehensive demographic analysis for the Global Burden of Disease Study 2021},
	volume = {403},
	year = {2024}}

\newpage

\appendix
\makeatletter
\renewcommand{\thefigure}{A.\arabic{figure}}
\makeatother
\setcounter{figure}{0}
\begin{center}
\LARGE{\bf Appendix}
\end{center}
\section{Gradual expansion of high-speed transport network in Japan}\label{app:network}
\begin{figure}[h!]
  \centering
  \captionsetup{width=\linewidth}
  \includegraphics[width=0.6\textwidth]{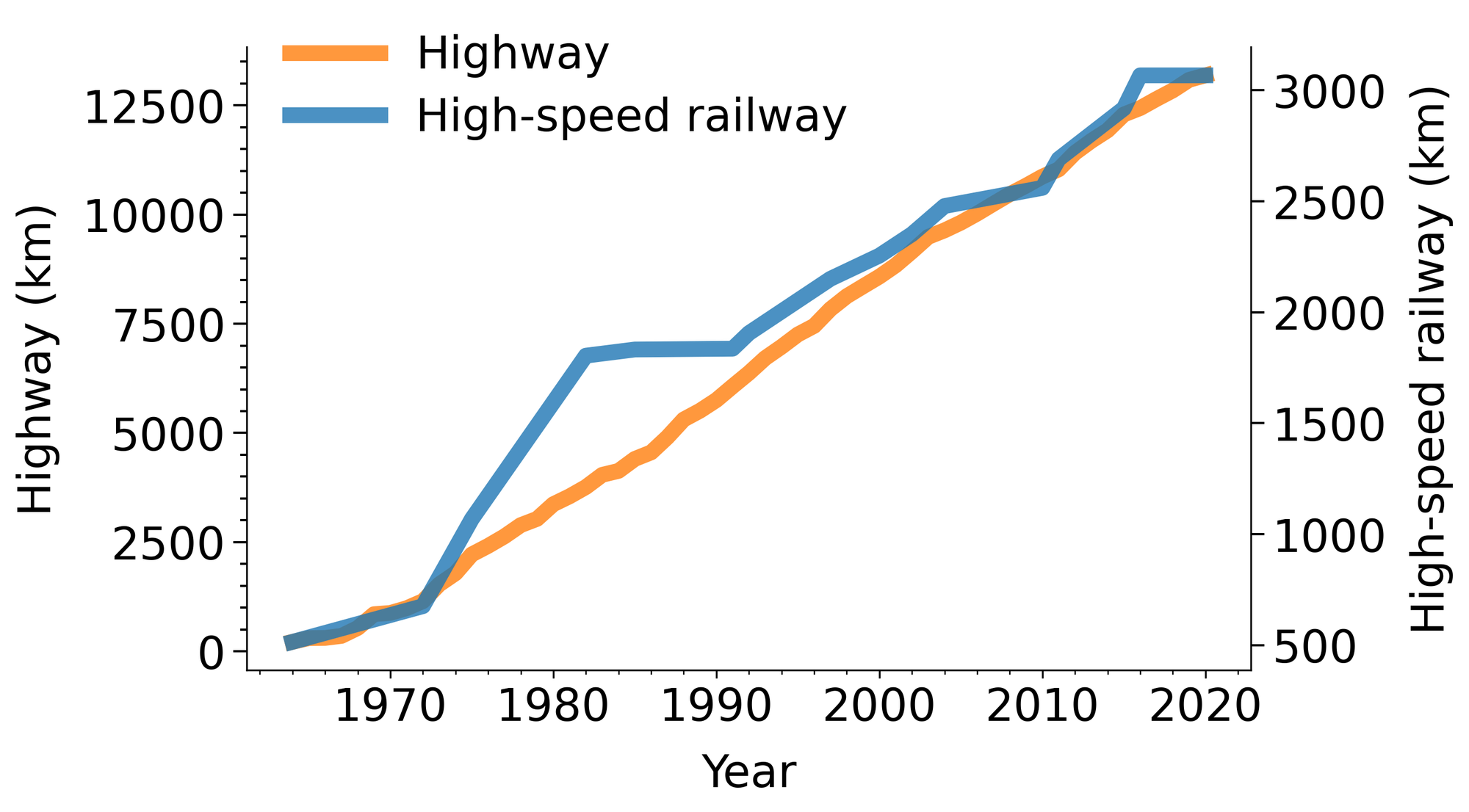}
  \caption{The total length of highway and high-speed rail lines in Japan between 1970 and 2020}
  \caption*{\footnotesize\textit{Note}: The line lengths are calculated based on the shapefiles of the transportation networks obtained from the NLNI of MLIT of Japan (\url{https://nlftp.mlit.go.jp/}).}
  \label{fig:length}
\end{figure}

The total length of both highways and high-speed railways in Japan has gradually increased during the period from 1970 to 2020 (Fig.\,\ref{fig:length}). 
This can be interpreted as a gradual decrease in transportation costs during this period.

\section{Population size and industrial diversity of a city\label{app:industrial-diversity}}
The decline in a city's population is closely related to the decline in its industrial diversity.
In this section, we compare changes in both industrial diversity and population size for each city between 2015 and 2020.
For this analysis, we use the 2020 urban agglomeration boundaries.
The population size of a city is calculated as the sum of the populations of the grid cells within the city.
A city's industrial diversity is measured as the number of industries with at least one establishment in the city.
The establishment data come from the NTT Townpage database, provided in cooperation with Grant-in-Aid for Scientific Research (S) 24H00012. The data includes 1,899 industries mapped to our grid cell system.

\begin{figure}[htbp!]
    \centering
    \begin{minipage}[c]{\textwidth}

    \end{minipage}
    \vspace{1cm}

     \begin{minipage}[c]{\textwidth}
        \centering
         \captionsetup{width=\linewidth} 
         \includegraphics[width=0.6\textwidth]{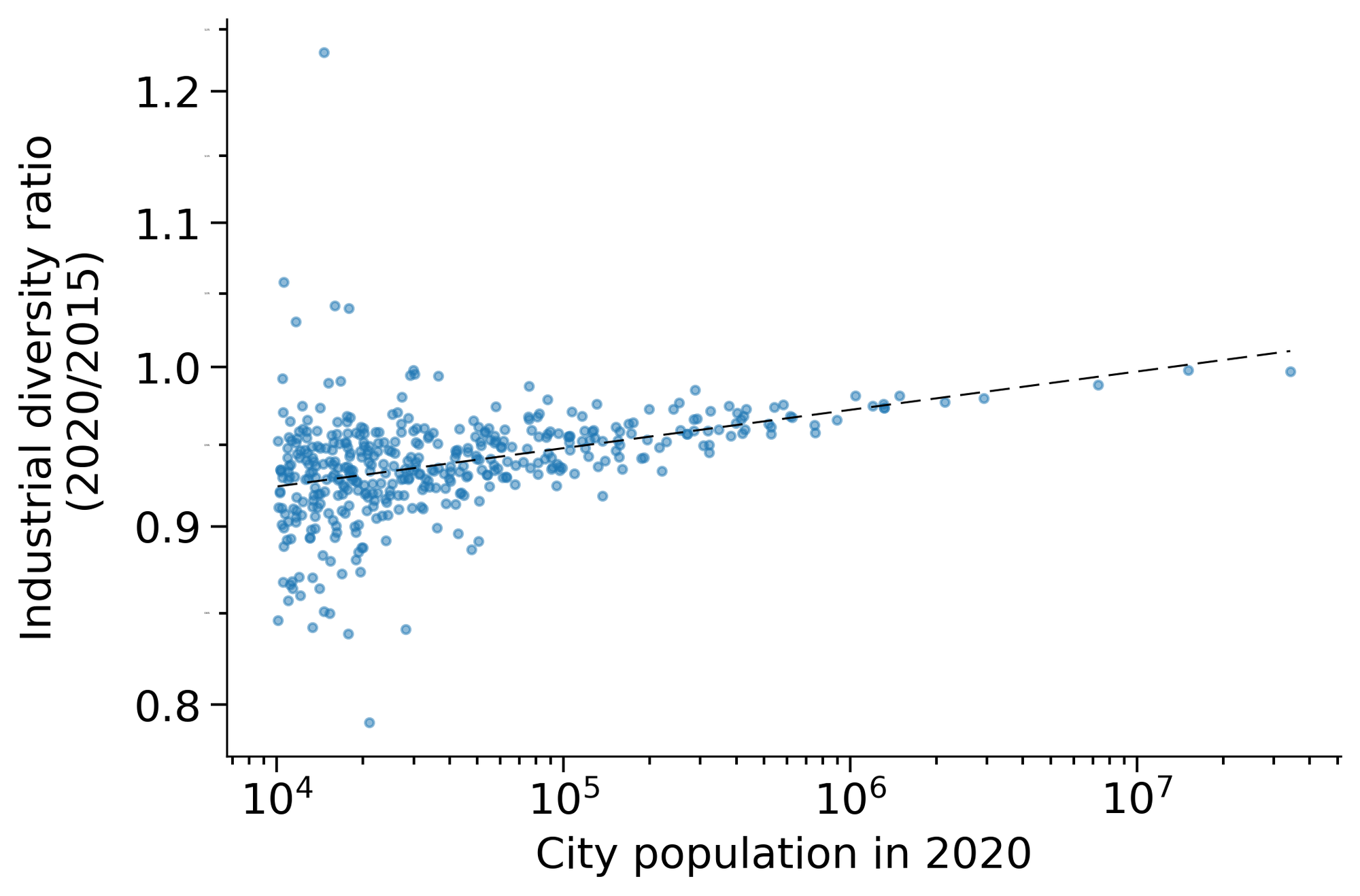}
         \caption{Change in the industrial diversity and the population size of a city}
         \caption*{\footnotesize\textit{Note}: Industrial diversity is defined as the number of industries with at least one establishment in a city, based on the highly disaggregated set of 1,899 industries recorded in the NTT Townpage Database. For both 2015 and 2020, the population size of each urban agglomeration (UA) is calculated as the sum of populations in all grid cells within the 2020 UA polygon. The dashed line represents the fitted line from a log-linear regression of the growth in industrial diversity on city population size in 2020.}
         \label{fig:industrial-diversity}
    \end{minipage}

    \vspace{1cm}
 
    \begin{minipage}[c]{\textwidth}
        \centering
         \captionsetup{width=\linewidth} 
         \includegraphics[width=.65\textwidth]{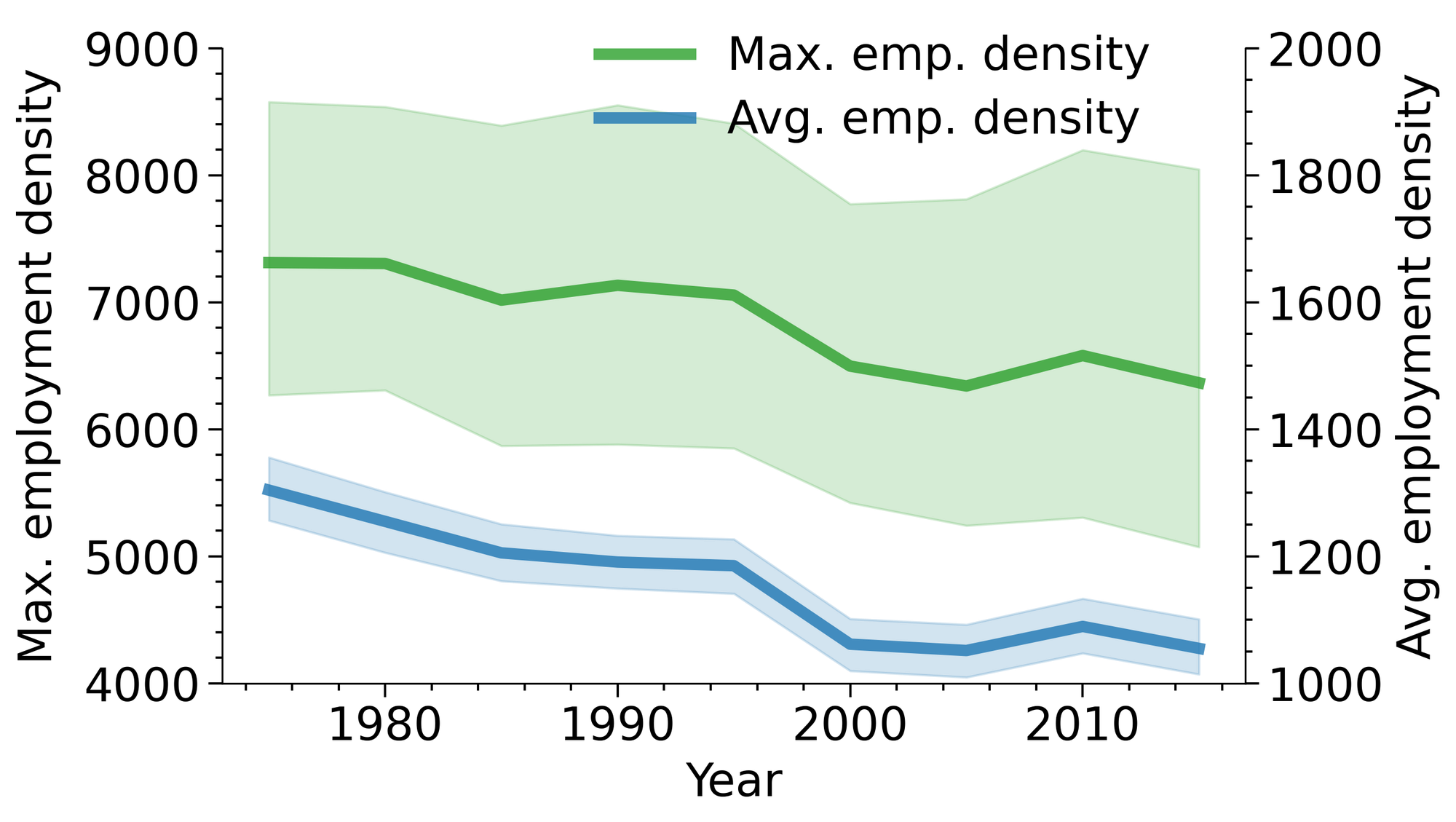}
         \caption{Dispersion of employment distribution at the city level}
         \caption*{\footnotesize\textit{Note}: The blue and green lines show the means of average and maximum employment densities of a city and the orange line the mean total area of a city for each year indicated along the horizontal axis, while the shaded area indicates the range covering 95\% of the values for individual cities. 
         Data: MIC \citeyearpar{ECensus-1975-2006,ECensus-2009-2016}.}
        
        \label{fig:employment-dispersion}
    \end{minipage}
\end{figure}

We saw in \cref{fig:global-concentration} that smaller cities became smaller while larger cities became larger in the recent past.
\cref{fig:industrial-diversity} compares the growth in industrial diversity from 2015 to 2020 with each city’s 2020 population size. 
Over this five-year period, smaller cities show a disproportionately large decline in industrial diversity.
The decline in population at the city level results not only in a decline in employment in a given industry (the intensive margin), but also in a decline in the range of job types available in terms of industrial diversity (the extensive margin).

\section{Spatial distribution of employment within a city\label{app:employment}}
Fig.\,\ref{fig:employment-dispersion} shows the maximum and average employment densities within a city for the period 1975--2015 MIC \citeyearpar{ECensus-1975-2006,ECensus-2009-2016}.
There is a clear tendency of flattening out of the employment distribution similar to that of the population distribution.

\section{Urban agglomerations\label{app:ua}}
Each \textit{Urban Agglomeration} (\textit{UA}) in a given time $t=\{1,\ldots,31\}$ (corresponding to years, $1970,1975,\ldots,2120$)  is identified as the set of 1km grid cells that (i) have a population density of at least 1,000 per square kilometer, (ii) are geographically contiguous, and (iii) have a total population of at least 10,000 in time $t$.

The set of UAs is updated every $t = \{1,\ldots,31\}$.
We keep track of each UA over time by assigning it a unique identifier (ID) throughout the study period as follows:
\begin{enumerate}[label=\arabic*.,parsep=0em]

    \item IDs of UAs in time $t = 1$ are set to be their city size ranking in $t$. For those with the same population size, UAs are ranked according to their average population density in descending order.

    \item A UA at time $t$ and a UA at time $t+1$ are considered the same if they share the largest population as of time $t$ in their areal intersection. (If there are multiple ties, the pair with the largest population density in their areal intersection takes the precedence.) In this case, their IDs at time $t+1$ are inherited from time $t$.
    
    \item If UA $i$ in time $t$ has the largest population at time $t$ in the areal intersection with UA $j$ in time $t+1$, and UA $j$ in time $t+1$ has the largest population in the areal intersection with a UA other than $i$ in time $t$, then UA $i$ is considered to be absorbed by UA $j$ in time $t+1$. 
    
    \item If UA $j$ in time $t+1$ has no intersection with any UA in time $t$, then UA $j$ is considered to be either a newly formed UA. 
    For a newly formed UA or a split UA with no predecessor in older time $<t$, a new ID is assigned in the descending order of their population size in time $t+1$ (in the case of a tie, the one with the highest average population density is assigned an ID first).
    
    \item If a UA is split from the existing UA at time $t+1$, but has a predecessor at time before $t$ (but not existing at time $t+1$), the ID of the latest predecessor is restored. 
    If there are multiple most recent predecessors, the one with the largest population in the areal intersection with the UA is chosen (again,  in the case of a tie, the predecessor with the largest average population density is chosen).
    Thus, a UA $i$ that was absorbed once by another UA $j$ and later split from UA $j$ will be named $i$ again.
\end{enumerate}

\section{Constraint least squares for the log-linear-in-time model \label{app:const-ls}}
The log-linear-in-time model for the observation period $t\in \{1,\ldots, 11\}$ is expressed in matrix notation as follows:
\begin{equation}
\mathbf{p}_u = \mathbf{X}_u \mathbf{a}_u + \bm{\epsilon}_u^{LL}, \hspace{1cm} 
\bm{\epsilon}_u^{LL} \sim N(\bm{0},\mathbf{V}_u),
\end{equation}
where
\begin{equation}
\mathbf{p} = \begin{bmatrix} P_{u,1} \\ \vdots \\ P_{u,11} \end{bmatrix}, \quad  \mathbf{X}_u = \begin{bmatrix} 1 &\mathrm{log}(1) \\ \vdots \\1 &\mathrm{log}(11) \end{bmatrix}, \quad
\mathbf{V}_u = \sigma^2_{u,\textit{LL}}\mathrm{diag}\left(\frac{1}{1^2}, \dots, \frac{1}{11^2} \right).
\end{equation}
The coefficients in $\mathbf{a}_u=[a_{0,u}^{LL},a_{1,u}^{LL}]^T$ are estimated under the constraint $\mathbf{x}_{u,11}^T \mathbf{a}_u= P_{u,11}$ where  $\mathbf{x}_{u,11}=[1,\mathrm{log}(11)]^T$ in year 2020 corresponding to $t=11$. 

Suppose that $\lambda$ is the Lagrange multiplier, the  least squares solutions $\hat{\mathbf{a}}_u=[\hat{a}_{0,u}^{LL},\hat{a}_{1,u}^{LL}]^T$ and $\hat{\lambda}$  are given by minimizing the following objective function:
\begin{equation}
L(\mathbf{a}_u, \lambda) = (\mathbf{p}_u - \mathbf{X}_u \mathbf{a}_u )^T \mathbf{V}_u (\mathbf{p}_u - \mathbf{X}_u \mathbf{a}_u) + 2\lambda(\mathbf{x}_{u,11}^T \mathbf{a}_u- P_{u,11}),
\end{equation}
Let us differentiate $L(\mathbf{a}_u, \lambda)$ with respect to $\mathbf{a}_u$ and $\lambda$. Then, we have
\begin{equation}    
\begin{split}    
\frac{\partial L(\mathbf{a}_u, \lambda)}{\partial \mathbf{a}_u} &= -2 \mathbf{X}_u ^T \mathbf{V}_u \mathbf{p}_u + 2 \mathbf{X}_u ^T \mathbf{V}_u \mathbf{X}_u \mathbf{a}_u + 2 \lambda \mathbf{x}^T_{u,11}. \\
\frac{\partial L(\mathbf{a}_u, \lambda)}{\partial \lambda} &= 2(\mathbf{x}^T_{u,11} \mathbf{a}_u - P_{u,t}).
\end{split}
\end{equation}
Setting these derivatives equal to zero gives the following equality:
\begin{equation}
\begin{split}    
\hat{\mathbf{a}}_u&=(\mathbf{X}_u ^T \mathbf{V}_u \mathbf{X}_u)^{-1}(\mathbf{X}_u ^T \mathbf{V}_u\mathbf{p}_u - \hat{\lambda} \mathbf{x}_{u,11}^T) \\
\mathbf{x}_{u,11}^T \hat{\mathbf{a}}_u &= P_{u,11}.
\end{split}    
\end{equation}
Substituting the first equation into the second equation,
\begin{equation}
\begin{split}    
\mathbf{x}_{u,11}^T (\mathbf{X}_u ^T \mathbf{V}_u \mathbf{X}_u)^{-1}(\mathbf{X}_u ^T \mathbf{V}_u\mathbf{y}_u - \hat{\lambda} \mathbf{x}_{u,11}^T) &= P_{u,11},\\
\hat{\lambda} \mathbf{x}_{u,11}^T (\mathbf{X}_u ^T \mathbf{V}_u \mathbf{X}_u)^{-1} \mathbf{x}_{u,11}&= \mathbf{x}_{u,11}^T  (\mathbf{X}_u ^T \mathbf{V}_u \mathbf{X}_u)^{-1}  \mathbf{X}_u^T \mathbf{p}_u -   P_{u,11},\\
\hat{\lambda} &= \frac{ \mathbf{x}_{u,11}^T (\mathbf{X}_u^T\mathbf{V}_u \mathbf{X}_u)^{-1} \mathbf{X}_u^T \mathbf{V}_u \mathbf{p}_u - P_{u,11} }{ \mathbf{x}_{u,11}^T (\mathbf{X}_u^T\mathbf{V}_u \mathbf{X}_u)^{-1} \mathbf{x}_{u,11} }.
\end{split}
\end{equation}
By substituting $\hat{\lambda}$, the constrained estimator $\hat{\mathbf{a}}_u$ is given by:
\begin{equation}    
\hat{\mathbf{a}}_u = \hat{\mathbf{a}}^0_u- (\mathbf{X}_u ^T \mathbf{V}_u \mathbf{X}_u)^{-1} \mathbf{x}_{u,11} \hat{\lambda},
\end{equation}
where $\hat{\mathbf{a}}^0_u= (\mathbf{X}_u ^T \mathbf{V}_u \mathbf{X}_u)^{-1}\mathbf{X}_u ^T \mathbf{V}_u \mathbf{p}_u$ is the conventional unconstrained estimator. The our log-linear-in-time model projects $P_{u,t+1}$ as $\hat{P}_{u,t+1} = \hat{a}_{0,u}^\textit{LL} + \hat{a}_{1,u}^\textit{LL} \log (t+1)$.

\section{Predictive variances}
\subsection{City-level model\label{app:variance-city}}

The predictive variance at time $t_0 = 12, \ldots, 31$ for the AR1, AR2 and LL models are given as follows. 
\begin{itemize}
    \item AR1: $V(\epsilon^{AR1}_{u,t_0}) = \hat{\sigma}_{u,AR1}^2\frac{1-\hat{\rho}_u^{2t_0}}{1-\hat{\rho}_u^2}$ where $\hat{\sigma}_{u,AR1}^2=\frac{\sum_{t=1}^{11}(\Delta P_{u,t+1}-\hat{\rho}_u \Delta P_{u,t})^2}{10-1}$.
    
    \item AR2: $V(\epsilon^{AR2}_{u,t_0}) = \hat{\sigma}^2_{u,AR2} \sum_{t=12}^{t_0} \psi_{u,t}^2$ where $\psi_{u,t}^2$ represents the impulse response depending on $\hat{\rho}_{1,u}$ and $\hat{\rho}_{2,u}$, and $\hat{\sigma}_{u,AR2}^2=\frac{\sum_{t=1}^{11}(\Delta P_{u,t+1}-\hat{\rho}_{1,u} \Delta P_{u,t} -\hat{\rho}_{2,u} \Delta P_{u,t-1}  )^2}{9-1}$.
    
    \item LL: $V(\epsilon^{LL}_{t_0}) = \hat{\sigma}_{u,LL}^2 \left( 1 + x'_{t0} (X'WX)^{-1} x_{t_0} \right)$ where $X$ is a matrix whose $u$-th row equals $[1,\log t]$, $x_{t_0}'=[1, \log t_0]$, $W$ is a diagonal matrix whose elements are $1,\cdots,11$, and $\hat{\sigma}_{u,LL}^2=\frac{\sum_{t=1}^{11}(P_{u,t}-\hat{a}_{u,0}^{LL} -\hat{a}_{u,1}^{LL}\log t)^2}{11-1}$.    
\end{itemize}

\subsection{The country and city level ensembling\label{app:variance-pl}}
The variance $V[\hat{P}^{PL}_{u,t+1}]$ at time $t+1$ is evaluated by using the following bootstrap procedure:
\begin{enumerate}
    \item Iterate the following procedure for each $l\in \{1,\cdots,200\}$:
\begin{enumerate}
    \item Sample $A^{(l)}_t \sim N(\hat{A}_t, V[\hat{A}_t])$ and $B^{(l)}_t \sim N(\hat{B}_t, V[\hat{B}_t])$ for each $t \in \{1,\cdots,11\}$ independently.
  
    \item Fit eq.\,\eqref{eq:PL_A} on  $A^{(l)}_t$ and estimate $\hat{a}^{A(l)}_1$ and $V[\hat{a}^{A(l)}_1]$. Estimate $\hat{a}^{B(l)}_1$ and $V[\hat{a}^{B(l)}_1]$ in the same manner using $B^{(l)}_t$.
    
    \item Sample $a^{A(l)}_1 \sim N(\hat{a}^{A(l)}_1, V[\hat{a}^{A(l)}_1])$ and $a^{B(l)}_t \sim N(\hat{a}^{B(l)}_1, V[\hat{a}^{B(l)}_1])$.
    
    \item Sample $P^l_{u,t} \sim N(\hat{P}_{u,t},V[\hat{P}_{u,t}])$ in which $\hat{P}_{u,t}$ and $V[\hat{P}_{u,t}]$ are evaluated using eqs.\,\eqref{eq:PL_TS_avg} and \eqref{eq:V_PL_TS_avg}.
    \item Evaluate $P^{PL(l)}_{u,t+1}$ by substituting $a^{A(l)}_1,a^{B(l)}_1,P^l_{u,t}$  into eq.\,\eqref{eq:condexp}.
\end{enumerate}
    \item Evaluate $V[\hat{P}^{PL}_{u,t+1}]=\frac{\sum^{200}_{l=1}(P^{PL(l)}_{u,t+1}-\bar{P}^{PL}_{u,t})^2}{200-1}$ where $\bar{P}^{PL}_{u,t}$ is the mean of the 200 ensembles.
\end{enumerate}

\subsection{Grid-level model for the neighboring populations\label{app:variance-grid}}
The variances for $\textit{ARI}1_N,\textit{ARI}2_N, \textit{LL}_N$, the variances are evaluated in the following steps.
    \begin{enumerate}
       \item Project the population $q_{i,t}$ in the neighborhoods by applying the \textit{ARI1, ARI2, LL} models until 2120, and obtain the expectation $\hat{q}_{i,t}$ and variance $V[\hat{q}_{i,t}]$ following the conventional procedures. 
       \item Fit equation (\ref{eq:8nn}) using the population data up to 2020, and obtain the estimate $\hat{b}_i$ of the coefficient and variance $V[\hat{b}_i]$ , as well as the error variance $\hat{s}^2$. 
       \item For $t \in \{2025, \ldots, 2120\}$, evaluate the predictive variance $V[\hat{p}_{i,t}]=V[\hat{q}_{i,t}\hat{b}_{i}]+\hat{s}^2$ where $V[\hat{q}_{i,t}\hat{b}_i]=V[\hat{q}_{i,t}]V[\hat{b}_i]+\hat{q}_{i,t}^2 V[\hat{b}_i] + V[\hat{q}_{i,t}]\hat{b}_i^2$ whose elements were evaluate in Steps 1 and 2.
    \end{enumerate}

\section{Smoothing of city boundaries\label{app:smoothing}}
The projected population $\hat{p}_{i,t+1}$ may be smoothed to avoid discontinuity at the city borders. 
The smoothing can be done by taking the weighted average of the projected population $\hat{p}_{i,t+1}$ and the average of $\hat{p}_{i,t+1}$ and the population $\hat{q}_{i,t+1}$ in the neighboring grids. 
The weight for $\hat{p}_{t+1,i}$ is given by $w^{ARI1}_{i,t}+w^{ARI2}_{i,t}+w^{LL}_{i,t}$  while the weight for $\hat{q}_{i,t+1}$ is given by  $w^{ARI1_N}_{i,t}+w^{ARI2_N}_{i,t}+w^{LL_N}_{i,t}$ which are defined in Section \ref{sec:grid-level}.

Although we have attempted to project with smoothing, the projected populations tend to be overly smoothed, resulting in a large gap between the observed population in 2020 and the projected population after 2025. Since the population is assumed to decrease in the target period, population smoothing, which is intended to model urban expansion, would be less important. For these reasons, we decided not to use smoothing.

Using this procedure, the weights of the 8 models are given by grids. However, since only 12 observations ($1970,1975,\ldots,2020$) are available in each grid, the weight estimation can be unstable. To stabilize the weights and the subsequent population projection, we used the weight averaged across the grids (i.e., for each year, each model has only one weight) in each year.

\section{Model performance\label{app:performance}}

Using 1970--2015 data, we estimate the evolution of the coefficients of the power law model using \cref{eq:PL,eq:PL_A,eq:PL_B}, the city and grid level time series model \cref{eq:ensemble_city,eq:ensemble_grid} to predict the population distribution across cities and grid cells in 2020.
In Sections \ref{app:performance-city} and \ref{app:performance-grid} we evaluate the performance of our model at the city and grid levels, respectively.

\subsection{The city level}
\label{app:performance-city}
\begin{figure}[htbp!]
    \centering
    \captionsetup{width=\linewidth}

    \includegraphics[width=\textwidth]{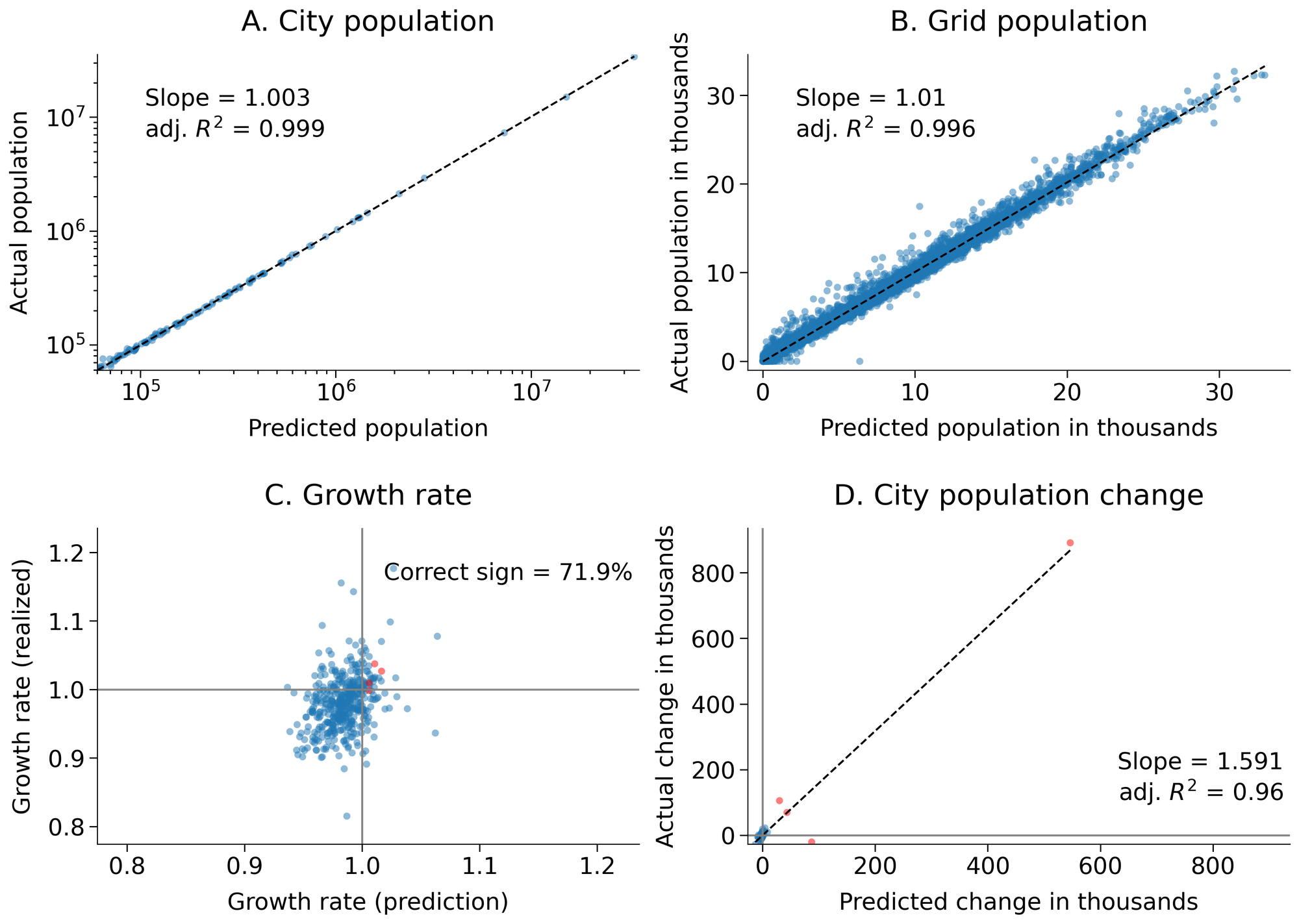}
     \caption{Model performance for the prediction of city size in 2020 from the 1970--2015 data (full model)}
     \caption*{\footnotesize\textit{Note}: (A) Actual versus predicted city size in 2020. (B) Actual versus predicted grid-cell population. (C) Actual versus predicted population growth rates for cities. (D) Actual versus predicted changes in city population}
     \label{fig:model-fit}
\end{figure}

\cref{fig:model-fit} summarizes the model performance at the city level.
Not surprisingly, the model fit with respect to the population size of each city as well as the population size of each grid cell is almost perfect for this short period of prediction, as shown in \cref{fig:model-fit}A and B, respectively. Thus, here we consider the model fit in terms of the non-targeted moments, the population growth rates, and the population growth of individual cities.

\cref{fig:model-fit}C and D compare the actual versus predicted growth rates and differences in population size of cities, respectively.
The model predicts the growth patterns of cities reasonably well, with the predicted signs of growth being correct for 72\% of all 427 cities identified in both 2015 and 2020. 
The model performs well in predicting the actual spatial redistribution of population, in particular the change in the population of Tokyo (an outlier in \cref{fig:model-fit}D), which accounts for most of the migration over the 5-year period.
The results for the four largest cities are shown in red in both panels, C and D.
The predictions are successful except for Osaka, the second largest city.
This is mainly because we have only a few time points to capture the declining trend of Osaka, which deviates from the power law.
As we can see in our projection (\cref{fig:large-city-growth}), however, once we include the data in 2020 for the model estimation, Osaka's recent declining trend relative to other large cities are captured.

\begin{figure}[htbp!]
     \centering
     \captionsetup{width=\linewidth}
     
     \includegraphics[width=\textwidth]{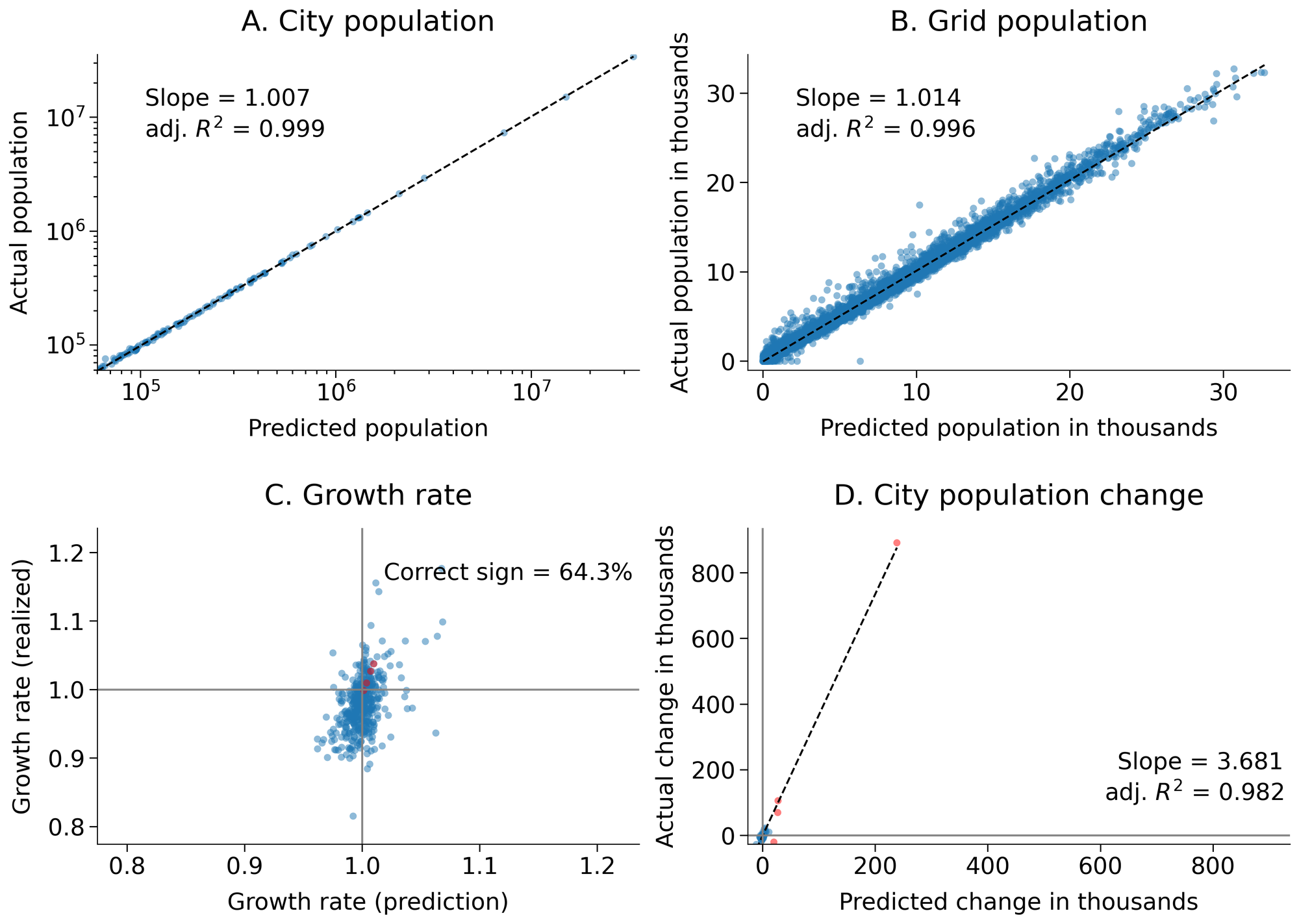}
     \caption{Model performance for the prediction of city size in 2020 from the 1970--2015 data (time-series model)}
     \caption*{\footnotesize\textit{Note}: The results in this figure are based on the prediction using the time series model only. (A) Actual versus predicted city size in 2020. (B) Actual versus predicted grid-cell population. (C) Actual versus predicted population growth rates for cities. (D) Actual versus predicted changes in city population}
     \label{fig:model-fit-ts-only}
\end{figure}

To gauge the importance of the PL model, \cref{fig:model-fit-ts-only} shows the model performance for predicting the 2020 city sizes using the 1970--2015 data using only TS model.
Again, the overall fit for the city and grid level population sizes is high (\cref{fig:model-fit-ts-only}A and B).
However, the performance of the models deteriorates significantly, as the predicted signs of population growth of cities are correct for only 64\% (\cref{fig:model-fit-ts-only}C).
Moreover, the time-series models alone cannot reasonably predict the actual population migration towards the largest cities, and the predictions generally underestimate the relative growth of the largest cities (\cref{fig:model-fit-ts-only}D).
That is, even in the 5 years from 2015, the concentration that takes place at the country level is better captured by the PL model.

\begin{figure}[htbp!]
     \centering
     \captionsetup{width=\linewidth}
     \includegraphics[width=\textwidth]{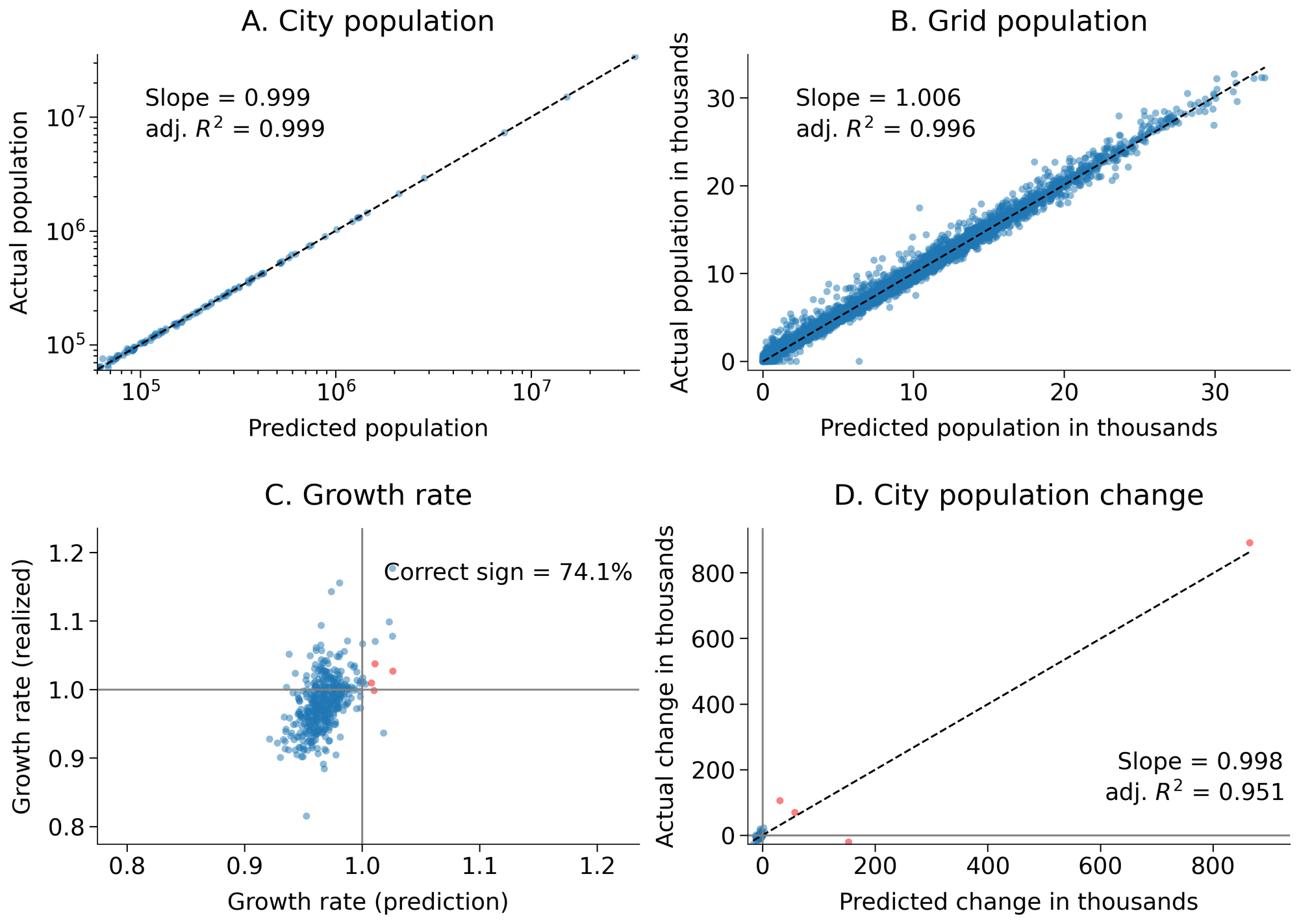}
     \caption{Model performance for the prediction of city size in 2020 from the 1970--2015 data (power-law model)}
     \caption*{\footnotesize\textit{Note}: The results in this figure are based on the prediction using the power-law model only. (A) Actual versus predicted city size in 2020. (B) Actual versus predicted grid-cell population. (C) Actual versus predicted population growth rates for cities. (D) Actual versus predicted changes in city population}
     \label{fig:model-fit-pl-only}
\end{figure}

\cref{fig:model-fit-pl-only} shows the model performance using only the PL model. 
The overall model fit is actually better than the full model shown in \cref {fig:model-fit}.
In addition to the targeted moments, the city and grid-level populations (panels A and B), the model predicts untargeted moments relatively well, correctly predicting 74\% of the signs of individual city population growth (panel C). Migration to Tokyo is also well captured (panel D).

The reason why the PL model outperforms the TS model in our validation is most likely that the time series trend of the total population has changed only in recent years (after 2010), which cannot be captured well by the TS model. 
On the other hand, the recent acceleration of concentration from rural cities to Tokyo can be well captured by the PL model due to the decline of most rural cities and the substantial growth of Tokyo.

Nevertheless, we believe that the full model is more appropriate for long-term projection. In particular, under the PL model \eqref{eq:condexp}, larger cities grow faster when the city size distribution becomes more skewed toward larger cities (i.e., $\hat{a}^B_1 <0$), as predicted by past data.
However, it is important to predict the impact of recent heterogeneity in the growth process of individual cities on their future growth and decline.

\subsection{The grid level}
\label{app:performance-grid}
\begin{figure}[p]
 \centering
 \begin{minipage}[c]{\textwidth}
     \captionsetup{width=\linewidth}
     
     \includegraphics[width=\textwidth]{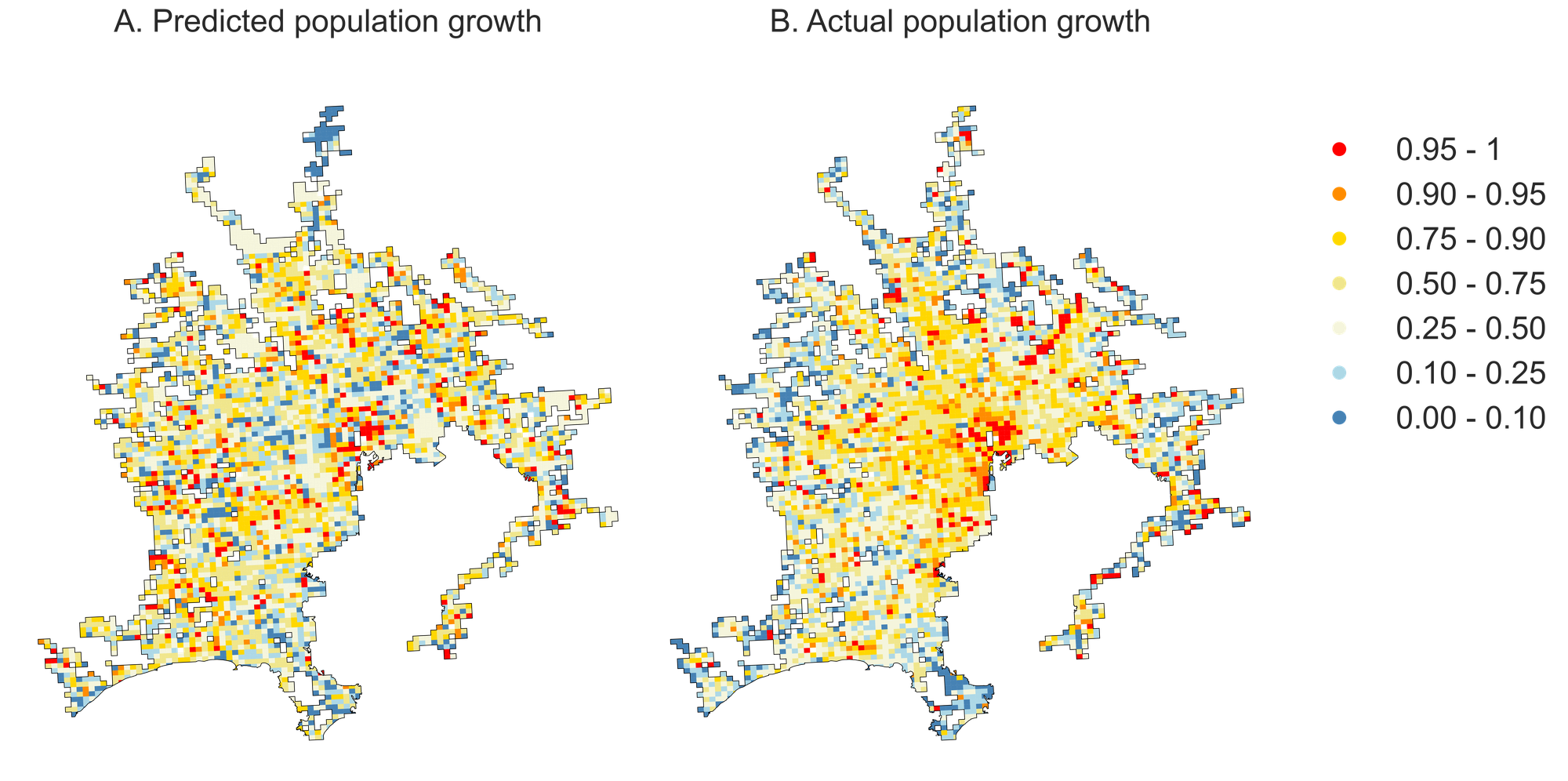}
     \caption{Predicted and actual growth of grid-cell population between 2015 and 2020 in Tokyo UA}
     \caption*{\footnotesize\textit{Note}: (A) and (B) show the percentiles of the predicted and the actual population growth rates between 2015 and 2020 of grid cells in grid cells in Tokyo UA, respectively.}
     \label{fig:valdation-density-ua1}
    \end{minipage}

    \vspace{1cm}
    
     \begin{minipage}[c]{\textwidth}
     \captionsetup{width=\linewidth}
     
     \includegraphics[width=\textwidth]{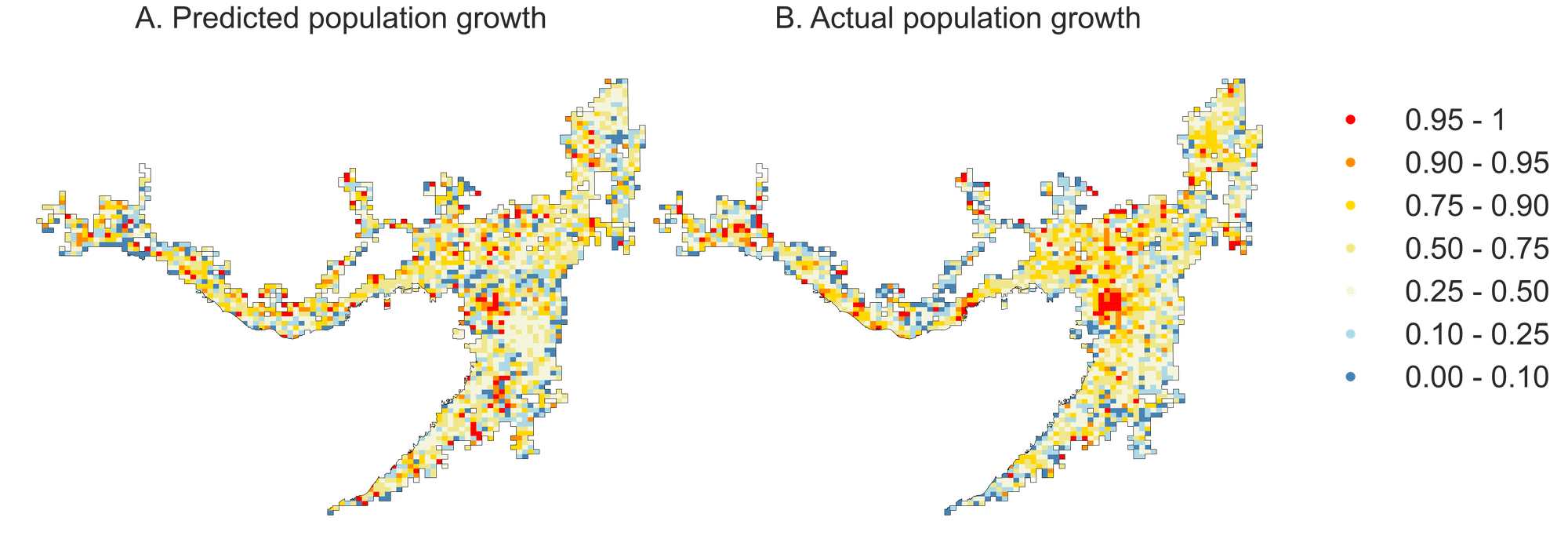}
     \caption{Predicted and actual growth of grid-cell population between 2015 and 2020 in Osaka UA}
     \caption*{\footnotesize\textit{Note}: (A) and (B) show the percentiles of the predicted and the actual population growth rates between 2015 and 2020 of grid cells in the 2nd largest Osaka UA, respectively.}
     \label{fig:valdation-density-ua2}
    \end{minipage}

\end{figure}

\begin{figure}[p]
 \centering

    \begin{minipage}[c]{\textwidth}
     
     \includegraphics[width=.95\textwidth]{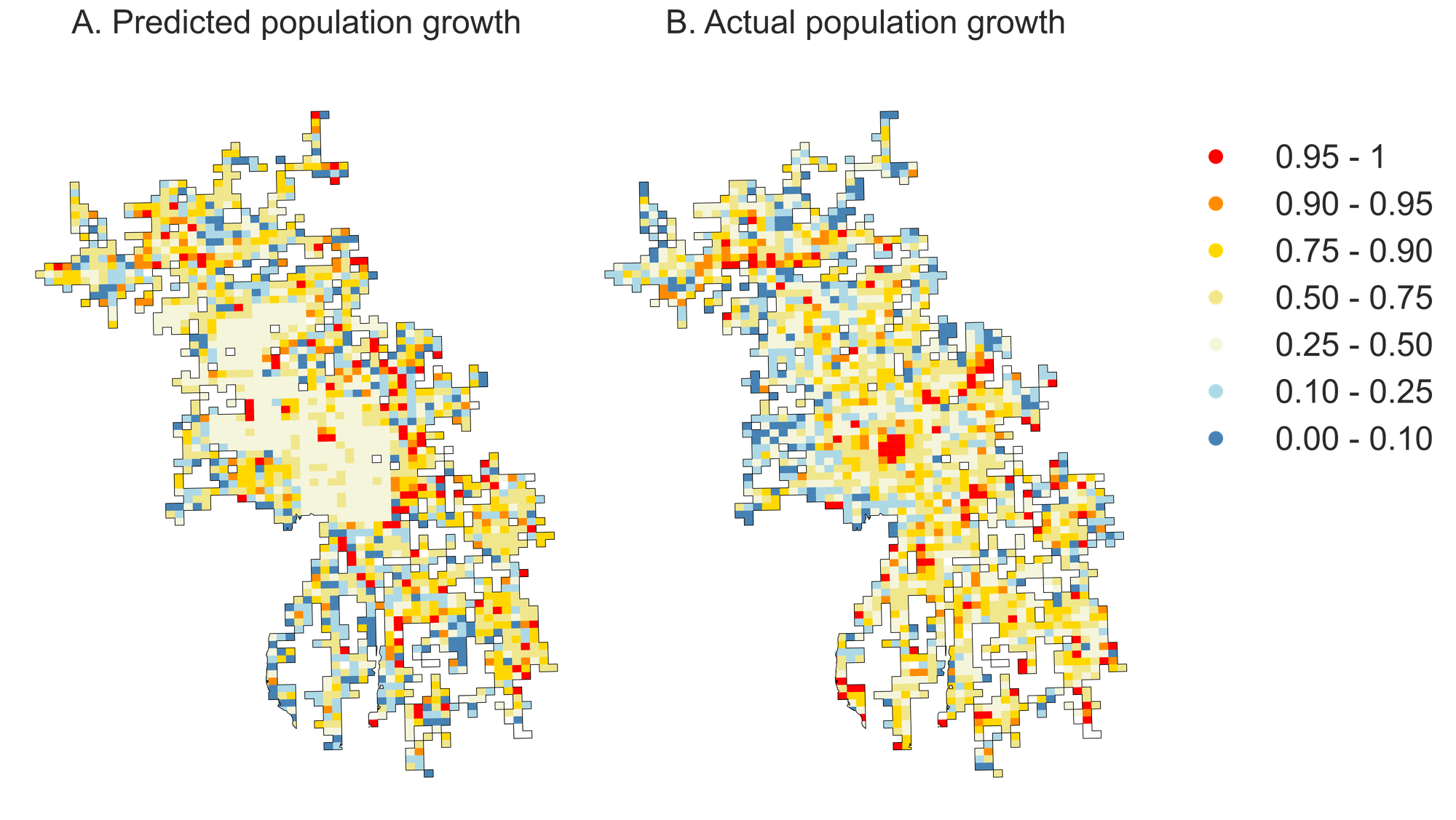}
     \caption{Predicted and actual growth of grid-cell population between 2015 and 2020 in Nagoya UA}
     \caption*{\footnotesize\textit{Note}: (A) and (B) show the percentiles of the predicted and the actual population growth rates between 2015 and 2020 of grid cells in the 3rd largest Nagoya UA, respectively.}
     \label{fig:valdation-density-ua3}
    \end{minipage}    

   \vspace{.5cm}
    
 \begin{minipage}[c]{\textwidth}
     \captionsetup{width=\linewidth}
     
     \includegraphics[width=.98\textwidth]{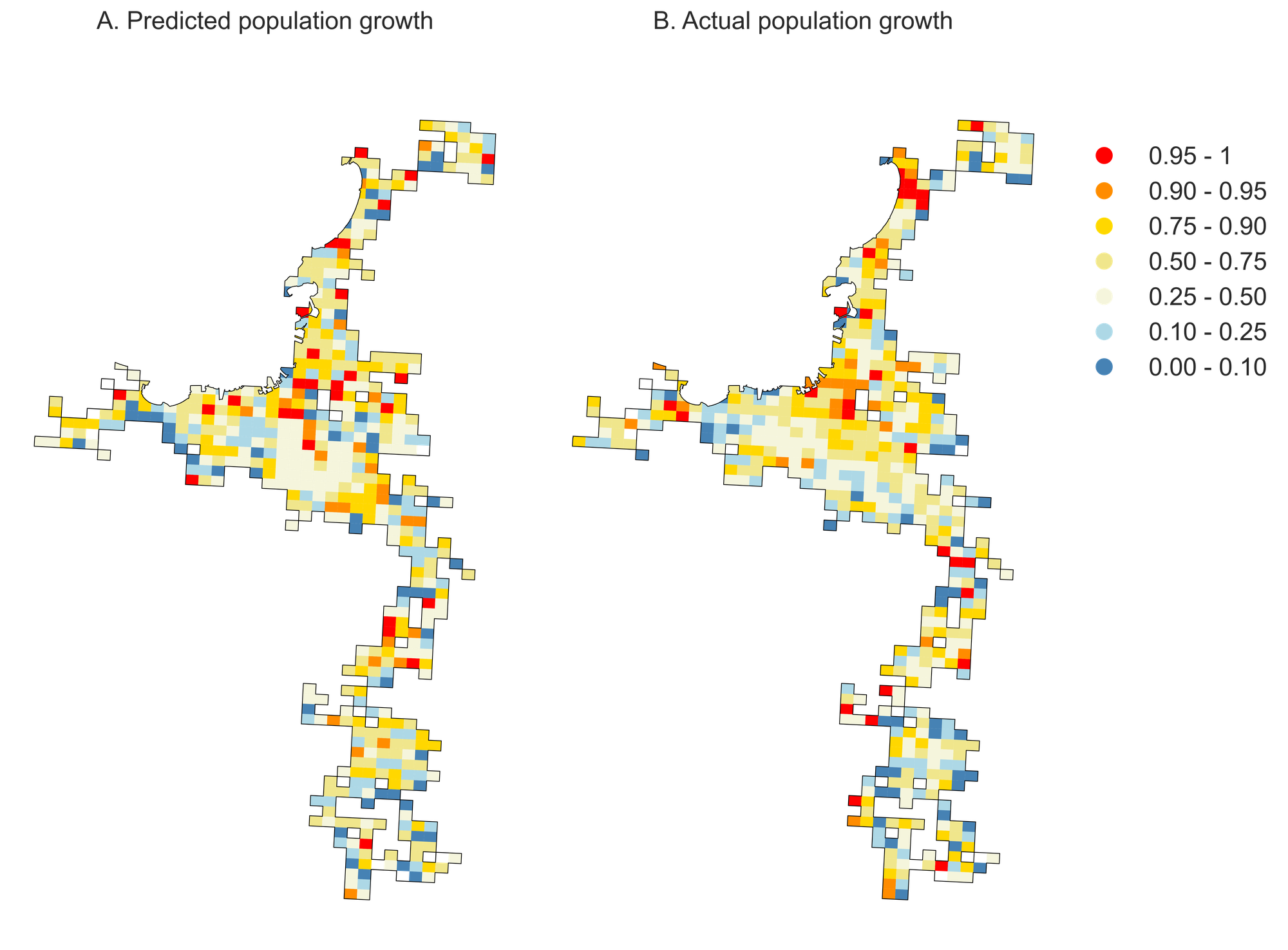}
     \caption{Predicted and actual growth of grid-cell population between 2015 and 2020 in Fukuoka UA}
     \caption*{\footnotesize\textit{Note}: (A) and (B) show the percentiles of the predicted and the actual population growth rates between 2015 and 2020 of grid cells in grid cells in the 4th largest Fukuoka UA, respectively.}
     \label{fig:valdation-density-ua5}
    \end{minipage}
\end{figure}

\begin{figure}[htbp!]
    
    \begin{minipage}[c]{\textwidth}
    \centering
    \captionsetup{width=\linewidth}
     
     \includegraphics[width=\textwidth]{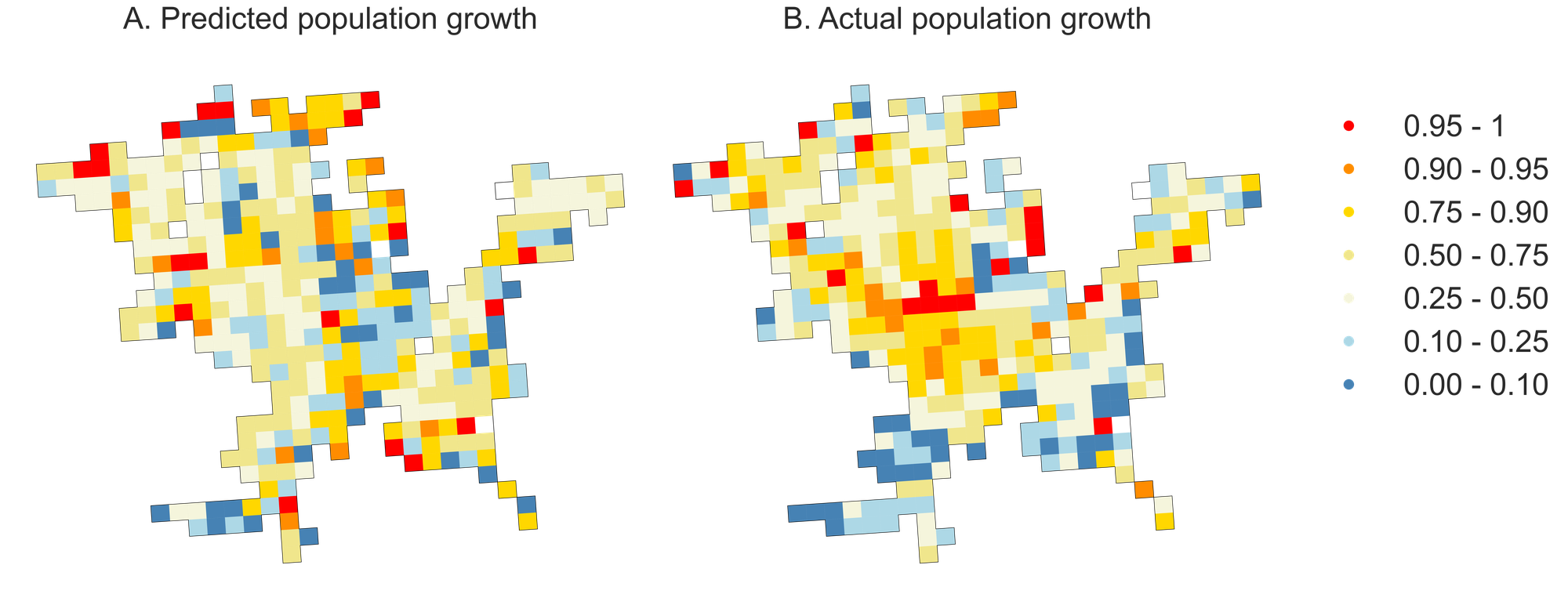}
     \caption{Predicted and actual growth of grid-cell population between 2015 and 2020 in Sapporo UA}
     \caption*{\footnotesize\textit{Note}: (A) and (B) show the percentiles of the predicted and the actual population growth rates between 2015 and 2020 of grid cells in the 5th largest Sapporo UA, respectively.}
     \label{fig:valdation-density-ua6}
    \end{minipage}

    \vspace{1cm}

    \begin{minipage}[c]{\textwidth}
    \centering
    \captionsetup{width=\linewidth}
     
     \includegraphics[width=\textwidth]{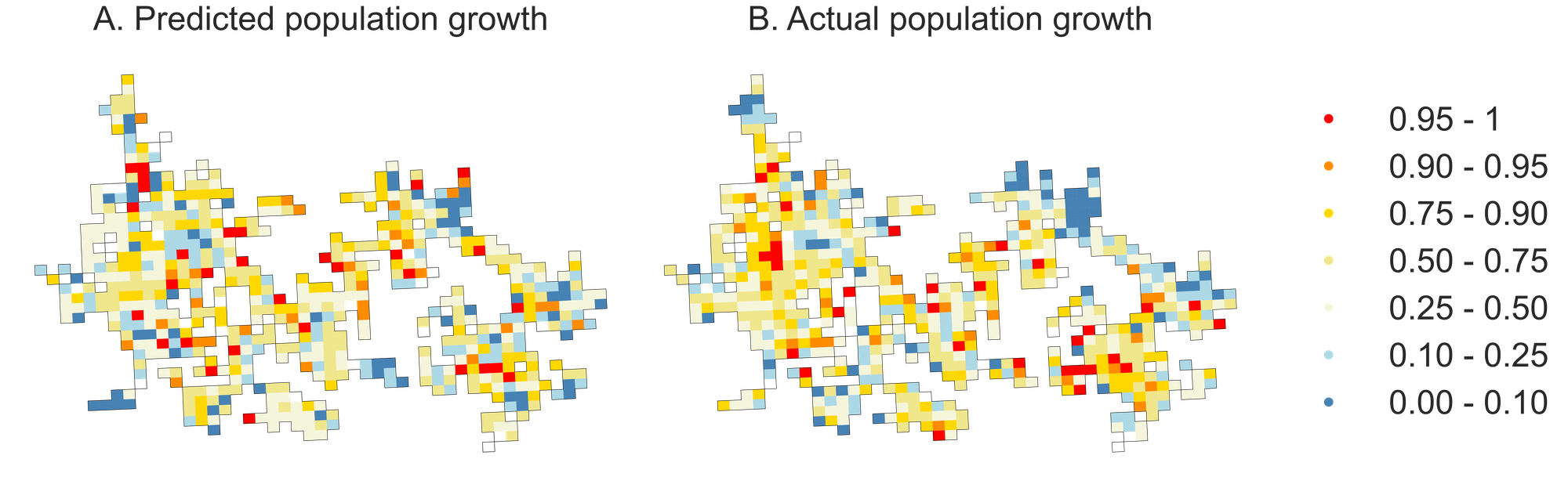}
     \caption{Predicted and actual growth of grid-cell population between 2015 and 2020 in Takasaki UA}
     \caption*{\footnotesize\textit{Note}: (A) and (B) show the percentiles of the predicted and the actual population growth rates between 2015 and 2020 of grid cells in the 6th largest Takasaki UA, respectively.}
     \label{fig:valdation-density-ua15}
    \end{minipage}    
\end{figure}

\begin{figure}[p]
 \centering
 \begin{minipage}[c]{\textwidth}
   \centering
     \captionsetup{width=\linewidth}
     
     \includegraphics[width=\textwidth]{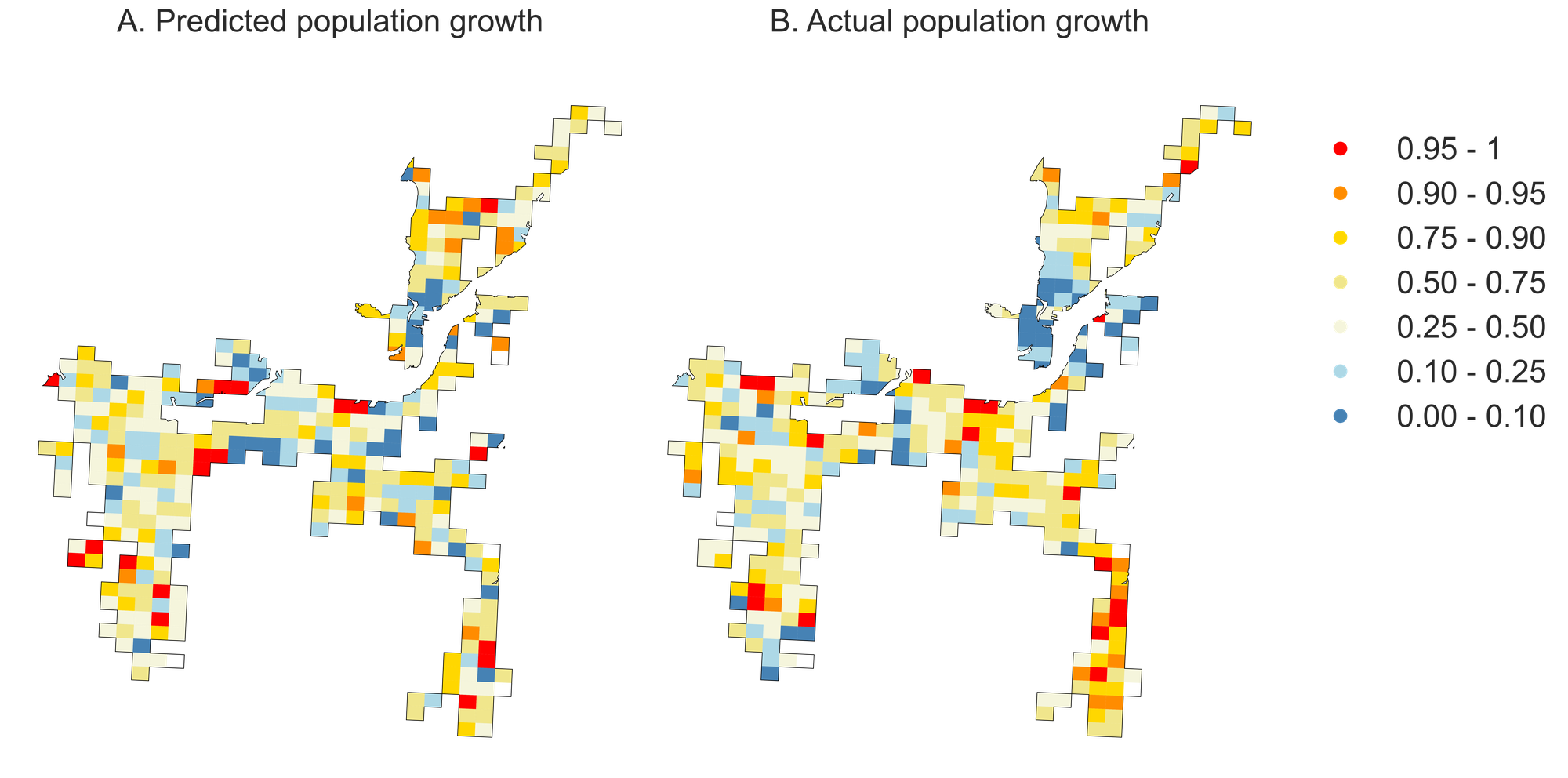}
     \caption{Predicted and actual growth of grid-cell population between 2015 and 2020 in Kitakyushu UA}
     \caption*{\footnotesize\textit{Note}: (A) and (B) show the percentiles of the predicted and the actual population growth rates between 2015 and 2020 of grid cells in grid cells in the 7th largest Kitakyushu UA, respectively.}
     \label{fig:valdation-density-ua4}
    \end{minipage}

    \vspace{1cm}
    
    \begin{minipage}[c]{\textwidth}
    \centering
    \captionsetup{width=\linewidth}
     
     \includegraphics[width=\textwidth]{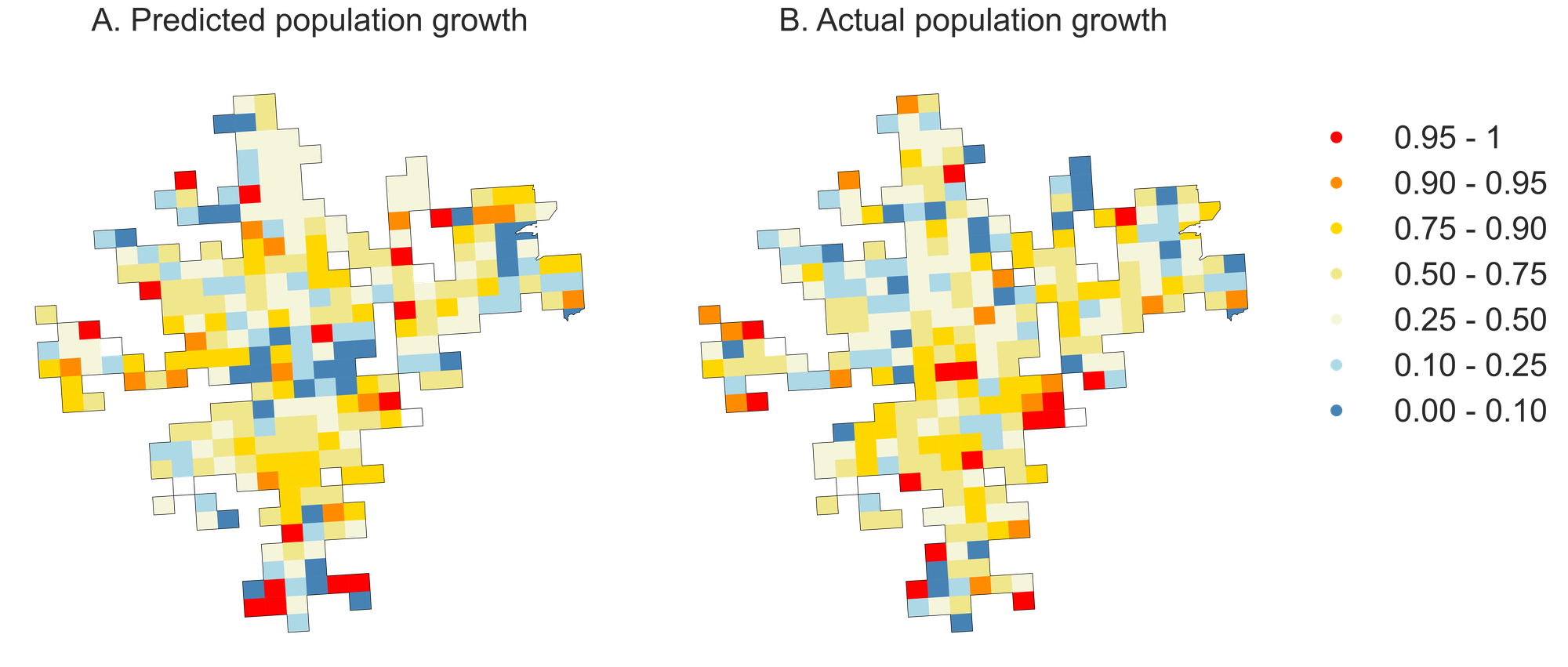}
     \caption{Predicted and actual growth of grid-cell population between 2015 and 2020 in Sendai UA}
     \caption*{\footnotesize\textit{Note}: (A) and (B) show the percentiles of the predicted and the actual population growth rates between 2015 and 2020 of grid cells in the 8th largest Sendai UA, respectively.}
     \label{fig:valdation-density-ua10}
    \end{minipage}

\end{figure}

\begin{figure}[htbp]
    \centering
    \captionsetup{width=\linewidth}

    \begin{minipage}[c]{\textwidth}
    \centering
    \captionsetup{width=\linewidth}
     
     \includegraphics[width=\textwidth]{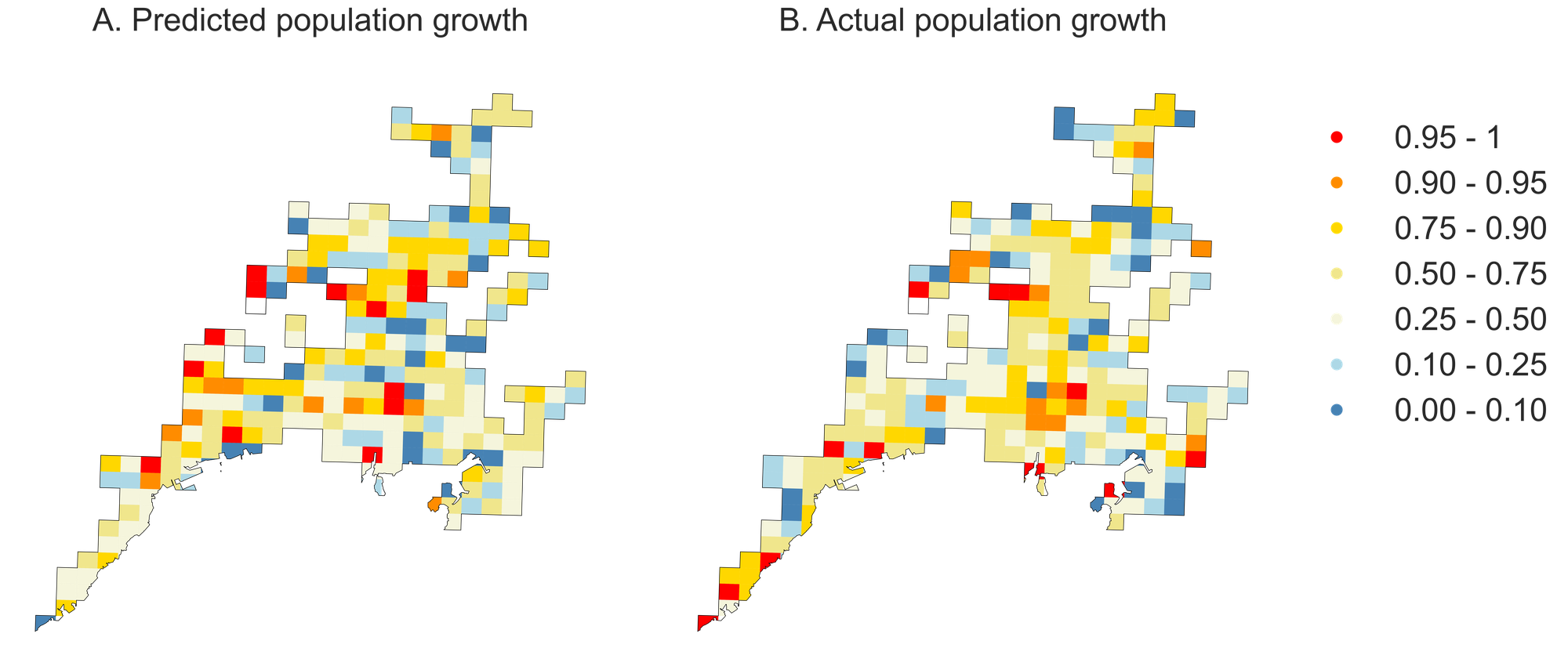}
     \caption{Predicted and actual growth of grid-cell population between 2015 and 2020 in Hiroshima UA}
     \caption*{\footnotesize\textit{Note}: (A) and (B) show the percentiles of the predicted and the actual population growth rates between 2015 and 2020 of grid cells in the 9th largest Hiroshima UA, respectively.}
     \label{fig:valdation-density-ua7}
    \end{minipage}    
    
    \vspace{1cm}

    \begin{minipage}[c]{\textwidth}
    \centering
    \captionsetup{width=\linewidth}

     \includegraphics[width=\textwidth]{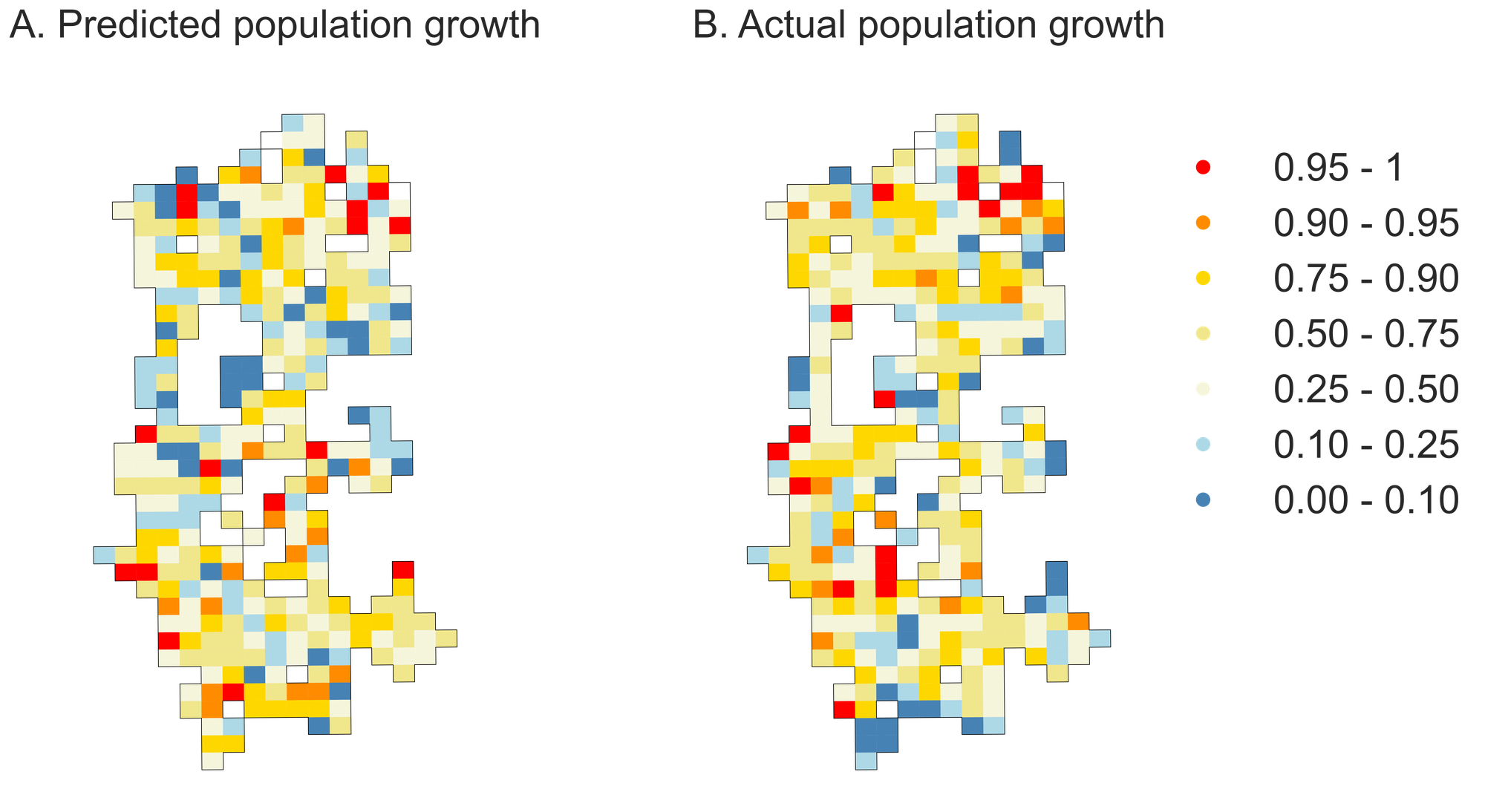}
     \caption{Predicted and actual growth of grid-cell population between 2015 and 2020 in Nara UA}
     \caption*{\footnotesize\textit{Note}: (A) and (B) show the percentiles of the predicted and the actual population growth rates between 2015 and 2020 of grid cells in grid cells in the 10th largest Nara UA, respectively.}
     \label{fig:valdation-density-ua19}
    \end{minipage}
     
\end{figure}

\cref{fig:valdation-density-ua1,fig:valdation-density-ua2,fig:valdation-density-ua3,fig:valdation-density-ua5,fig:valdation-density-ua6,fig:valdation-density-ua15,fig:valdation-density-ua4,fig:valdation-density-ua10,fig:valdation-density-ua7,fig:valdation-density-ua19} compare the predicted and actual population growth in each grid cell within the 10 largest cities in 2020.
While the overall model fit is reasonable as we saw in Fig.\,\ref{fig:model-fit}B, there are certain discrepancy between the actual and predicted populations at the grid level.

In general, the city centers in these largest cities have grown more between 2015 and 2020 than our projection.
For example, in Tokyo (Fig.\,\ref{fig:valdation-density-ua1}), there is a clear underprediction of population growth in around the CBD  (the mass of red area near the bay in Fig.\,\ref{fig:valdation-density-ua1}). 
Similar results are found in most of the other largest cities.
It reflects the recent boom in urban redevelopment, with the rush to build high-rise buildings in the urban centers associated with urban gentrification in Section \ref{sec:urban-redevelopment}.

Nevertheless, we do not consider the low model fit for these urban centers in the short term to be problematic for the long term projection, as these relatively recent urban redevelopment trends are unlikely to continue in the long term given the apparent population decline.


\section{Constrained versus unconstrained LL models}
\label{app:bias-correction}

In this section, we examine the effect of constraining the LL models for city and grid levels, eqs., \eqref{eq:LL_city} and \eqref{eq:LL_grid}, in Section \ref{sec:model} to match the projected population size of each city in 2020 to its actual size.
In our case, we estimated the LL models for the population size of each city and grid cell, as well as the 8NN of each grid cell, while constraining them to pass through the actual population size of these regional units in 2020.

\begin{figure}[htbp!]
 \centering

    \begin{minipage}[c]{\textwidth}
     \centering

    \captionsetup{width=\linewidth} 

     
     \includegraphics[width=\textwidth]{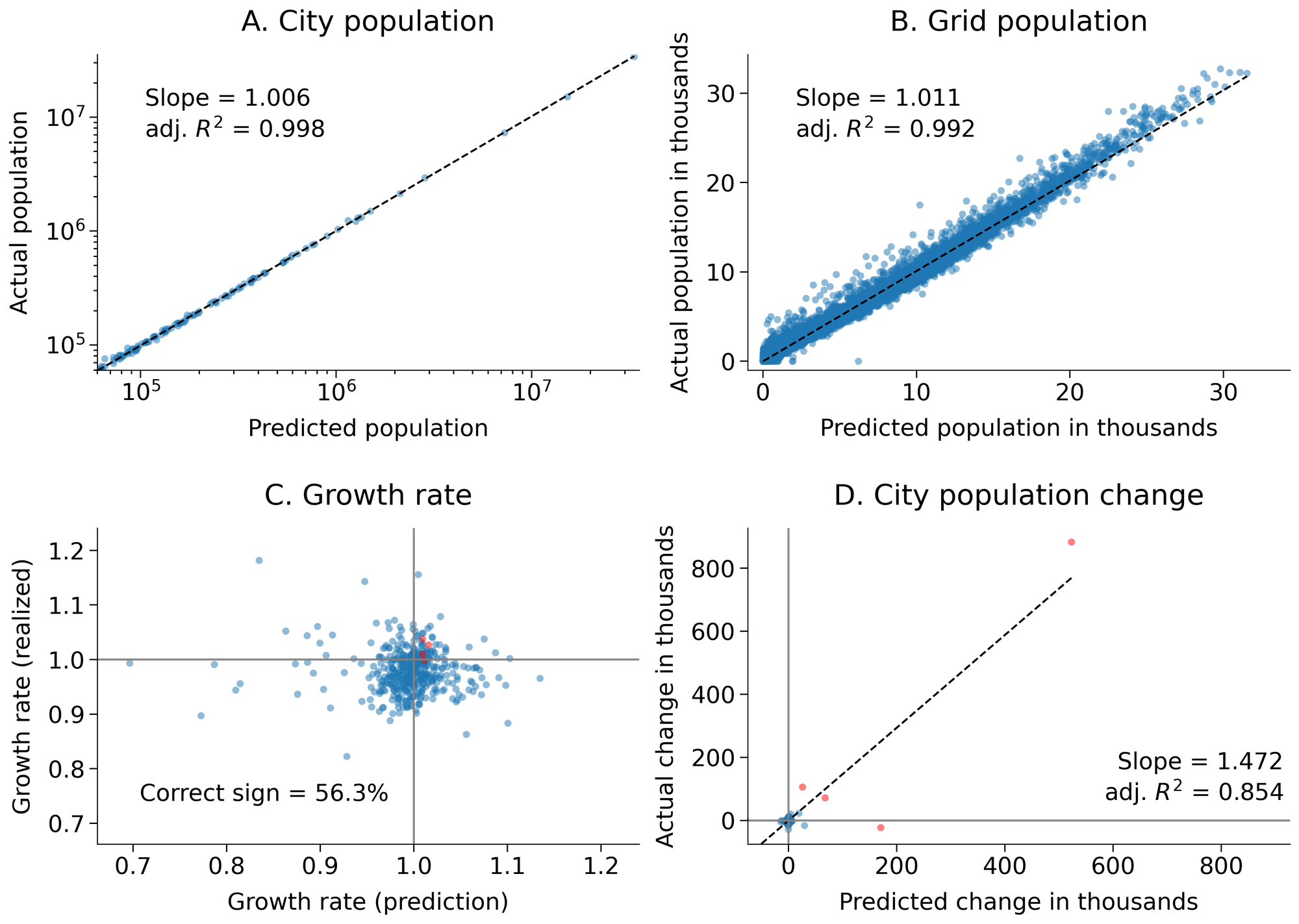}
     \caption{Model performance for the prediction of city size in 2020 from the 1970--2015 data (full model, unconstrained)}
     \caption*{\footnotesize\textit{Note}: The city- and grid-level LL models, as well as the grid-level 8NN LL models, are estimated without constraints to match the actual and projected population size of each city at the 2020 level. (A) Actual versus predicted city size in 2020. (B) Actual versus predicted grid-cell population. (C) Actual versus predicted population growth rates for cities. (D) Actual versus predicted changes in city population}
     \label{fig:model-fit-unconstrained}

    \end{minipage}

    \vspace{1cm}

    \begin{minipage}[c]{\textwidth}
     \centering

    \captionsetup{width=\linewidth}
    
        \includegraphics[width=0.6\textwidth]{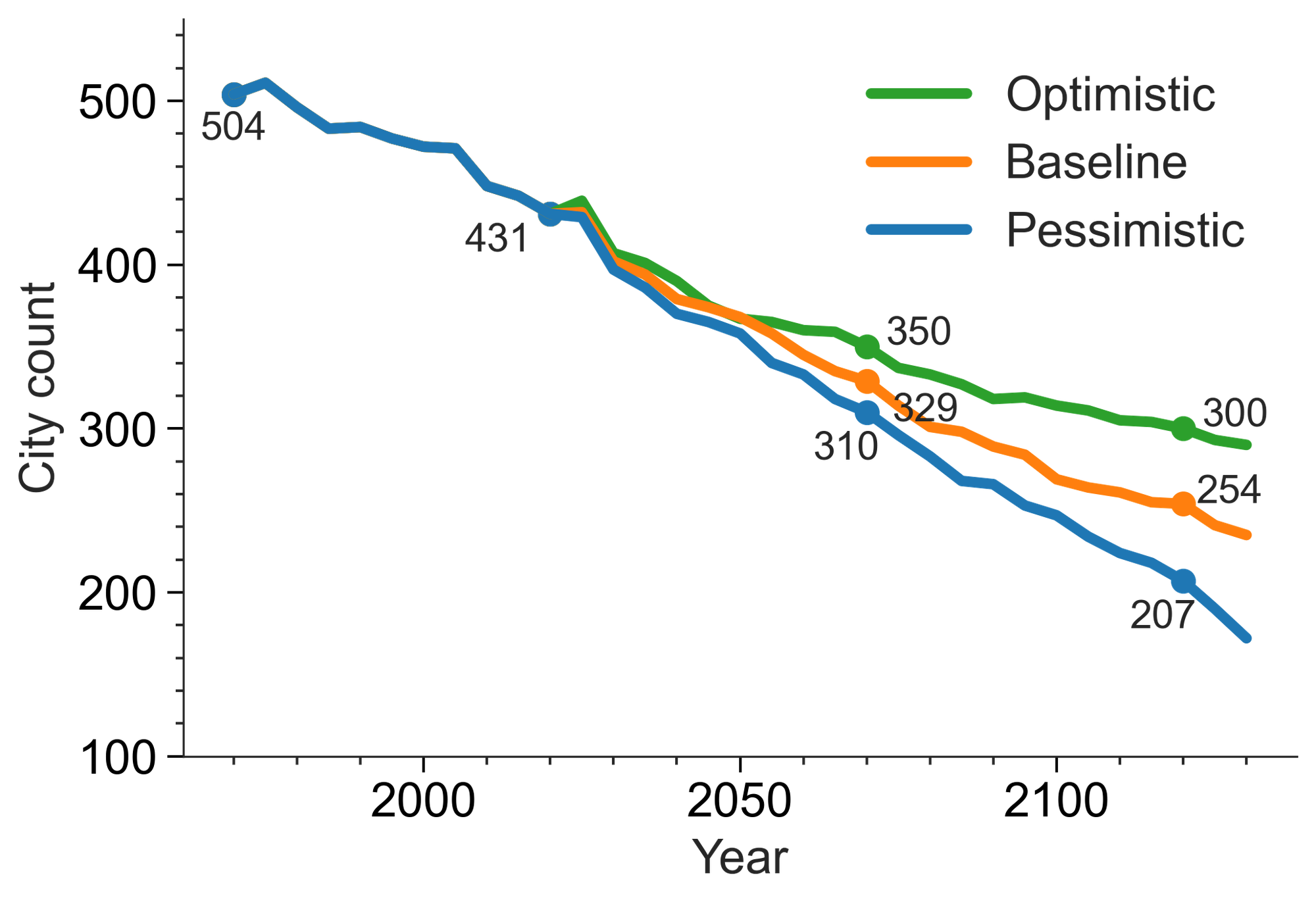}
        \bigskip
        
        \caption{City counts in 1970--2120 under the three scenarios of population decline in Japan (unconstrained model)}
        \caption*{\footnotesize\textit{Note}: The city counts in 1970--2020 are realized values, and those in 2025-2120 are our projections under three scenarios of population decline in Japan. The city- and grid-level LL models, as well as the grid-level 8NN LL models, are estimated without constraints to match the actual and projected population size of each city at the 2020 level.}
        \label{fig:city-count-unconstrained}    
    \end{minipage}
    \end{figure}

Fig.\,\ref{fig:model-fit-unconstrained} shows the validation result of the unconstrained model under the same condition considered in Appendix \ref{app:performance}.
While the model fit for the targeted moments (panels A and B) is reasonable, that for untargeted growth rates and population changes are not. 
In particular, the share of the sign agreement between the actual and predicted city population growth is 56\%, which is considerably worse than the constrained model (72\%). 
The unconstrained model overstates the population size of 80\% of cities, essentially rural cities, while correctly predicting the population growth of Tokyo. 
Consequently, the predicted number of surviving cities is uniformly larger under the unconstrained model (Fig.\,\ref{fig:city-count-unconstrained}) than the constrained model (Fig.\,\ref{fig:city-count}) in each of the future years.

\section{Model ensemble}

In this section we discuss the model ensemble.
As we have seen in Appendix \ref{app:performance}, even for the short-run projection, the PL model plays a crucial role when there is a significant change in the skewness of the city size distribution.
In the long run, while the time series models capture city-specific factors, their predictive variances, estimated from past observations, increase over years in the future for which no data are available. 
This characteristic makes them less suitable for long-term forecasting. 
In contrast, the PL model is less sensitive to such an influence of past observations because the relationship between population and city rank is stable over years due to the persistence of the power law for city size distribution. 
Thus, the PL model tends to be more reliable for long-term projections of city size.

\begin{figure}[htbp!]
 \centering
 \begin{minipage}[c]{\textwidth}
     
     \includegraphics[width=\textwidth]{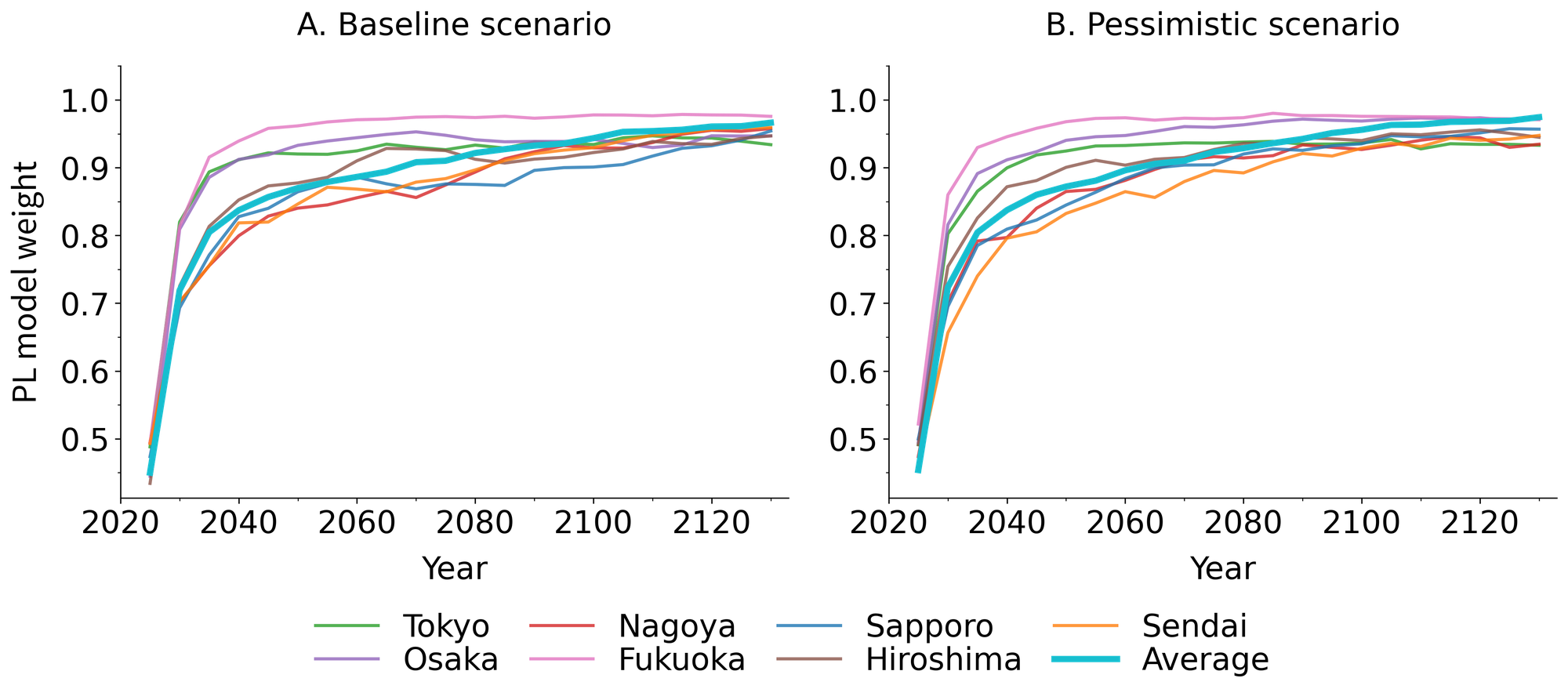}
     \caption{Weights of the power-law and time-series models for city size prediction}
     \caption*{\footnotesize\textit{Note}: (A) and (B) show under the baseline and pessimistic scenarios, respectively, the model weights for ensembling the power-law and time-series models for the predicting city sizes in the future. 
     Each panel shows the model weights for the power law model for the largest cities in the seven traditional regional divisions shown in Fig.\,\ref{fig:7regions} and the mean weights for all the cities in each year in 2025--2120.}
     \label{fig:model-weights--ua}
    \end{minipage}

    \vspace{1.5cm}
    
    \begin{minipage}[c]{\textwidth}
     
     \includegraphics[width=\textwidth]{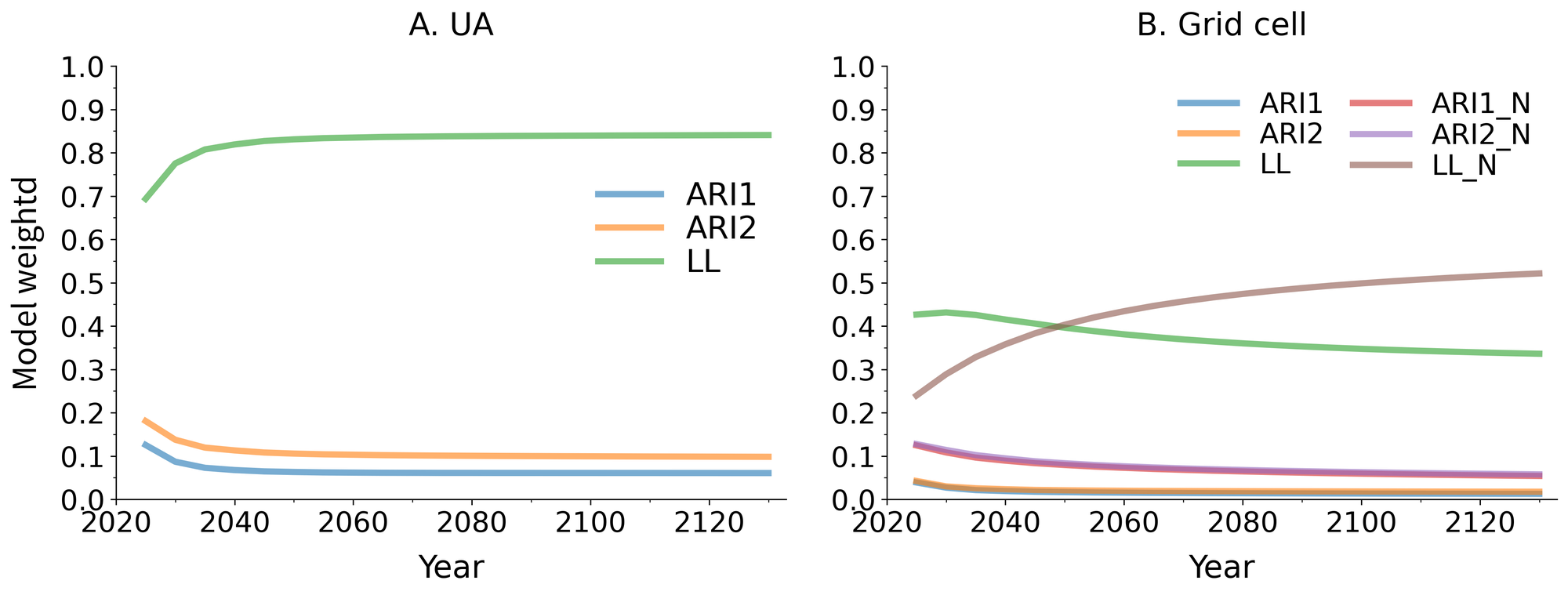}
     \caption{Weights for the time series models for the prediction of the grid cell and city population size in 2025--2120.}
     \caption*{\footnotesize\textit{Note}: (A) Average weights for the six time series models for the prediction of grid cell population. (B) Average weights for the three time series models for the prediction of city population size.}
     \label{fig:model-weights-ts}
     \end{minipage}
\end{figure}

\cref{fig:model-weights--ua}A and B show the weights of the PL model in \cref{eq:ensemble_city} at the city level in the baseline and pessimistic scenarios, respectively, for the largest cities in the seven regional divisions (\cref{fig:7regions}) and the average weights for all cities over the years 2020-2120.
As expected, Fig.\,\ref{fig:model-weights--ua} shows that the weight of the PL model is increasing over years. 

\cref{fig:model-weights-ts}A shows the weights of the three time series models at the city level. 
The LL model performs better than ARI1 and ARI2. 
This may be due to the long term trend or inertia in urban population change. 
In other words, ARI1, which only considers short-term dependency, is relatively less accurate in our case. For a similar reason, LL, which implicitly models the long-term trend according to a log-linear function, is more accurate than ARI2. As shown in \cref{fig:model-weights-ts}B, the LL and LL\_N models at the grid level together have a weight comparable to the LL model at the city level.
Note also that the grid-level ARI1 and ARI1\_N models together, and the grid-level ARI2 and ARI2\_N models together, have weights comparable to the city-level ARI1 and ARI2 models, respectively.

\section{The distribution of population over grid cells\label{app:pop-dist}}
This section lists the web links to the maps of the population distributions over the grid cells in the study period, 1970-2120.%
\footnote{The 1970 map lacks data for part of Akita Prefecture in northern Honshu.}

\subsubsection*{The realized population distribution in the past}

\href{https://www.mori.kier.kyoto-u.ac.jp/data/forecast/Mori_Murakami/250301/fig_pop-dist_with_ua_intermediate_min-pop_100_1970_en.html}{1970},
\href{https://www.mori.kier.kyoto-u.ac.jp/data/forecast/Mori_Murakami/250301/fig_pop-dist_with_ua_intermediate_min-pop_100_1975_en.html}{1975},
\href{https://www.mori.kier.kyoto-u.ac.jp/data/forecast/Mori_Murakami/250301/fig_pop-dist_with_ua_intermediate_min-pop_100_1980_en.html}{1980},
\href{https://www.mori.kier.kyoto-u.ac.jp/data/forecast/Mori_Murakami/250301/fig_pop-dist_with_ua_intermediate_min-pop_100_1985_en.html}{1985},
\href{https://www.mori.kier.kyoto-u.ac.jp/data/forecast/Mori_Murakami/250301/fig_pop-dist_with_ua_intermediate_min-pop_100_1990_en.html}{1990},
\href{https://www.mori.kier.kyoto-u.ac.jp/data/forecast/Mori_Murakami/250301/fig_pop-dist_with_ua_intermediate_min-pop_100_1995_en.html}{1995},
\href{https://www.mori.kier.kyoto-u.ac.jp/data/forecast/Mori_Murakami/250301/fig_pop-dist_with_ua_intermediate_min-pop_100_2000_en.html}{2000},
\href{https://www.mori.kier.kyoto-u.ac.jp/data/forecast/Mori_Murakami/250301/fig_pop-dist_with_ua_intermediate_min-pop_100_2005_en.html}{2005},
\href{https://www.mori.kier.kyoto-u.ac.jp/data/forecast/Mori_Murakami/250301/fig_pop-dist_with_ua_intermediate_min-pop_100_2010_en.html}{2010},
\href{https://www.mori.kier.kyoto-u.ac.jp/data/forecast/Mori_Murakami/250301/fig_pop-dist_with_ua_intermediate_min-pop_100_2015_en.html}{2015},
\href{https://www.mori.kier.kyoto-u.ac.jp/data/forecast/Mori_Murakami/250301/fig_pop-dist_with_ua_intermediate_min-pop_100_2020_en.html}{2020}

\subsubsection*{The future projection under the baseline scenario}
\href{https://www.mori.kier.kyoto-u.ac.jp/data/forecast/Mori_Murakami/250301/fig_pop-dist_with_ua_intermediate_min-pop_100_2030_en.html}{2030},
\href{https://www.mori.kier.kyoto-u.ac.jp/data/forecast/Mori_Murakami/250301/fig_pop-dist_with_ua_intermediate_min-pop_100_2040_en.html}{2040},
\href{https://www.mori.kier.kyoto-u.ac.jp/data/forecast/Mori_Murakami/250301/fig_pop-dist_with_ua_intermediate_min-pop_100_2050_en.html}{2050},
\href{https://www.mori.kier.kyoto-u.ac.jp/data/forecast/Mori_Murakami/250301/fig_pop-dist_with_ua_intermediate_min-pop_100_2060_en.html}{2060},
\href{https://www.mori.kier.kyoto-u.ac.jp/data/forecast/Mori_Murakami/250301/fig_pop-dist_with_ua_intermediate_min-pop_100_2070_en.html}{2070},
\href{https://www.mori.kier.kyoto-u.ac.jp/data/forecast/Mori_Murakami/250301/fig_pop-dist_with_ua_intermediate_min-pop_100_2080_en.html}{2080},
\href{https://www.mori.kier.kyoto-u.ac.jp/data/forecast/Mori_Murakami/250301/fig_pop-dist_with_ua_intermediate_min-pop_100_2080_en.html}{2090},
\href{https://www.mori.kier.kyoto-u.ac.jp/data/forecast/Mori_Murakami/250301/fig_pop-dist_with_ua_intermediate_min-pop_100_2100_en.html}{2100},
\href{https://www.mori.kier.kyoto-u.ac.jp/data/forecast/Mori_Murakami/250301/fig_pop-dist_with_ua_intermediate_min-pop_100_2110_en.html}{2110},
\href{https://www.mori.kier.kyoto-u.ac.jp/data/forecast/Mori_Murakami/250301/fig_pop-dist_with_ua_intermediate_min-pop_100_2120_en.html}{2120}

\subsubsection*{The future projection under the pessimistic scenario}
\href{https://www.mori.kier.kyoto-u.ac.jp/data/forecast/Mori_Murakami/250301/fig_pop-dist_with_ua_pessimistic_min-pop_100_2030_en.html}{2030},
\href{https://www.mori.kier.kyoto-u.ac.jp/data/forecast/Mori_Murakami/250301/fig_pop-dist_with_ua_pessimistic_min-pop_100_2040_en.html}{2040},
\href{https://www.mori.kier.kyoto-u.ac.jp/data/forecast/Mori_Murakami/250301/fig_pop-dist_with_ua_pessimistic_min-pop_100_2050_en.html}{2050},
\href{https://www.mori.kier.kyoto-u.ac.jp/data/forecast/Mori_Murakami/250301/fig_pop-dist_with_ua_pessimistic_min-pop_100_2060_en.html}{2060},
\href{https://www.mori.kier.kyoto-u.ac.jp/data/forecast/Mori_Murakami/250301/fig_pop-dist_with_ua_pessimistic_min-pop_100_2070_en.html}{2070},
\href{https://www.mori.kier.kyoto-u.ac.jp/data/forecast/Mori_Murakami/250301/fig_pop-dist_with_ua_pessimistic_min-pop_100_2080_en.html}{2080},
\href{https://www.mori.kier.kyoto-u.ac.jp/data/forecast/Mori_Murakami/250301/fig_pop-dist_with_ua_pessimistic_min-pop_100_2080_en.html}{2090},
\href{https://www.mori.kier.kyoto-u.ac.jp/data/forecast/Mori_Murakami/250301/fig_pop-dist_with_ua_pessimistic_min-pop_100_2100_en.html}{2100},
\href{https://www.mori.kier.kyoto-u.ac.jp/data/forecast/Mori_Murakami/250301/fig_pop-dist_with_ua_pessimistic_min-pop_100_2110_en.html}{2110},
\href{https://www.mori.kier.kyoto-u.ac.jp/data/forecast/Mori_Murakami/250301/fig_pop-dist_with_ua_pessimistic_min-pop_100_2120_en.html}{2120}\\

%
%
\section{Projection using annual population data\label{app:proj-annual}}
In this section, we assess the reliability of our projection result, which is based on census populations that are only available for 5 years.
In particular, since the population decline is a recent phenomenon observed between 2010 and 2015 and between 2015 and 2020, our projection may be driven by some idiosyncratic factor observed at these few time points.

We adopt an alternative projection based on the annual municipal population data reported in the Basic Resident Register (BRR) published by the MIC of Japan (\citeyear{BRR}).
Since the foreign population is taken into account only after 2013, we use the BRR municipal population data between 2013 and 2020, which covers most of the declining phase of Japan's population.
If the main and BRR-based projections show a similar trend, our main result would be reliable.

For simplicity, the city boundaries at the time $t=11$ (year 2020) are used for all years.
The BRR-based population in city $u$ at time $t=1,\ldots,11$ is given by the total population of municipalities where more than half of the residents lived within the grid-based city.
The gridded population data is the same as the main projection.

Given these data, populations are projected at five-year intervals, modifying steps 0 - 7 in Section \ref{sec:procedure}.
The changes are as follows.
In Step 0, the parameters of the city-level model are estimated by fitting the city-level model to the BRR-based annual city populations.
In Step 2, a five-year-ahead projection is performed using the model to estimate $P^{TS}_{i,t+1}$.
In Step 3, the power law model is estimated using the BRR city population data, and another five-year-ahead projection is performed to obtain $\hat{P}^{TS}_{i,t+1}$.
The other parts are identical to the main projection.

\begin{figure}[h!]
 \centering
 \includegraphics[width=0.65\textwidth]{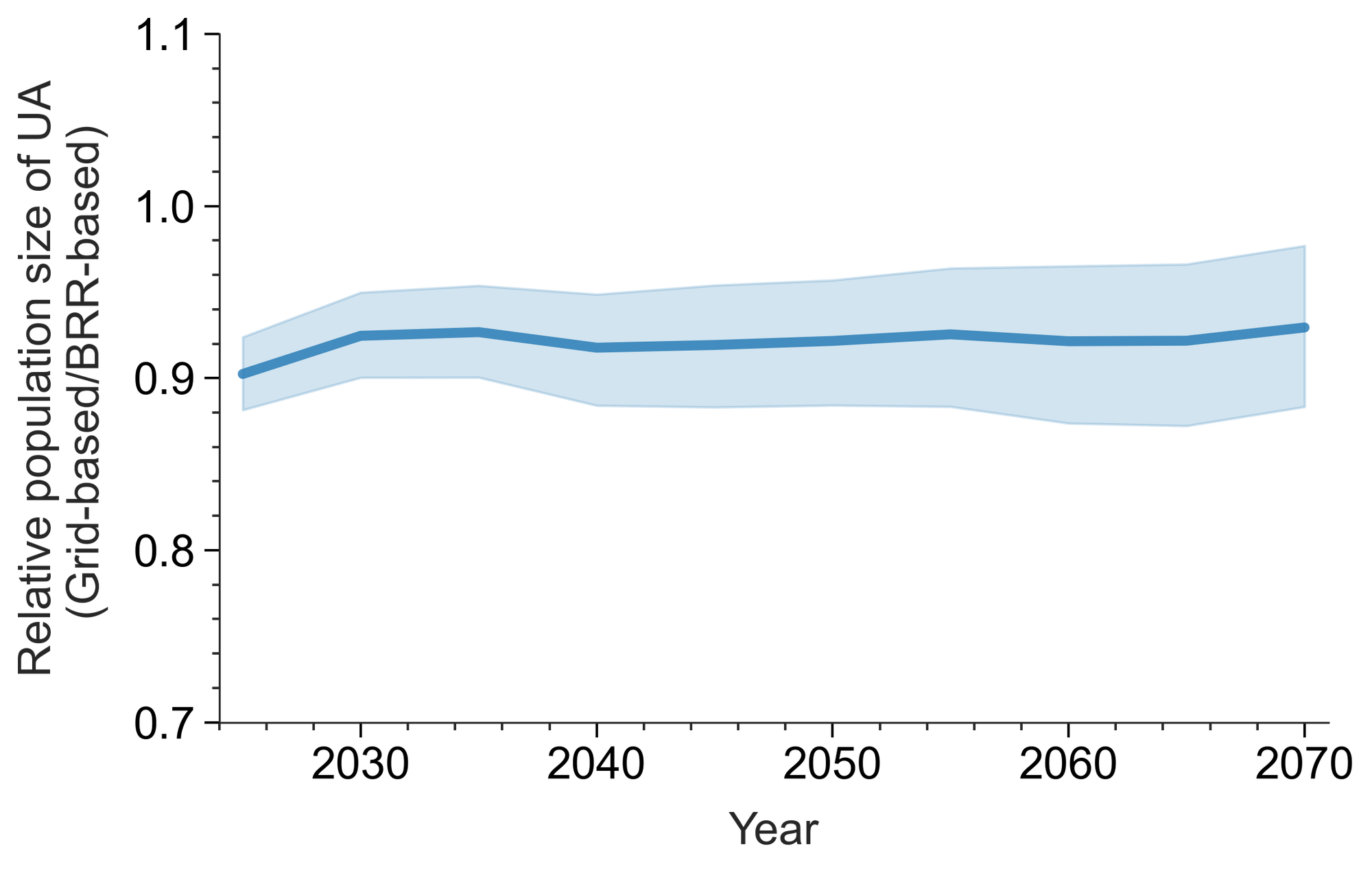}

 \caption{Population size of the grid-based city relative to the BRR-based city}
 \caption*{\footnotesize\textit{Note}: The solid line indicates the average of the ratio of the population size of the grid-based city to that of the BRR-based city under the baseline scenario for the largest 50 cities as of 2020. 
 The shaded area around the average indicates the 95\% range of the ratio among all cities.}
 \label{fig:bbr-ua}
\end{figure}

Fig.\,\ref{fig:bbr-ua} shows the ratio of the population size of the grid-based city to that of the BRR-based city for the largest 50 cities as of 2020, over the future 50 years, 2025--2070.
We focus here on the relatively large cities, as small cities tend to be consolidated into large cities under the BRR-based city definition, since the municipal polygons tend to include several small cities.

The grid-based cities are generally smaller than the BRR-based cities because the BRR-based cities tend to cover a larger geographic area by construction.
Otherwise, the population sizes of cities under these alternative city definitions are generally in good agreement.
We therefore conclude that the evolution of cities during the period of national population decline after 2010 is well captured by our
projection.

\section{Land price prediction\label{app:land-price}}
The land price $z_{i,t}$ of the center of the $i$-th grid at $t$-th year is projected until 2120 using the following log-linear additive model \citep{Wood-Book2017}:
\begin{equation}
    log (z_{i,t})=c_0 + c(i \in u) + c_1 log(\hat{p}_{i,t}) + c_2 log(\hat{q}_{i,t}) + f_{\tau_1}(x_{i,1})+ f_{\tau_2}(x_{i,2}) +\epsilon^z_{i,t} \hspace{0.5cm}\epsilon^z_{i,t}\sim N(0,\sigma_z^2),\label{eq:GAM}
\end{equation}
where $c_0$ is an intercept while $c(i \in u)$ denotes intercepts by urban area that is introduced to capture the heterogeneity across cities. $c_1$ and $c_2$ are coefficients estimating the influence of the logged population $\hat{p}_{i,t}$ and the logged neighboring population $\hat{q}_{i,t}$. $f_{\tau_1}(x_{i,1})$ is a two dimensional spline function of spatial coordinates $x_{i,1}$ estimating the latent map pattern. $f_{\tau_2}(x_{i,2})$ is another spline function estimating the relationship between elevation $x_{i,2}$ and population. The spline functions are assumed because spatial coordinates and elevation are likely to have non-linear associations with grid populations. The variance parameters $\tau_1, \tau_2, \sigma_z^2$ controls the amount of variations in each term. 

The model is estimated by fitting it on the residential land price data between 1985 - 2020 ($N=$316,171) provided by the NLNI. The fast marginal likelihood maximization method of \cite{Wood-JRSS2011} is used for the estimation. Then, land price by 2120 is projected by substituting the parameter estimates, projected populations, and other covariates into eq. \eqref{eq:GAM}.

The estimated coefficients are as follows: $\hat{c}_0=9.60, \hat{c}_1=9.38 \times 10^{-2}, \hat{c}_2 = 7.79 \times 10^{-5}$, which are all statistically significant at the 0.1 $\%$ level. These estimates confirmed that population growth in own and neighboring grids increases land prices. The adjusted $R^2$ value equals 0.744, meaning that this model explains 74.4 $\%$ of variations in populations.

\section{Fukuoka's Tenjin Big Bang area\label{app:tenjin}}

\begin{figure}[h!]
 \centering
 \captionsetup{width=\linewidth}
 
 \includegraphics[width=\textwidth]{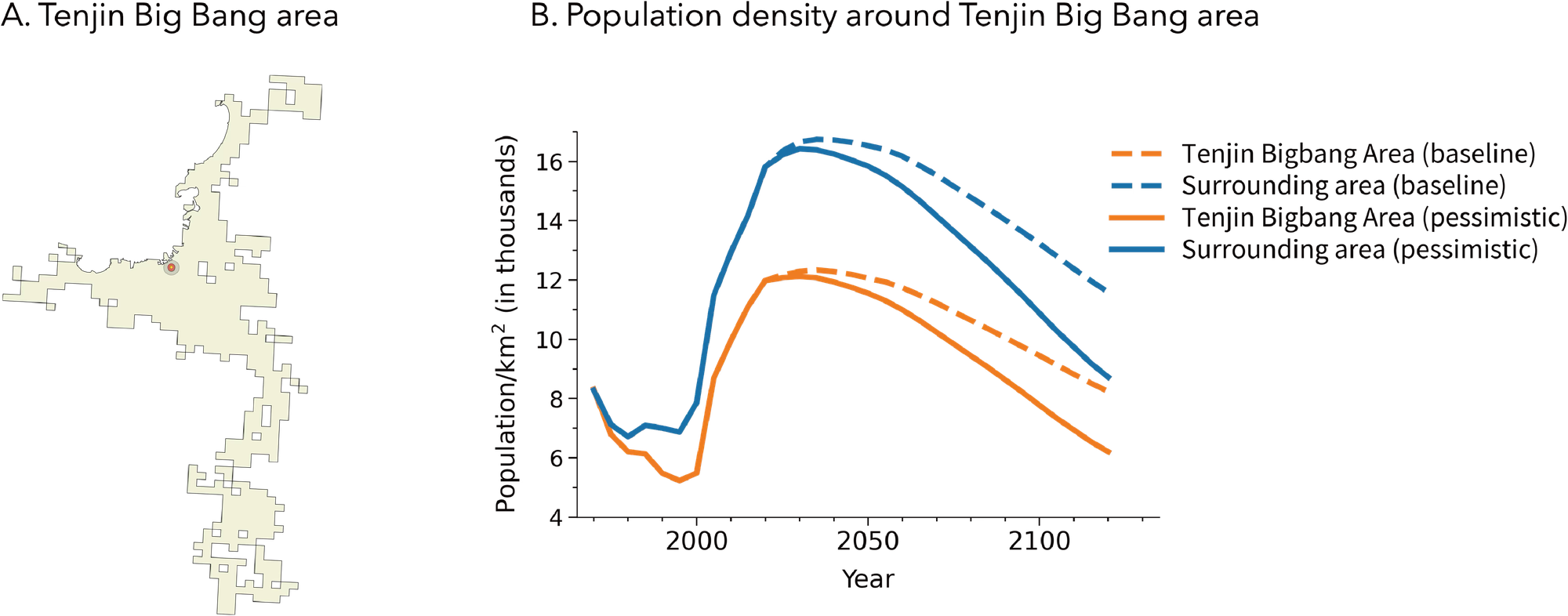}

 \caption{Fukuoka's Tenjin Big Bang area and the population density around it}
 \caption*{\footnotesize\textit{Note}:(A) The outer polygon is the spatial coverage of Fukuoka UA in 2020. The Tenjin Big Bang area of Fukuoka is the 500-meter radius red circle around the city center indicated by the yellow dot. The gray circle indicates the 500 to 1,000 radius circle around the city center. (B) Average population density in the Tenjin Big Bang area (the orange line) and that in the surrounding area (the blue line) in the baseline and pessimistic scenarios for 1970--2120.}
 \label{fig:tenjin-bigbang}
\end{figure}

Fukuoka is unique among the largest cities, as it has the major airport right at the city center.
Fukuoka Airport began operations as a public airport in 1972, and with the opening of the airport, there were restrictions on building heights in accordance with the Civil Aeronautics Law.
To support the high growth of the city, the “Tenjin Big Bang” redevelopment project is underway since 2015 with bold height deregulation that will allow construction of buildings over 100 meters high within a 500-meter radius (the red circle in  \cref{fig:tenjin-bigbang}A) of the city center (the yellow circle in \cref{fig:tenjin-bigbang}A).
It is expected to increase total floor space and employment by 1.7 times and 2.4 times, respectively, in 10 years.%
\footnote{Data source: \href{https://www.city.fukuoka.lg.jp/data/open/cnt/3/56223/1/tenjinbb0.pdf}{The official announcement by Fukuoka city government} on February 24, 2015.}\ 

Fig.\,\ref{fig:tenjin-bigbang}B shows the population density in the Big Bang area (the orange lines) and that in its surrounding area of the 500-to-1000-meter bound around the city center (the gray circle in Fig.\,\ref{fig:tenjin-bigbang}A).
Because of the building height restriction, the population density at the very center of the city is apparently lower than its immediate surrounding area \citep[][]{Nakajima-Takano-RSUE2023}.
But, the projected population densities in the Big Bang area in 2070 and 2120 are 94\% and 69\%, respectively, of the 2020 level in the baseline scenario, and 85\% and 52\%, respectively, in the pessimistic scenario.
While the Big Bang redevelopment may make the central area more resilient to the population decline than our projection, the population shrinkage seems too rapid to catch up.
The surrounding area of the Big Bang area also exhibits population decline than growth.
Their projected population densities in 2070 and 2120 are 98\% and 73\%, respectively, in the baseline scenario, and 90\% and 55\% in the pessimistic scenario.

\clearpage


\end{document}